\documentclass[psfig]{aa}                                                    
\usepackage{longtable}
\input psfig.tex

\begin{document}                                                                
\def\et{et al.}                                                                 
\def\egs{erg s$^{-1}$}                                                          
\def\egsc{erg s$^{-1}$ cm$^{-2}$}                                               
\def\msu{M$_{\odot}$\ }                                                         
\def\kms{km s$^{-1}$ }                                                          
\def\kmsM{km s$^{-1}$ Mpc$^{-1}$ }                                              
                                                                                
   \title{The ROSAT-ESO Flux Limited X-ray (REFLEX) Galaxy Cluster
   Survey. V. The cluster catalogue\thanks{
   Based on observations at the European Southern Observatory La Silla,
   Chile}}

   \titlerunning{The REFLEX Cluster Catalogue}

   \author{H. B\"ohringer\inst{1}, P. Schuecker\inst{1}, L. Guzzo\inst{2},
   C.A. Collins\inst{3}, W. Voges\inst{1}, R.G. Cruddace \inst{4},
   A. Ortiz-Gil\inst{5}, G. Chincarini \inst{2,6},
   S. De Grandi \inst{2}, A.C. Edge \inst{7}, 
   H.T. MacGillivray \inst{8}, D.M. Neumann \inst{9}, S. Schindler \inst{10},
   P. Shaver \inst{11}}

   \authorrunning{B\"ohringer et al.}
                                             
   \offprints{H. B\"ohringer \\ hxb@mpe.mpg.de}                                 
                                                                                
   \institute{$^1$ Max-Planck-Institut f\"ur extraterrestrische Physik,
                 D 85748 Garching, Germany\\
              $^2$ Osservatorio Astronomico di Brera, via Bianchi 46, I-22055 Merate, Italy\\
              $^3$ Astrophysics Research Institute, Liverpool John Moores University, Liverpool CH41 1LD,U.K.\\
              $^4$ E. O. Hulburt Center for Space Research, Naval Research Laboratory, Washington, DC 20375., USA\\
              $^5$ Observatorio Astron\'omico, Universidad de Valencia, 22085 Valencia, Spain\\
              $^6$ Dipartimento di Fisica, Universita degli Studi di Milano, Italy\\
              $^7$ Physics Department, University of Durham, South Road, Durham DH1 3LE, U.K.\\
              $^8$ Institute for Astronomy, University of Edinburgh, Blackford Hill, Edinburgh EH9 3HJ, U.K.\\
              $^9$ CEA Saclay, Service d`Astrophysique, Gif-sur-Yvette, France\\
              $^{10}$ Institute for Astrophysics, Universit\"at Innsbruck, 6020 Innsbruck, Austria\\
              $^{11}$ European Southern Observatory, D 85748 Garching, Germany }

   \date{Received .... ; accepted ....}
                                                                                
   \markboth {The REFLEX Cluster Catalogue}{}                                  
                                                                                
\abstract{We present the catalogue of the REFLEX Cluster Survey
providing information on the X-ray properties, redshifts, and some
identification details of the clusters in the REFLEX sample. The
catalogue describes a statistically complete X-ray flux-limited sample
of 447 galaxy clusters above an X-ray flux of $3\ 10^{-12}$ erg s$^{-1}$
cm$^{-2}$ (0.1 to 2.4 keV) in an area of 4.24 ster in the southern sky. 
The cluster candidates were first selected by their
X-ray emission in the ROSAT-All Sky Survey and subsequently spectroscopically
identified in the frame of an ESO key programme. Previously described tests 
have shown that the sample is more
than 90\% complete and there is a conservative upper limit of 9\% on the
fraction of clusters with a dominant X-ray contamination from AGN. 
In addition to the cluster
catalogue we also describe the complete selection criteria as a
function of the sky position and the conversion functions used
to analyse the X-ray data. These are essential for the precise
statistical analysis of the large-scale cluster distribution. 
This data set is at present the largest,
statistically complete X-ray galaxy cluster sample. 
Together with these data set we also provide for the first time 
the full three-dimensional selection function. The sample forms
the basis of several cosmological studies, one of the most
important applications being the assessment of the statistics of the
large-scale structure of the universe and the test of cosmological 
models. Part of these cosmological results have
already been published.
\thanks{The full versions of Tables 2 through 9 will be available
in electronic form at the CDS via anonymous ftp to cdsarc.u-strasbg.fr (130.79.128.5)
or via http://cdsweb.u-strasbg.fr/cgi-bin/qcat?J/A+A/ as well as on our home page
http://www.xray.mpe.mpg.de/theorie/REFLEX/DATA}
}
      
\maketitle
                                                                                
%
                                                                                
\section{Introduction}                         

Clusters of galaxies are the largest building blocks of our Universe that can
still reasonably well be characterized as unique objects. This makes
them on one hand very important large-scale astrophysical laboratories
in which a large variety of astrophysical processes can
be studied in well characterized environments. For these laboratories
we can measure for example their total gravitational
mass, their matter composition, the internal gas density, temperature,
and pressure of the intergalactic medium, their distance, and other
important properties. The best basis for such
astrophysical studies is a well documented catalogue of galaxy
clusters to choose the best suited objects for the prospective study
(e.g. B\"ohringer et al. 2001b).

On the other hand X-ray selected galaxy clusters are very good tracers
of the large-scale structure of the Universe. Since there is a quite
well understood relation between the distribution of galaxy clusters with
known mass and the dark matter distribution, the
statistics of the large-scale matter distribution in the Universe can
be derived from the distribution of clusters in a well selected, 
statistically complete sample. This study of the large-scale
structure was the main objective for the construction of the REFLEX
sample. Several results on the construction
of the sample (B\"ohringer et al. 2001a, Paper I), the assessment 
of the large-scale structure (Collins et
al. 2000, Paper II; Schuecker et al. 2001a, Paper III; B\"ohringer et
al. 2002, Paper IV; Schuecker et al. 2002, 2003a (Papers VI and VII), 2003b,
Kerscher et al. 2001), on the statistics of substructure in REFLEX
clusters Schuecker et al. (2001b), on the statistics of the
cluster galaxy velocity dispersions (Ortiz-Gil et al. 2003), and on
the X-ray temperatures of the most luminous, distant REFLEX clusters
(Zhang et al. 2003) have already been published. 
Several further papers are in preparation.

Due to the close correlation of X-ray luminosity and mass for clusters
of galaxies (e.g. Reiprich \& B\"ohringer 2002) the detection and selection 
of the sample clusters is currently best performed through the cluster
X-ray emission. The ROSAT All-Sky Survey (RASS), which is still the only all-sky
or wide-angle X-ray survey performed with an imaging X-ray telescope,
is by far the best basis for such cosmological studies. It has been
used previously in several projects to construct statistical galaxy
cluster samples (Pierre et al. 1994, Romer et al. 1994, Ebeling et
al. 1996, 1998, 2000, Burns et al. 1996, De Grandi et al. 1999, Ledlow
et al. 1999, B\"ohringer et al. 2000, Henry et al. 2001,
Cruddace et al. 2002, 2003, Ebeling et al. 2001, 2002, Gioia et al. 2003).   
Part of these projects were studies connected
to and profiting from the REFLEX survey program. None of the previous
projects covers an area in the southern sky as large as REFLEX, except
for the XBACS Abell cluster survey (Ebeling et al. 1996), which is shallower
and restricted to those clusters previously identified by Abell (1958)
and Abell, Corwin, and Olowin (1989).

The REFLEX catalogue of 447 clusters provides presently the
largest statistically complete X-ray cluster sample. The volume of
the Universe that is probed is larger than that covered by any present 
galaxy redshift survey except for the Sloan Digital Sky
Survey, which goes to a slightly larger depth but 
will only cover about half the sky area of that covered by REFLEX,
when completed. 

The paper is organized as follows. In section 2 we describe the survey
and the selection characteristics. Section 3 provides a brief description
of the X-ray data reduction and section 4 describes the redshift
determination and the cluster galaxy redshift statistics. 
The main catalogue is presented in
section 5 and some of its properties are reviewed in section 6. 
In the latter section we also provide the numerical data and the 
recipe to construct the survey selection function in one and two
dimensions for any flux limit equal to or above the nominal REFLEX
flux limit. Section
7 gives some further information on the identification and the
properties of some individual clusters. Section 8 lists close
cluster pairs and clusters with double or multiple X-ray maxima found
in the REFLEX catalogue, and we describe in more detail those clusters
where multiple redshift clustering is observed in the line-of-sight of the
X-ray source.
In section 9 we compare the results with
the previously derived survey samples and, finally, 
in section 10 we provide a summary and
conclusions. Table \ref{tab0} gives an overview of
the information presented in this paper in tabular form.

   \begin{table}
      \caption{Overview on the data presented in this paper in tabular form}
         \label{tab0}
      \[
         \begin{array}{ll}
            \hline
            \noalign{\smallskip}
 {\rm Table} & {\rm Content} \\
            \noalign{\smallskip}
            \hline
            \noalign{\smallskip}
 \ref{tab1} & {\rm count~ rate~ to~ flux~ conversion~ (for~ z=0)} \\ 
 \ref{tab2} & {\rm K-correction~ as~ a~ function~ of~ z~ and~ T_x} \\
 \ref{tab3} & {\rm Flux~ conversion~ from~ the~ 0.1 - 2.4~ keV~ to~}\\
   & {\rm the~ 0.5~ to~ 2.0~ keV~ band} \\
 \ref{tab4} & {\rm Flux~ conversion~ from~ the~ 0.1 - 2.4~ keV~ band~}\\
   & {\rm  to~ bolometric~ flux} \\
 \ref{tab5} \& \ref{tab6} & {\rm REFLEX~ cluster~ catalogue~ for~ h = 0.7~ and~ \Lambda -cosmology} \\
 \ref{tab7} & {\rm Further~ X-ray~ parameters~ of~ the~ REFLEX~ clusters} \\
 \ref{tab8} & {\rm Sky~ coverage~ as~ a~ function~ of~ the~ flux~ limit} \\
 \ref{tab9} & {\rm Angular~ modulation~ of~ the~ survey~ selection~ function} \\
\ref{tab10} & {\rm Close~ cluster~ pairs~ in~ the~ REFLEX~ sample} \\
\ref{tab11} & {\rm Clusters~ with~ multiple~ maxima~ in~ the~ REFLEX~ sample} \\
\ref{tab12} & {\rm Line-of-sight~ redshift~ clustering~ at~ the~}\\
            & {\rm  position~ of~ REFLEX~ clusters} \\
           \noalign{\smallskip}
           \hline
      \end{array}\]
   \end{table}

The luminosities and other cluster parameters which
depend on the distance scale are derived for a Hubble
constant of $H_0 = 70$ km s$^{-1}$ Mpc$^{-1}$ and a cosmological model
with $\Omega_m = 0.3$ and $\Omega_{\Lambda} =0.7$ in the main tables of the
paper. We also give in complementary tables provided only in electronic form
the cluster properties for the previously most
often used Einstein-de Sitter model with $H_0 = 50$ km s$^{-1}$ Mpc$^{-1}$,
$\Omega_m = 1.0$ and $\Omega_{\Lambda} =0$ 
for an easier comparison with previous literature results.

\section{The REFLEX Survey}

The construction of the REFLEX cluster sample is described
in detail in paper I.
The survey area covers the southern sky up to declination
$\delta = +2.5^o $, avoiding the band of the Milky Way
($|b_{II}| \le 20^o $) and the regions of the Magellanic clouds.
The total survey area is 13924 $\deg^2$ or 4.24 sr.
The regions that have been excised are defined in Table~\ref{tab1} in
paper I.

   \begin{table*}
      \caption{Count rate to flux conversion factors for different
	temperatures (for $Z = 0.3$ solar, $z = 0$) as a function of column density. 
        The values quoted give the 0.1 - 2.4 keV flux per counts in the 0.5 to 2 keV band 
         (channel 52 to 201) in units of $10^{-12}$ erg s$^{-1}$ cm$^{-2}$ counts$^{-1}$.
        The last column gives the hardness ratio for an assumed temperature of 5 keV,
         defined as  (counts(0.5 - 2.0keV)-counts(0.1-0.4 keV))/(counts(0.5-2keV)+counts(0.1-0.4 keV)).
         }
         \label{tab1}
      \[
         \begin{array}{llllllllllllll}
            \hline
            \noalign{\smallskip}
 N_H  & \multicolumn{12}{c}{\rm temperature} & {\rm HR} \\
      & 0.5  & 1.0 & 1.5 & 2.0 & 3.0 & 4.0 & 5.0 & 6.0 & 7.0 & 8.0 & 9.0 & 10.0  &{\rm (for} \\
10^{20} {\rm cm}^{-2} & {\rm keV} & {\rm keV} & {\rm keV} & {\rm keV} &
{\rm keV} & {\rm keV} & {\rm keV} & {\rm keV} & {\rm keV} & {\rm keV}& {\rm keV} & {\rm keV}&{\rm 5~ keV} \\
            \noalign{\smallskip} 
            \noalign{\smallskip}
            \hline
            \noalign{\smallskip}
  0.10 & 1.281& 1.413& 1.751& 1.831& 1.868& 1.880& 1.887& 1.893& 1.897& 1.900& 1.901& 1.902&0.003\\ 
  0.30 & 1.291& 1.422& 1.761& 1.842& 1.879& 1.891& 1.898& 1.904& 1.908& 1.910& 1.912& 1.913&0.089\\
  1.00 & 1.325& 1.453& 1.796& 1.880& 1.917& 1.929& 1.936& 1.942& 1.946& 1.948& 1.950& 1.951&0.323\\
  3.02 & 1.429& 1.543& 1.900& 1.989& 2.028& 2.040& 2.046& 2.052& 2.055& 2.057& 2.059& 2.059&0.691\\
 10.00 & 1.840& 1.886& 2.278& 2.383& 2.427& 2.438& 2.442& 2.446& 2.449& 2.450& 2.451& 2.450&0.943\\
 30.20 & 3.654& 3.213& 3.583& 3.718& 3.766& 3.766& 3.762& 3.760& 3.756& 3.753& 3.749& 3.744&0.978\\

          \noalign{\smallskip}
            \hline
         \end{array}
      \]
An extended version of this table is given in electronic form at CDS and our home page.
   \end{table*}
%
%


   \begin{table*}
      \caption{K-correction table for different temperatures and redshifts. The given value is to be 
      multiplied with the luminosity in the observed band to obtain the luminosity in the rest frame band.}
         \label{tab2}
      \[
         \begin{array}{lllllllllllll}
            \hline
            \noalign{\smallskip}
  {\rm redshift} & \multicolumn{12}{c}{\rm temperature} \\
 & 0.5  & 1.0 & 1.5 & 2.0 & 3.0 & 4.0 & 5.0 & 6.0 & 7.0 & 8.0 & 9.0 & 10.0 \\
 & {\rm keV} & {\rm keV} & {\rm keV} & {\rm keV} &
{\rm keV} & {\rm keV} & {\rm keV} & {\rm keV} & {\rm keV} & {\rm keV} & {\rm keV} & {\rm keV} \\
            \noalign{\smallskip}
            \hline
            \noalign{\smallskip}
0.0500 & 1.0026 & 0.9935 & 0.9867 & 0.9838 & 0.9800 & 0.9771 & 0.9750 & 0.9733 & 0.9720 & 0.9709 & 0.9700 & 0.9693\\
0.1000 & 1.0086 & 1.0253 & 0.9852 & 0.9700 & 0.9596 & 0.9540 & 0.9502 & 0.9472 & 0.9449 & 0.9431 & 0.9415 & 0.9402\\
0.1500 & 1.0126 & 1.0258 & 0.9806 & 0.9611 & 0.9450 & 0.9359 & 0.9299 & 0.9253 & 0.9217 & 0.9189 & 0.9166 & 0.9147\\
0.2000 & 1.0273 & 1.0416 & 0.9771 & 0.9528 & 0.9314 & 0.9192 & 0.9112 & 0.9050 & 0.9003 & 0.8966 & 0.8936 & 0.8911\\
0.2500 & 1.0452 & 1.0799 & 0.9880 & 0.9489 & 0.9197 & 0.9041 & 0.8940 & 0.8864 & 0.8806 & 0.8760 & 0.8724 & 0.8693\\
0.3000 & 1.0497 & 1.0820 & 0.9833 & 0.9401 & 0.9070 & 0.8891 & 0.8775 & 0.8686 & 0.8619 & 0.8566 & 0.8523 & 0.8488\\
0.4000 & 1.0584 & 1.0850 & 0.9768 & 0.9254 & 0.8837 & 0.8614 & 0.8469 & 0.8359 & 0.8276 & 0.8211 & 0.8159 & 0.8115\\
            \noalign{\smallskip}
            \hline
         \end{array}
      \]
An extended version of this table is given in electronic form at CDS and our home page.
   \end{table*}
%
%


The X-ray  detection of the clusters is based on the second
processing of the RASS (RASS II, Voges et al. 1999), providing
54076 sources in the REFLEX area. Note that the public RASS catalogue
available through the internet
(http://www.xray.mpe.mpg.de/rosat/survey/rass-bsc/; 
http://www.xray.mpe.mpg.de/rosat/survey/rass-fsc/) 
is based on the third
processing of the RASS which has a more complete sky coverage.
Since we have frozen the cluster candidate list in 1998 near the
end of the optical follow-up program, the cluster selection of the
present cluster catalogue is based on RASS II. We have, however,
checked that the X-ray properties of the individual catalogued objects 
are still consistent with the results obtained with the new
processing. We have in particular redetermined all cluster centers 
by applying our analysis software to the RASS III data base
(to ensure that the naming convention will be stable
when finally basing all the analysis on RASS III).
The positional changes were small in relation to the extent   
of the X-ray source. The second, 
enlarged version of the REFLEX sample with a lower flux limit will
be based completely on RASS III.

All sources were reanalysed by means of
the growth curve analysis (GCA) method (B\"ohringer et al. 2000)
and the results are used to produce a flux-limited sample of RASS
sources with a nominal flux of $F_n \ge 3 \cdot 10^{-12}$
erg s$^{-1}$ cm$^{-2}$ (with $F_n$ as defined below).
This redetermination of the fluxes has been shown to be crucial
for a precise measure of the fluxes of extended sources, as are the
majority of the REFLEX clusters (Ebeling et al. 1996, De Grandi et
al. 1997, B\"ohringer et al. 2000).
Cluster candidates were found using a machine based correlation of these
X-ray sources with galaxy density enhancements in
the COSMOS optical data base (derived from digital scans of the
UK Schmidt survey plates by COSMOS at the Royal Observatory Edinburgh
(MacGillivray \& Stobie 1984, Heydon-Dumbleton et al. 1989).
The resulting candidate list was carefully screened based on
X-ray and optical information, literature data, and results
from the optical follow-up observation program. The selection
process was designed to provide a completeness in the final cluster catalogue
in excess of 90\% with respect to the flux-limited sample of GCA
selected RASS sources. This high completeness of the cluster identification
of the RASS sources ensures that the selection effects
introduced by the optical identification
process are minimized and negligible for our purpose (see also
the statistics given in paper I).
Further tests provide support that this value of $> 90\%$ also describes the
the overall detection completeness of the flux-limited cluster sample in the survey
area. For example an independent search for X-ray emission from the
clusters catalogued by Abell, Corwin, \& Olowin (1989) returns only
one supplementary cluster, S567, with a flux above the flux-limit 
that has not
been included in the REFLEX sample. In addition tests
based on the Galactic latitude, redshift and photon count
distribution, and on an independent screening of all
significantly extended RASS X-ray sources in the survey region, are
consistent with this claim (B\"ohringer et al. 2001,
Schuecker et al. 2001). Based on the X-ray spectral
properties of the REFLEX cluster sources we can also estimate that
at most 9\% of the X-ray cluster sources may have a strong X-ray flux
contribution from AGN (see paper I).

\begin{figure*}                                                                  
\psfig{figure=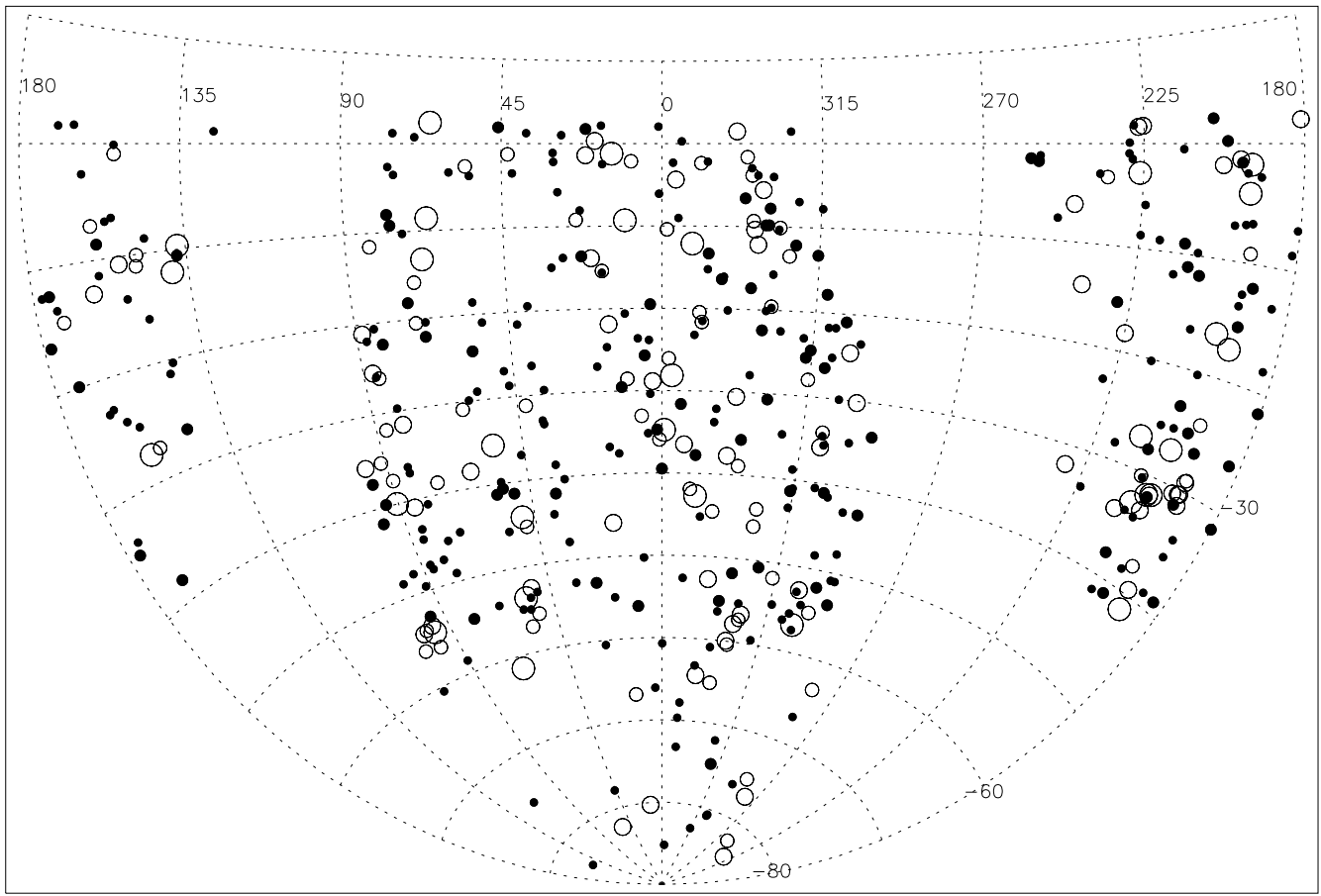,height=13cm}
\caption{Sky distribution in $\alpha$ and $\delta$ of the galaxy clusters in the REFLEX sample. The symbols give 
an indication of the cluster flux. The clusters are sorted into five flux bins: 
$3 - 5 \cdot 10^{-12}$, $5 - 7 \cdot 10^{-12}$, $7 - 10 \cdot 10^{-12}$, $\ 1 - 2 \cdot 10^{-11}$,
and $\ge 2 \cdot 10^{-11}$ erg s$^{-1}$ cm$^{-2}$ and indicated by increasing symbol size, respectively.
The three largest symbol classes are shown as open circles to avoid shading of other clusters.
}
\label{fig1}
\end{figure*}

To test further the identification of cluster candidates prior
to the follow-up observations in La Silla, we adopted a
very conservative scheme, as described in more detail in paper
I. To reject a source, we required it to have at least two 
(in some combinations three)
properties which are incompatible with a cluster identification. 
The properties were drawn from the following list:
X-ray source is point-like, X-ray source is too soft, no
optical cluster visible on digitized optical images, known optical
or radio AGN at the center of the X-ray emission. This conservative
scheme made the follow-up observations somewhat more expensive, 
but ensured a high completeness of the final catalogue.  

The final cluster sample comprises 447 objects. The distribution
of these sources in the sky is shown in Figs. 1 and 2

\section{Determination of the X-ray parameters}
  
The X-ray count rates, fluxes, and luminosities 
of the REFLEX clusters are determined from the
count rate measurements provided by the GCA
(B\"ohringer et al. 2000). In the first step the source count rates are integrated
out to the radius where an effectively flat plateau of
the cumulative source count curve is reached. The radius at which the
plateau is reached is documented.

\begin{figure*}                                                                  
\psfig{figure=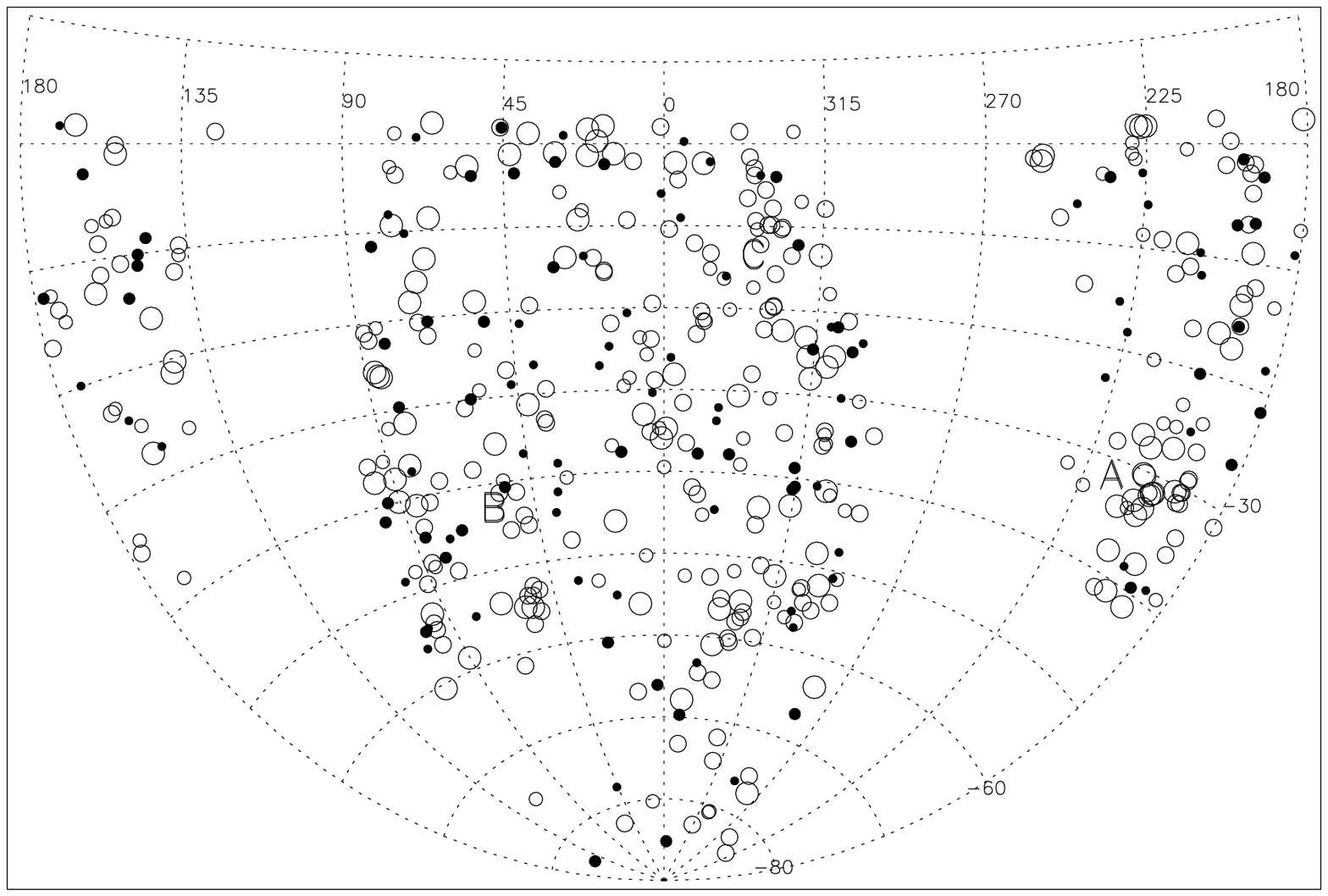,height=13cm}
\caption{The same sky distribution of REFLEX clusters as shown in
Fig. 2, but now the symbols indicate the cluster distance. The
clusters are sorted into five redshift classes: $z = 0 - 0.05$, $z =
0.05 - 0.1$, $z = 0.1 - 0.15$, $z = 0.15 - 0.2$, and $z \ge 0.2$
indicated by decreasing symbol size with increasing redshift. The
three largest symbol classes are shown as open circles to avoid
shading of other clusters. In this plot superstructures can be recognized. Three
of the most prominent superclusters have been marked: A is the Shapley 
concentration partly overlapping with Hydra-Centaurus in the foreground, 
B is the Horlogium-Reticulum complex, and C refers to the Aquarius-Cetus and 
Aquarius-Capricornus superclusters. For more details on these superstructures 
see Einasto et al. (2001).}
\label{fig1a}
\end{figure*}

To determine the cluster X-ray flux we convert the measured
count rate into a unabsorbed ``nominal'' X-ray flux for the ROSAT band (0.1 to
2.4 keV), $F_n$, by assuming a Raymond-Smith type spectrum
(Raymond \& Smith 1977) for a temperature of 5 keV, a metallicity $\alpha$ of
0.3 of the solar value (Anders \& Grevesse 1989), a redshift of zero,
and an interstellar hydrogen column density given for the line-of-sight
in the compilation by Dickey \& Lockman (1990),
as provided within EXSAS (Zimmermann et al. 1994).
The value of $F_n$ is used to make the flux cut independent of
any redshift information (since the redshift is not available
for all objects at the start of the survey).
With the redshift value at hand, the  
unabsorbed X-ray flux is redetermined ($F_x$)
with an improved spectral model, where the temperature is now
estimated (iteratively) from the preliminarily derived X-ray luminosity
and the luminosity-temperature relation (uncorrected for
cooling flow effects) derived by Markevitch (1998).
The estimated temperature of the cluster is now taken into account,
by folding the appropriate thermal spectrum with the instrument
response and the interstellar absorption, leading to a revised flux,
$F_x$ of the source (this correction is less than 5\% for sources with an
X-ray luminosity above $4~ 10^{43}$ erg s$^{-1}$). 
To obtain the cluster rest-frame luminosity
from the flux, $F_x$ we use the usual conversion with the cosmological
luminosity distance and further scale the luminosity by the ratio of
the luminosity integrated in the observed, redshifted and rest frame
0.1 - 2.4 keV band. The latter is equivalent to the 
K-correction. We note one simplification made in the above
transformation. In calculating the revised flux, $F_x$, we should have
additionally taken the redshift effect into account by folding a
redshifted source spectrum with the instrument response matrix and
interstellar absorption effect to calculate the exact countrate-to-flux 
conversion factor. We found, however, that this correction is
largest for the low temperature objects at high redshift and since we
observe these objects only at closer distances, we established that
this effect never becomes larger than 2\% for any source in our
catalogue and therefore we neglected this effect for the present
analysis. In the various applications of our project we have
used the results in Tables 2 and 3 for all the necessary 
conversion factors. Therefore in neglecting
this effect we limit the necessary interpolations between observed and
tabulated values to two-dimensional interpolations. This helps in the
theoretical modeling when large parameter grids have to be evaluated.   

Thus, luminosities are calculated for the rest frame 
energy band 0.1 to 2.4 keV. We also account for the X-ray flux missed outside 
the detection aperture by the following correction. We correct the
fluxes and luminosities based on a self-similar cluster model as
described in B\"ohringer et al. (2000): a $\beta$-model
(Cavaliere \& Fusco-Femiano 1976) with a $\beta$-value of 2/3,
a core radius that scales with mass,
and an assumed extent of the X-ray halo out to
12 times the core radius. The correction procedure has been
successfully tested by simulations based on the same cluster model.
The typical mean correction factor is about 8\% with the largest corrections of up
to 30\% occuring for the nearby groups which are extended and have a low
surface brightness (B\"ohringer et al. 2000). Note that in contrast to
our convention of extrapolating the corrected total flux out to the
estimated virial radius, Ebeling et al. (1996, 1998) and De Grandi et
al. (1999) extrapolate their flux corrections to infinity. The
difference between the two approaches - which uses the same
$\beta$-model with a slope of 2/3  - is 8.3\%. Nevertheless
the agreement between the results of, for example, Ebeling et al. (1998) and the
results of B\"ohringer et al. (2000) have been found to show a bias smaller
than this difference. We attribute this to the fact that the GCA method is
capturing slightly more of the cluster flux than the other methods. 

To allow the reader to easily and fully
reproduce our present results and the results on the cosmological 
implications of the REFLEX project published in other papers of this
series, without having to resort to ROSAT instrument specific
calculations, we provide here all the conversion tables used
in the flux and luminosity determination
described above. In the underlaying calculations we were aiming for an
accuracy of better than about 2-3\% (e.g. in the energy dependent
vignetting correction, the tested difference between calculations with
different radiation codes), so that the errors from the conversion
factors is in any case negligible compared to the measurement errors.
Table~\ref{tab1} gives the conversion factor from the observed
count rate in the ROSAT hard band (defined by the range from energy
channel 52 to 201) to the flux in the ROSAT band (0.1 - 2.4 keV) for
various temperatures covering the relevant temperature range,
$\alpha= 0.3$ solar, and $z = 0$, as a function of
the interstellar column density. A graphical representation of this
function for three selected temperatures can be found in B\"ohringer
et al. (2000, Fig. 8a) together with the column density dependence of
the hardness ratio for the same temperatures (Fig. 8b). 
This table is used to interpolate the count rate conversion factor for
the first step described above, where a fixed temperature of 5 keV is
used. In the second step the calculation is improved by interpolating
to the estimated temperature. Table~\ref{tab2} then provides the
K-correction factors to convert the redshifted luminosity measure
into rest-frame luminosity as a function of redshift and temperature of
the cluster. 

For an easy comparison of the catalogue with data given in other
frequently used energy bands, we provide in Table~\ref{tab3} the conversion
factor from the 0.1 - 2.4 keV to the 0.5 - 2.0 keV band as a function
of temperature. Table~\ref{tab4} provides the conversion from the ROSAT band
(0.1 - 2.4 keV) to the bolometric system. The print version of the
paper gives only a few example lines for Tables~\ref{tab1}  through \ref{tab4}. The
full tables can be obtained from the electronic version of the paper.  

   \begin{table}
      \caption{Conversion factor to be multiplied to the flux in the 0.5 to 2.0 keV
      band to obtain the flux in the 0.1 to 2.4 keV band as a function of temperature.}
         \label{tab3}
      \[
         \begin{array}{ll}
            \hline
            \noalign{\smallskip}
 {\rm temperature} & {\rm flux} \\
 {\rm  (keV)     } & {\rm ratio} \\
            \noalign{\smallskip}
            \hline
            \noalign{\smallskip}
  0.5 & 1.387 \\
  1.0 & 1.496 \\
  1.5 & 1.615 \\
  2.0 & 1.623 \\
  3.0 & 1.618 \\
  4.0 & 1.611 \\
  5.0 & 1.607 \\
  6.0 & 1.605 \\
  7.0 & 1.603 \\
  8.0 & 1.599 \\ 
  9.0 & 1.598 \\
 10.0 & 1.597 \\ 
            \noalign{\smallskip}
            \hline
         \end{array}
      \]
   \end{table}
%
   \begin{table}
   \begin{center}
      \caption{Conversion factor of the flux/luminosity in the 0.1 to 2.4 keV
      band to bolometric flux/luminosity as a function of temperature}
         \label{tab4}
      \[
         \begin{array}{ll}
            \hline
            \noalign{\smallskip}
 {\rm temperature} & {\rm flux} \\
 {\rm (keV)}      &  {\rm ratio} \\
            \noalign{\smallskip}
            \hline
            \noalign{\smallskip}
  0.30 & 1.443 \\
  0.51 & 1.187 \\
  1.02 & 1.269 \\
  1.98 & 1.448 \\
  3.00 & 1.660 \\
  5.10 & 2.122 \\
  8.08 & 2.743 \\
 10.17 & 3.152 \\
 15.04 & 4.053 \\
            \noalign{\smallskip}
            \hline
         \end{array}
      \]
An extended version of this table is given in electronic form at CDS and our home page.
   \end{center}
   \end{table}

\section{Redshift determination}

For cluster candidates with published redshifts we adopted the following
procedure to obtain the final redshift. The search for redshift
information in the literature was conducted with an aperture radius of 7
arcmin for more distant clusters and the search radius was increased
to half the virial radius (half the virial radius was taken
to be 5 times the core radius estimated from $L_X$ 
as described in B\"ohringer et al. 2000) if this radius was 
larger than 7 arcmin. This is of course an iterative process in which a first
redshift result was chosen for which the search radius was calculated and then
a refined search for cluster galaxy members was performed in NED.
This was found to be very important for the nearby clusters where
an increasing number of redshifts becomes available from large redshift surveys
which help greatly in improving the cluster redshifts. 
The final redshift taken was the cluster redshift
that involved the largest number of galaxy redshifts. If no
reliable cluster redshift was available, the cluster redshift was
computed from the median of individual galaxy redshifts located within
the aperture radius around the central target position after the
rejection of obvious non-members of the cluster. The secondary line-of-sight clustering
found for a number of REFLEX clusters is further described in section
8. The present literature redshift compilation is largely based on
information requested from NED in July 2003.

Cluster candidates without published redshifts or reliable
identification were observed from 1992 to 1999 within a large ESO key
programme (B\"ohringer et al. 1998, Guzzo et al. 1999). For a detailed
spectroscopic follow-up we used for the brighter objects the ESO\,1.5m
with the Boller \& Chivens spectrograph and the ESO\,2.2m with the
EFOSC-2 instrument. For the more distant and thus fainter objects the
ESO\,3.6m telescope with the EFOSC-1 (later EFOSC-2) instrument was
used. This instrument was preferentially used in multi-object
spectroscopic mode, but also long-slit spectra were taken with EFOSC 
for some of the clusters.
If necessary we also performed CCD imaging in order to test for
galaxy overdensities in the target direction. A detailed description
of the observing strategy, the reduction of the spectra, redshift
measurements, astrometry, and morphological classification is in
preparation (Guzzo et al. 2003, in preparation). 
A brief summary is given below.

The main goal of the follow-up observations was to get for each X-ray
cluster candidate at least five spectra, which we found was enough for a
first identification of the X-ray source and for a good estimate of
the cluster redshift. This comparatively small number of spectra
results from the fact that the optical follow-up could be restricted
to the location of the X-ray peak, that is, to the region of the
expected deepest cluster potential well where contamination by
background galaxies is minimal. In order to optimize the efficiency of
cluster identification and redshift measurement, for clusters with a
few bright (nearby) galaxies, spectra were observed in the single-slit
mode, whereas for clusters with many faint galaxies the multi-slit mode
was chosen. The multi-slit mode gives spectra for about 10-25 galaxies,
so that for many clusters we also have information about the velocity
dispersion (Ortiz-Gil et al. 2003).

\begin{figure}                                                                  
\psfig{figure=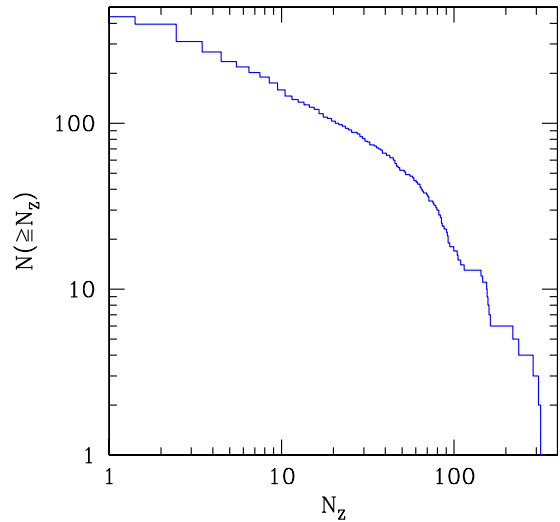,height=8.5cm}
\caption{Cumulative distribution, $N_{cl}(\ge N_z)$, of the number of galaxy redshifts known for the 
clusters in the REFLEX sample.}
\label{fig3}
\end{figure}

Fig. 3 shows the cumulative distribution of the REFLEX clusters with a 
number of member galaxy redshifts greater than the given limit. 
Indeed about half of the clusters have 5 or more galaxy redshifts
and the derived redshift is fairly secure. However, about
42 clusters feature only one galaxy redshift, mostly values coming
from the literature. The available observation time budget for this
project did not allow for a redetermination of all the literature
values. Still most of these data yield a reliable, as they
refer to the central
brightest cluster galaxy at the center of the X-ray emission.   
For 8 clusters in the catalogue we were so far not able to find
the information in the literature on the number of galaxies from 
which the cluster redshift was determined. 

The majority of spectra were reduced in a standard manner (bias and
flat field correction, wavelength calibration) using procedures from
the IRAF software package. Cosmics were rejected from the combination
of at least two spectral exposures obtained for each spectral
observation. All spectra were visually classified into the categories
elliptical galaxy, spiral galaxy, Seyfert 1/2 galaxy, quasar,
starforming galaxy (tentatively), and stellar. Heliocentric redshifts
were determined from cross-correlation with a 
sample of 17 template spectra with spectral types covering the large
range usually found in clusters of galaxies and in the
field. In addition, redshifts were also determined from emission lines
(if present), including a measurement of their equivalent widths. The
final redshifts of the clusters were obtained from the median of the
measured galaxy redshifts after the rejection of obvious non-members.
The criterion for the assignment of the cluster membership was
a maximum velocity deviation of 3000 km s$^{-1}$ from the median.

All optical and X-ray information is collected and organized in a
data base which we are planing to make publicly available. 
A more detailed account of the optical data will be
given in a forthcoming paper listing in particular the individual
galaxy redshifts determined in this ESO key program (Guzzo et al.)

\section{The catalogue}                      

Table~\ref{tab5} and \ref{tab6} list the 
X-ray properties and redshifts of the 447 galaxy clusters of the
REFLEX sample (compared to the previously used sample of 449 clusters,
7 clusters have been removed as detailed in section 7, 4 clusters
considered before as being contaminated by non-cluster emission have
been added, and one double cluster has been split into two sources).  
{\it Please note that the table is artificially split
here in two parts, as a result of memory limitation problems with the
current A\&A LATEX implementation.  This should be corrected with the
help of the editor before going into print.}  The columns of the table
provide: (1) the REFLEX name, (2) a previous catalogue name, where for
some of the smaller galaxy groups the name refers to the central
dominant galaxy in the group (e.g. the NGC name of this galaxy), (3)
and (4) the right ascension and declination for the epoch J2000 in
hours (degrees), minutes, and seconds, (5) the redshift, 
(6) the number of galaxies with
which the redshift has been determined after the rejection of
non-members (a zero means that no information is available about the
number of galaxies used to determine the cluster redshift), (7) and
(8) the measured, unabsorbed X-ray flux, $F_x$, in units of $10^{-12}$ erg s$^{-1}$
cm$^{-2}$ for the 0.1 - 2.4 keV energy band and the fractional error
in percent (note that this is not the nominal flux, $F_n$, used
to define the flux limit of REFLEX; $F_n$ is given in Table~\ref{tab7}), 
(9) the X-ray luminosity uncorrected for missing flux in
units of $10^{44}$\,erg/s in the rest frame 0.1 to 2.4 keV band, (10)
the aperture radius in arcmin within which the X-ray count rate and flux were
determined (the radius where the plateau value is reached in the
cumulative count rate curve of the GCA), (11) the 0.1 - 2.4 keV
luminosity corrected for the estimated flux lost outside the
measurement aperture, (12) the interstellar column density in units of
$10^{20}\,{\rm cm}^{-2}$ 
as obtained from the 21 cm observations by Dickey \& Lockman (1990)
and Stark et al. (1992), (13) comment flags as described below, and
(14) references to literature values for the cluster redshifts. The
comment flags refer to: (L) clusters which are too extended for the
standard GCA count rate determination in $2 \times 2$ degree$^2$
fields, and therefore were analysed in $4 \times 4$ or $8 \times 8$ degree$^2$
fields, (B) clusters blended with point sources or double clusters
which were deblended, (X) clusters where the details of the source
identification are commented in Section 7, (P) parts of close
cluster pairs or groups, and (D) double clusters discussed in section
8. For 12 objects marked with an astrisk in column (13) the cluster
origin of the X-ray emission is not completely certain as detailed
in section 7.  

{\footnotesize
\begin{table*}
      \caption{The REFLEX cluster catalogue}
         \label{tab5}
      \[
         \begin{array}{llrrrrrrrrrrrl}
            \hline
            \noalign{\smallskip}
 {\rm name}& {\rm alt. name}& RA(2000) & DEC(2000) & z & N_{gal}.& F_x & Error 
& L_x & R_{ap}& L_x^* & N_H & Cm&Ref.  \\
(1) & (2) & (3) & (4) & (5) & (6) & (7) & (8) & (9) & (10) & (11) & (12) & (13) & (14) \\
            \noalign{\smallskip}
            \hline
            \noalign{\smallskip}
 {\rm RXCJ0003.1-0605}&{\rm A2697       }&00~03~11.8& -06~05~10&0.2320&  0&  4.497&  12.4&  6.395&  7.5&  6.876&    3.1&  &139                             \\ 
 {\rm RXCJ0003.2-3555}&{\rm A2717       }&00~03~12.1& -35~55~38&0.0490& 40&  7.537&  17.8&  0.421& 10.0&  0.478&    1.1&  &2                               \\ 
 {\rm RXCJ0003.8+0203}&{\rm A2700       }&00~03~50.6& +02~03~48&0.0924&  9&  4.155&  18.8&  0.855&  8.5&  0.929&    3.0&  &S,10                            \\ 
 {\rm RXCJ0006.0-3443}&{\rm A2721       }&00~06~03.0& -34~43~27&0.1147& 75&  5.832&  13.6&  1.875& 10.0&  1.995&    1.2&  &1                               \\ 
 {\rm RXCJ0011.3-2851}&{\rm A2734       }&00~11~20.7& -28~51~18&0.0620& 83& 12.014&   9.0&  1.089& 12.5&  1.197&    1.8&  &2                               \\ 
 {\rm RXCJ0013.6-1930}&{\rm A0013       }&00~13~38.3& -19~30~08&0.0940& 37&  6.071&  11.3&  1.285& 11.5&  1.353&    2.0&  &S,2,140,141                     \\ 
 {\rm RXCJ0014.3-6604}&{\rm A2746       }&00~14~18.4& -66~04~39&0.1599&  5&  4.485&  13.8&  2.907&  7.5&  3.160&    2.8&X &E                               \\ 
 {\rm RXCJ0014.3-3023}&{\rm A2744       }&00~14~18.8& -30~23~00&0.3066& 65&  4.964&  18.7& 12.787& 12.0& 12.916&    1.6&  &142,143                         \\ 
 {\rm RXCJ0015.4-2350}&{\rm             }&00~15~24.0& -23~50~42&0.0645&  4&  3.426&  18.3&  0.336& 11.5&  0.354&    2.5&X &33                              \\ 
 {\rm RXCJ0017.5-3509}&{\rm A2755       }&00~17~33.7& -35~09~54&0.0968& 23&  3.211&  20.0&  0.729& 11.5&  0.752&    1.3&  &E,3,33,140,141                  \\ 
 {\rm RXCJ0020.7-2542}&{\rm A0022       }&00~20~42.8& -25~42~37&0.1410&  3&  5.910&  12.1&  2.909&  7.5&  3.232&    2.3&  &1                               \\ 
 {\rm RXCJ0025.5-3302}&{\rm S0041       }&00~25~32.4& -33~02~50&0.0491&  3&  8.818&   9.7&  0.494& 13.0&  0.537&    1.7&  &26,144                          \\ 
 {\rm RXCJ0027.3-5015}&{\rm A2777       }&00~27~21.3& -50~15~04&0.1448&  1&  4.286&  17.1&  2.247&  8.5&  2.390&    1.7&  &E,145                           \\ 
 {\rm RXCJ0028.6-2338}&{\rm A0042       }&00~28~39.3& -23~38~14&0.1120&  5&  4.836&  13.3&  1.491& 13.0&  1.521&    1.8&  &33                              \\ 
 {\rm RXCJ0034.6-0208}&{\rm             }&00~34~36.0& -02~08~24&0.0812&  2&  8.629&   7.9&  1.360& 28.0&  1.388&    2.8&DL&S,139                           \\ 
 {\rm RXCJ0040.1-5607}&{\rm A2806       }&00~40~06.5& -56~07~00&0.0277& 20&  4.068&  19.3&  0.071& 13.5&  0.077&    2.2&  &12,18,146                       \\ 
 {\rm RXCJ0041.8-0918}&{\rm A0085       }&00~41~50.1& -09~18~07&0.0555&308& 74.215&   3.2&  5.293& 25.0&  5.631&    3.6&L &130,148,149                     \\ 
 {\rm RXCJ0042.1-2832}&{\rm A2811       }&00~42~08.7& -28~32~09&0.1082& 29&  9.774&   9.8&  2.788& 10.0&  3.030&    1.5&  &E,33,96                         \\ 
 {\rm RXCJ0043.4-2037}&{\rm A2813       }&00~43~24.4& -20~37~17&0.2924&  7&  3.186&  15.6&  7.387&  8.5&  7.615&    1.5&  &E                               \\ 
 {\rm RXCJ0049.4-2931}&{\rm S0084       }&00~49~24.0& -29~31~28&0.1084& 18&  5.228&  16.0&  1.503& 11.0&  1.566&    1.8&  &33,150                          \\ 
 {\rm RXCJ0052.7-8015}&{\rm A2837       }&00~52~44.9& -80~15~59&0.1141&  7& 11.264&  21.2&  3.547& 12.5&  3.734&    6.6&  &E                               \\ 
 {\rm RXCJ0055.9-3732}&{\rm Cl0053^a    }&00~55~59.2& -37~32~51&0.1653& 21&  3.985&  17.3&  2.764& 13.0&  2.792&    2.6&B &E,1                             \\ 
 {\rm RXCJ0056.3-0112}&{\rm A0119       }&00~56~18.3& -01~12~60&0.0442&104& 33.613&   5.3&  1.505& 26.5&  1.568&    3.1&L &2                               \\ 
 {\rm RXCJ0057.8-6648}&{\rm S0112       }&00~57~48.1& -66~48~44&0.0661&  8&  7.225&  36.0&  0.743& 11.5&  0.799&    2.6&  &30,151                          \\ 
 {\rm RXCJ0102.7-2152}&{\rm A0133       }&01~02~42.1& -21~52~25&0.0569&  9& 19.031&   6.7&  1.439& 13.0&  1.617&    1.6&  &5,24,152,153                    \\ 
 {\rm RXCJ0105.5-2439}&{\rm A0141       }&01~05~34.8& -24~39~17&0.2300&  0&  3.884&  15.7&  5.416&  7.5&  5.762&    1.6&  &1                               \\ 
 {\rm RXCJ0106.8-0229}&{\rm A0145       }&01~06~52.4& -02~29~24&0.1909&  1&  3.209&  21.1&  3.010& 14.0&  3.040&    4.1&  &19,154                          \\ 
 {\rm RXCJ0107.8-3643}&{\rm A2871       }&01~07~49.1& -36~43~38&0.1186& 19&  3.557&  13.6&  1.229& 12.0&  1.254&    1.9&  &E,2,3,33                        \\ 
 {\rm RXCJ0108.1+0210}&{\rm A0147       }&01~08~11.5& +02~10~34&0.0436&  8&  4.369&  16.0&  0.192& 12.0&  0.206&    3.0&  &120,123,156,157                 \\ 
 {\rm RXCJ0108.8-1524}&{\rm A0151N      }&01~08~50.1& -15~24~36&0.0533& 63&  6.862&   9.8&  0.456&  8.5&  0.530&    1.7&B &1,2                             \\ 
 {\rm RXCJ0108.9-1537}&{\rm A0151S      }&01~08~55.2& -15~37~44&0.0970& 13&  3.703&  15.3&  0.845& 10.0&  0.889&    1.8&B &3                               \\ 
 {\rm RXCJ0110.0-4555}&{\rm A2877       }&01~10~00.4& -45~55~22&0.0238& 58& 14.005&   8.0&  0.179& 20.0&  0.195&    2.1&L &12,130,131,146                  \\ 
 {\rm RXCJ0115.2+0019}&{\rm A0168       }&01~15~12.0& +00~19~48&0.0450& 76& 10.429&   9.3&  0.488& 20.0&  0.503&    3.3&L &S,2                             \\ 
 {\rm RXCJ0117.8-5455}&{\rm             }&01~17~50.5& -54~55~26&0.2510&  6&  3.054&  33.4&  5.155&  6.5&  5.543&    2.7&X\ast &E                           \\ 
 {\rm RXCJ0118.1-2658}&{\rm A2895       }&01~18~11.1& -26~58~23&0.2275&  4&  3.822&  14.0&  5.225&  7.5&  5.559&    1.6&  &S,96                            \\ 
 {\rm RXCJ0120.9-1351}&{\rm CAN 010^b   }&01~20~58.9& -13~51~31&0.0519&  7& 12.430&   7.9&  0.778& 15.5&  0.828&    1.9&B &S,120                           \\ 
 {\rm RXCJ0125.5+0145}&{\rm NGC533      }&01~25~30.2& +01~45~44&0.0174& 19&  4.712&  15.9&  0.032& 15.0&  0.036&    3.1&  &E,120                           \\ 
 {\rm RXCJ0125.6-0124}&{\rm A0194       }&01~25~40.8& -01~24~26&0.0180&146&  9.713&  14.1&  0.070& 24.0&  0.074&    4.1&L &158                             \\ 
 {\rm RXCJ0131.8-1336}&{\rm A0209       }&01~31~53.0& -13~36~34&0.2060&  2&  5.481&  10.4&  6.037&  9.5&  6.289&    1.6&  &1                               \\ 
 {\rm RXCJ0132.6-0804}&{\rm             }&01~32~40.9& -08~04~20&0.1489&  3&  3.784&  11.9&  2.108& 10.0&  2.173&    3.4&X\ast &E                           \\ 
 {\rm RXCJ0137.2-0912}&{\rm             }&01~37~15.4& -09~12~10&0.0409&  5&  7.090&   8.4&  0.272&  9.5&  0.316&    2.8&  &S,120                           \\ 
 {\rm RXCJ0145.0-5300}&{\rm A2941       }&01~45~02.3& -53~00~50&0.1168&  4&  6.028&  16.0&  2.005&  7.5&  2.253&    2.3&  &E                               \\ 
 {\rm RXCJ0145.2-6033}&{\rm RBS 0238^c  }&01~45~16.7& -60~33~54&0.1805& 11&  4.811&  13.4&  4.010&  4.5&  4.890&    3.4&  &E                               \\ 
 {\rm RXCJ0152.7+0100}&{\rm A0267       }&01~52~42.3& +01~00~45&0.2300&  1&  3.090&  14.6&  4.329&  5.0&  4.919&    2.8&DB&55,147,159                      \\ 
 {\rm RXCJ0152.9-1345}&{\rm NGC0720     }&01~52~59.0& -13~45~12&0.0050&  3&  2.284&  16.4&  0.001& 10.5&  0.001&    1.7&  &12,160,161                      \\ 
 {\rm RXCJ0157.4-0550}&{\rm A0281       }&01~57~24.3& -05~50~24&0.1289&  4&  3.277&  14.7&  1.355& 10.0&  1.397&    2.2&D &E,1                             \\ 
 {\rm RXCJ0201.7-0212}&{\rm A0291       }&02~01~44.2& -02~12~03&0.1960&  2&  4.244&  11.5&  4.199&  5.0&  4.883&    2.6&  &S,53,54                         \\ 
 {\rm RXCJ0202.3-0107}&{\rm A0295       }&02~02~19.9& -01~07~13&0.0427& 47&  3.586&  13.7&  0.151& 14.0&  0.157&    2.6&  &2,148,162                       \\ 
 {\rm RXCJ0206.4-1453}&{\rm A0305       }&02~06~30.0& -14~53~38&0.1529&  2&  2.993&  17.9&  1.778&  6.0&  1.976&    2.5&  &S                               \\ 
 {\rm RXCJ0211.4-4017}&{\rm A2984       }&02~11~25.5& -40~17~12&0.1008&  6&  3.222&  11.5&  0.798&  8.0&  0.858&    1.4&  &33                              \\ 
 {\rm RXCJ0216.7-4749}&{\rm S0239       }&02~16~42.3& -47~49~24&0.0640&  1&  3.574&  16.9&  0.346& 11.5&  0.364&    3.0&  &5,9,29                          \\ 
 {\rm RXCJ0217.2-5244}&{\rm             }&02~17~12.6& -52~44~49&0.3432&  2&  3.641&  15.4& 11.914& 11.5& 12.034&    3.2&  &E                               \\ 
 {\rm RXCJ0220.9-3829}&{\rm             }&02~20~56.6& -38~29~05&0.2280&  5&  3.406&  11.8&  4.679&  7.0&  5.031&    1.9&  &E                               \\ 
 {\rm RXCJ0225.1-2928}&{\rm             }&02~25~10.5& -29~28~26&0.0604& 17&  4.736&  23.4&  0.408& 12.0&  0.434&    1.7&  &S,96                            \\ 
 {\rm RXCJ0225.9-4154}&{\rm A3016       }&02~25~54.6& -41~54~35&0.2195&  1&  5.185&  11.2&  6.507& 11.0&  6.640&    2.1&  &S                               \\ 
 {\rm RXCJ0229.3-3332}&{\rm             }&02~29~22.3& -33~32~16&0.0792& 17&  3.912&  12.9&  0.588& 12.5&  0.606&    2.1&P &S,33                            \\ 
 {\rm RXCJ0230.7-3305}&{\rm A3027       }&02~30~43.5& -33~05~55&0.0760& 22&  3.207&  15.7&  0.445& 15.0&  0.449&    2.0&PB&33                              \\ 
 {\rm RXCJ0231.9+0114}&{\rm RCS145^d    }&02~31~57.1& +01~14~40&0.0221& 10&  2.618&  21.1&  0.029& 14.5&  0.031&    2.9&  &55,64,110                       \\ 
 {\rm RXCJ0232.2-4420}&{\rm             }&02~32~16.8& -44~20~51&0.2836&  2&  4.074&  17.7&  8.875&  6.5&  9.647&    2.6&B &S                               \\ 
 {\rm RXCJ0236.6-1923}&{\rm A0367       }&02~36~40.2& -19~23~13&0.0907& 27&  3.434&  14.6&  0.679&  8.5&  0.730&    2.7&  &140,141                         \\ 

            \noalign{\smallskip}
            \hline
         \end{array}
      \]
\end{table*}
%
%

}
{\footnotesize
\begin{table*}
         \label{TabR1b}
      \[
         \begin{array}{llrrrrrrrrrrrl}
            \hline
            \noalign{\smallskip}
 {\rm name}& {\rm alt. name}& R.A. & Decl. & z & N_{gal}& F_x & Error 
& L_x & R_{ap}& L_x^* & N_H & Cm&Ref.  \\
(1) & (2) & (3) & (4) & (5) & (6) & (7) & (8) & (9) & (10) & (11) & (12) & (13) & (14) \\
            \noalign{\smallskip}
            \hline
            \noalign{\smallskip}
 {\rm RXCJ0237.4-2630}&{\rm A0368       }&02~37~29.2& -26~30~17&0.2216& 10&  3.082&  15.4&  3.971&  6.5&  4.270&    1.6&  &E                               \\ 
 {\rm RXCJ0248.0-0332}&{\rm A0383       }&02~48~02.0& -03~32~15&0.1883&  1&  4.837&  18.3&  4.377& 10.0&  4.559&    4.1&  &S,7,147                         \\ 
 {\rm RXCJ0249.6-3111}&{\rm S0301       }&02~49~36.9& -31~11~19&0.0230& 28&  6.979&   8.6&  0.083& 17.0&  0.089&    1.8&L &14,28,72,96                     \\ 
 {\rm RXCJ0250.2-2129}&{\rm             }&02~50~17.2& -21~29~56&0.2070&  2&  3.526&  16.5&  3.942&  6.5&  4.285&    2.5&X\ast &E                           \\ 
 {\rm RXCJ0252.8-0116}&{\rm NGC1132     }&02~52~49.4& -01~16~27&0.0235&  6&  6.028&  16.2&  0.075&  9.5&  0.091&    5.2&  &55,163                          \\ 
 {\rm RXCJ0301.6+0155}&{\rm Zw0258.9^e  }&03~01~37.2& +01~55~11&0.1712& 14&  4.973&  15.8&  3.704&  6.0&  4.209&    7.7&X &E                               \\ 
 {\rm RXCJ0303.3+0155}&{\rm A0409       }&03~03~21.1& +01~55~35&0.1530&  1&  5.082&  15.1&  2.990&  6.0&  3.437&    7.8&  &53                              \\ 
 {\rm RXCJ0303.7-7752}&{\rm             }&03~03~46.4& -77~52~09&0.2742&  8&  3.243&  20.2&  6.593&  6.0&  7.166&    8.9&  &E                               \\ 
 {\rm RXCJ0304.1-3656}&{\rm A3084       }&03~04~07.2& -36~56~36&0.2192&  2&  3.010&  15.5&  3.806&  7.0&  4.049&    2.0&  &E                               \\ 
 {\rm RXCJ0307.0-2840}&{\rm A3088       }&03~07~04.1& -28~40~14&0.2537& 10&  3.877&  17.3&  6.675&  9.0&  6.953&    1.4&  &E                               \\ 
 {\rm RXCJ0311.4-2653}&{\rm A3094       }&03~11~25.0& -26~53~59&0.0685& 23&  2.942&  50.4&  0.329&  8.0&  0.362&    1.6&  &3,17,33,96                      \\ 
 {\rm RXCJ0314.3-4525}&{\rm A3104       }&03~14~19.8& -45~25~27&0.0718&  4&  8.255&   9.5&  1.009&  9.5&  1.134&    3.6&  &E                               \\ 
 {\rm RXCJ0317.9-4414}&{\rm A3112       }&03~17~58.5& -44~14~20&0.0752& 38& 29.301&   3.9&  3.904& 15.5&  4.243&    2.5&  &S,3,31                          \\ 
 {\rm RXCJ0322.2-5310}&{\rm APMCC391^f  }&03~22~12.7& -53~10~41&0.0797&  6&  3.099&  10.9&  0.476& 12.0&  0.491&    2.3&  &E                               \\ 
 {\rm RXCJ0322.3-4121}&{\rm A3122       }&03~22~18.6& -41~21~34&0.0643& 87&  6.355&   9.7&  0.616& 14.0&  0.642&    2.1&  &3,140,141                       \\ 
 {\rm RXCJ0328.6-5542}&{\rm A3126       }&03~28~37.5& -55~42~46&0.0853& 42&  8.441&   7.6&  1.476& 10.0&  1.622&    3.1&  &38,164                          \\ 
 {\rm RXCJ0330.0-5235}&{\rm A3128       }&03~30~00.7& -52~35~46&0.0624& 40& 12.786&   5.5&  1.171& 22.0&  1.195&    1.5&DL&3,38,131,165                    \\ 
 {\rm RXCJ0331.1-2100}&{\rm             }&03~31~07.3& -21~00~12&0.1880&  9&  4.044&  24.5&  3.662&  4.5&  4.360&    2.5&X\ast &E                           \\ 
 {\rm RXCJ0334.9-5342}&{\rm APMCC421^f  }&03~34~56.2& -53~42~08&0.0619&  4&  3.323&  11.9&  0.301& 14.0&  0.307&    1.2&  &24,17,164                       \\ 
 {\rm RXCJ0336.3-4037}&{\rm A3140       }&03~36~18.7& -40~37~20&0.1729&  9&  5.631&  15.4&  4.267&  8.0&  4.588&    1.4&  &E                               \\ 
 {\rm RXCJ0336.3-0347}&{\rm             }&03~36~22.9& -03~47~29&0.1595&  3&  3.745&  15.9&  2.426&  9.5&  2.527&    4.9&X\ast &E                           \\ 
 {\rm RXCJ0337.0-3949}&{\rm A3142       }&03~36~60.0& -39~49~12&0.1030& 21&  3.298&  23.6&  0.855& 10.0&  0.891&    1.8&  &2                               \\ 
 {\rm RXCJ0338.4-3526}&{\rm FORNAX      }&03~38~27.9& -35~26~54&0.0051& 32& 21.272&  11.1&  0.012& 19.0&  0.016&    1.5&B &166,167,168,170,171             \\ 
 {\rm RXCJ0340.1-5503}&{\rm IC 1987     }&03~40~08.6& -55~03~14&0.0464&  1&  3.803&  12.0&  0.190& 17.0&  0.194&    2.2&L &17,164                          \\ 
 {\rm RXCJ0340.1-1835}&{\rm NGC1407^g   }&03~40~11.4& -18~35~15&0.0056&  4&  3.443&  14.0&  0.002& 14.5&  0.002&    5.2&  &172                             \\ 
 {\rm RXCJ0340.6-0239}&{\rm             }&03~40~41.8& -02~39~57&0.0352&  2&  7.923&  10.6&  0.225& 17.5&  0.237&    7.3&  &E                               \\ 
 {\rm RXCJ0340.8-4542}&{\rm RBS 0459^c  }&03~40~49.3& -45~42~19&0.0698&  9&  3.981&  14.0&  0.461& 10.5&  0.490&    1.6&  &E                               \\ 
 {\rm RXCJ0342.8-5338}&{\rm A3158       }&03~42~53.9& -53~38~07&0.0590&105& 36.344&   3.3&  2.951& 30.0&  3.011&    1.1&L &2                               \\ 
 {\rm RXCJ0345.7-4112}&{\rm S0384       }&03~45~45.7& -41~12~27&0.0603&  1&  5.763&  18.9&  0.495& 11.5&  0.532&    1.9&  &E,29,30                         \\ 
 {\rm RXCJ0345.9-2416}&{\rm A0458       }&03~45~56.0& -24~16~49&0.1057& 30&  5.510&  10.0&  1.501&  9.5&  1.614&    1.6&  &1                               \\ 
 {\rm RXCJ0346.1-5702}&{\rm A3164       }&03~46~09.6& -57~02~60&0.0570&  3&  7.465&  12.7&  0.570& 20.0&  0.576&    2.6&L &146                             \\ 
 {\rm RXCJ0347.0-2900}&{\rm A3165       }&03~47~00.2& -29~00~13&0.1419&  3&  4.054&  22.0&  2.033&  8.0&  2.186&    0.9&  &19,173                          \\ 
 {\rm RXCJ0351.1-8212}&{\rm S0405       }&03~51~08.9& -82~12~60&0.0613&  2& 14.078&   8.0&  1.243& 16.5&  1.308&    7.7&L &E,27                            \\ 
 {\rm RXCJ0352.3-5453}&{\rm RBS 0485^c  }&03~52~20.7& -54~53~09&0.0447&  2&  3.798&  25.9&  0.176& 13.0&  0.185&    3.0&  &E                               \\ 
 {\rm RXCJ0358.8-2955}&{\rm A3192       }&03~58~50.5& -29~55~18&0.1681& 13&  3.060&  13.3&  2.202&  7.5&  2.343&    1.0&  &E                               \\ 
 {\rm RXCJ0359.1-0320}&{\rm             }&03~59~09.2& -03~20~28&0.1220&  4&  3.546&  17.1&  1.302&  8.5&  1.385&    8.2&  &E                               \\ 
 {\rm RXCJ0408.2-3053}&{\rm A3223       }&04~08~16.2& -30~53~40&0.0600& 81&  8.186&   8.8&  0.693& 16.0&  0.722&    1.8&  &E,2                             \\ 
 {\rm RXCJ0413.9-3805}&{\rm             }&04~13~57.1& -38~05~60&0.0501& 22& 12.608&   7.7&  0.734& 15.0&  0.789&    1.4&B &E,31,96,144                     \\ 
 {\rm RXCJ0419.6+0224}&{\rm NGC1550     }&04~19~37.8& +02~24~50&0.0131&  7& 40.095&   6.1&  0.153& 22.0&  0.182&   11.6&L &11,64,123,174                   \\ 
 {\rm RXCJ0425.8-0833}&{\rm RBS 0540^c  }&04~25~51.4& -08~33~33&0.0397&  2& 28.011&   5.8&  1.008& 13.5&  1.186&    6.4&X &E,175,176                       \\ 
 {\rm RXCJ0429.1-5350}&{\rm S0463       }&04~29~07.8& -53~50~51&0.0400& 28&  4.018&  16.5&  0.148& 10.0&  0.164&    0.8&  &19,72,74,177                    \\ 
 {\rm RXCJ0431.4-6126}&{\rm A3266       }&04~31~24.1& -61~26~38&0.0589&317& 49.741&   1.8&  4.019& 18.0&  4.416&    1.5&  &155                             \\ 
 {\rm RXCJ0433.6-1315}&{\rm A0496       }&04~33~38.4& -13~15~33&0.0326&143& 72.075&   3.7&  1.746& 19.0&  2.054&    5.7&  &120,130,152,178,179             \\ 
 {\rm RXCJ0437.1-2027}&{\rm A0499       }&04~37~07.2& -20~27~26&0.1550&  2&  3.442&  56.5&  2.094& 13.0&  2.094&    2.6&X &E                               \\ 
 {\rm RXCJ0437.1+0043}&{\rm             }&04~37~10.1& +00~43~38&0.2842&  5&  3.942&  13.0&  8.629&  8.5&  8.989&    8.2&  &E                               \\ 
 {\rm RXCJ0438.9-2206}&{\rm A0500       }&04~38~54.7& -22~06~49&0.0670&  3&  6.168&  16.1&  0.652& 10.0&  0.716&    2.8&  &29                              \\ 
 {\rm RXCJ0445.1-1551}&{\rm NGC1650     }&04~45~10.0& -15~51~01&0.0360&  3&  8.634&  10.3&  0.255& 16.0&  0.274&    4.8&  &120                             \\ 
 {\rm RXCJ0448.2-2028}&{\rm A0514       }&04~48~12.2& -20~28~11&0.0720& 90&  8.583&   8.8&  1.052& 15.5&  1.096&    3.2&  &2                               \\ 
 {\rm RXCJ0449.9-4440}&{\rm A3292       }&04~49~55.2& -44~40~41&0.1501&  2&  4.418&  13.7&  2.501&  6.0&  2.842&    1.5&  &E                               \\ 
 {\rm RXCJ0454.1-1014}&{\rm A0521       }&04~54~09.1& -10~14~19&0.2475& 42&  4.944&  16.6&  8.014& 11.5&  8.178&    5.4&  &E,180                           \\ 
 {\rm RXCJ0454.8-1806}&{\rm CID 28^b    }&04~54~50.3& -18~06~33&0.0335& 17&  5.539&  11.8&  0.142& 12.5&  0.156&    4.2&  &E,120                           \\ 
 {\rm RXCJ0500.7-3840}&{\rm A3301       }&05~00~46.5& -38~40~41&0.0536&  5&  7.370&  10.0&  0.495& 13.5&  0.527&    3.1&  &40                              \\ 
 {\rm RXCJ0501.3-0332}&{\rm A0531       }&05~01~19.4& -03~32~33&0.0913& 10&  3.994&  23.4&  0.801& 10.0&  0.843&    6.0&  &E                               \\ 
 {\rm RXCJ0501.6+0110}&{\rm             }&05~01~39.1& +01~10~30&0.1248& 15&  4.398&  16.3&  1.696& 10.0&  1.785&    7.9&B &E                               \\ 
 {\rm RXCJ0507.6-0238}&{\rm A0535       }&05~07~36.0& -02~38~24&0.1241&  2&  3.341&  21.7&  1.273& 13.0&  1.286&    9.7&B &E                               \\ 
 {\rm RXCJ0507.7-0915}&{\rm A0536       }&05~07~45.7& -09~15~16&0.0398&  1&  5.056&  13.7&  0.184& 14.0&  0.196&    7.9&  &1                               \\ 
 {\rm RXCJ0510.2-4519}&{\rm A3322       }&05~10~13.9& -45~19~16&0.2000&  1&  4.792&  14.3&  4.960&  6.5&  5.511&    3.4&  &181                             \\ 
 {\rm RXCJ0510.7-0801}&{\rm             }&05~10~44.7& -08~01~06&0.2195& 14&  6.565&  10.7&  8.209& 10.0&  8.551&    8.3&  &E                               \\ 
 {\rm RXCJ0514.6-4903}&{\rm A3330       }&05~14~36.3& -49~03~18&0.0912&  2&  4.130&  17.9&  0.826& 14.0&  0.843&    2.4&  &17,32,182                       \\ 
 {\rm RXCJ0516.6-5430}&{\rm S0520       }&05~16~38.0& -54~30~51&0.2952&  8&  5.816&  21.5& 13.727& 16.5& 13.866&    6.2&  &E                               \\ 

            \noalign{\smallskip}
            \hline
         \end{array}
      \]
\end{table*}
%
%

}
{\footnotesize
\begin{table*}
         \label{TabR1c}
      \[
         \begin{array}{llrrrrrrrrrrrl}
            \hline
            \noalign{\smallskip}
 {\rm name}& {\rm alt. name}& R.A. & Decl. & z & N_{gal} & F_x & Error 
& L_x & R_{ap}& L_x^* & N_H & Cm&Ref.  \\
(1) & (2) & (3) & (4) & (5) & (6) & (7) & (8) & (9) & (10) & (11) & (12) & (13) & (14) \\
            \noalign{\smallskip}
            \hline
            \noalign{\smallskip}
 {\rm RXCJ0521.4-4049}&{\rm A3336       }&05~21~29.6& -40~49~29&0.0756&  2&  4.899&  19.4&  0.667& 15.0&  0.681&    3.3&  &17,69                           \\ 
 {\rm RXCJ0525.5-3135}&{\rm A3341       }&05~25~32.8& -31~35~44&0.0380& 64& 11.350&   7.3&  0.375& 16.5&  0.403&    1.8&  &2                               \\ 
 {\rm RXCJ0525.8-4715}&{\rm A3343       }&05~25~51.6& -47~15~02&0.1913&  5&  3.814&  11.1&  3.594&  7.5&  3.865&    4.0&  &E                               \\ 
 {\rm RXCJ0528.2-2942}&{\rm             }&05~28~15.4& -29~42~57&0.1582&  2&  3.864&  37.6&  2.458& 12.0&  2.483&    1.8&B &E                               \\ 
 {\rm RXCJ0528.9-3927}&{\rm RBS 0653    }&05~28~56.3& -39~27~46&0.2839&  4&  5.888&   8.6& 12.856& 11.0& 13.118&    2.1&  &E                               \\ 
 {\rm RXCJ0530.6-2226}&{\rm A0543       }&05~30~38.4& -22~26~54&0.1706& 11&  5.828&  11.2&  4.288& 14.0&  4.331&    2.6&  &E,1                             \\ 
 {\rm RXCJ0532.3-1131}&{\rm A0545       }&05~32~23.1& -11~31~50&0.1540&  2&  8.490&   8.7&  5.047&  7.5&  5.671&   11.2&  &1                               \\ 
 {\rm RXCJ0532.9-3701}&{\rm             }&05~32~55.5& -37~01~28&0.2708&  9&  3.302&  11.7&  6.525&  7.0&  6.941&    2.7&  &E                               \\ 
 {\rm RXCJ0533.3-3619}&{\rm S0535       }&05~33~18.5& -36~19~32&0.0479&  4&  3.734&  11.9&  0.199& 13.0&  0.209&    2.9&  &E                               \\ 
 {\rm RXCJ0538.2-2037}&{\rm A3358       }&05~38~16.3& -20~37~23&0.0915&  0&  4.601&  10.4&  0.927&  6.5&  1.066&    4.0&  &106                             \\ 
 {\rm RXCJ0540.1-4050}&{\rm S0540       }&05~40~06.3& -40~50~32&0.0358&  1& 14.881&   5.2&  0.438& 15.5&  0.481&    3.5&  &5,9                             \\ 
 {\rm RXCJ0540.1-4322}&{\rm A3360       }&05~40~10.0& -43~22~56&0.0850& 36&  4.759&   9.6&  0.830& 11.0&  0.874&    3.8&B &38,140,141                      \\ 
 {\rm RXCJ0542.1-2607}&{\rm CID 36^b    }&05~42~09.3& -26~07~25&0.0390&  4&  7.375&   8.9&  0.257& 18.0&  0.268&    1.9&P &E,120                           \\ 
 {\rm RXCJ0543.4-4430}&{\rm             }&05~43~24.4& -44~30~19&0.1637&  3&  4.174&   9.4&  2.816& 12.0&  2.873&    4.6&  &E                               \\ 
 {\rm RXCJ0545.4-2556}&{\rm A0548W      }&05~45~27.2& -25~56~20&0.0424&155&  2.440&  13.5&  0.102&  9.5&  0.111&    1.8&P &2,72                            \\ 
 {\rm RXCJ0545.5-4756}&{\rm A3363       }&05~45~30.6& -47~56~26&0.1254&  2&  4.522&   9.4&  1.762& 10.0&  1.855&    6.2&B &E                               \\ 
 {\rm RXCJ0547.6-3152}&{\rm A3364       }&05~47~38.2& -31~52~31&0.1483& 10&  8.526&   7.5&  4.667& 17.5&  4.667&    2.0&  &E                               \\ 
 {\rm RXCJ0547.7-4723}&{\rm S0547       }&05~47~45.6& -47~23~24&0.0515&  2&  2.876&  29.6&  0.178& 15.0&  0.182&    6.5&  &E                               \\ 
 {\rm RXCJ0548.6-2527}&{\rm A0548E      }&05~48~39.5& -25~27~30&0.0420&237& 14.109&   6.2&  0.573& 21.0&  0.597&    1.9&PL&3                               \\ 
 {\rm RXCJ0548.8-2154}&{\rm             }&05~48~50.4& -21~54~43&0.0928&  9&  3.977&  12.3&  0.825& 11.5&  0.859&    3.0&  &E,3                             \\ 
 {\rm RXCJ0552.8-2103}&{\rm A0550       }&05~52~52.4& -21~03~25&0.0989& 25& 10.319&   6.8&  2.429& 13.5&  2.557&    4.3&  &E                               \\ 
 {\rm RXCJ0557.2-3727}&{\rm S0555       }&05~57~13.2& -37~27~58&0.0442&  4&  7.141&  10.9&  0.322& 17.5&  0.335&    4.0&  &E                               \\ 
 {\rm RXCJ0600.8-5835}&{\rm S0560       }&06~00~48.3& -58~35~14&0.0369&  2&  2.600&   7.6&  0.082& 12.5&  0.086&    4.6&  &E                               \\ 
 {\rm RXCJ0601.7-3959}&{\rm A3376       }&06~01~45.7& -39~59~34&0.0468& 32& 20.492&   4.8&  1.036& 21.5&  1.079&    5.0&L &72                              \\ 
 {\rm RXCJ0605.8-3518}&{\rm A3378       }&06~05~52.8& -35~18~02&0.1392&  2&  9.393&   6.2&  4.478& 12.5&  4.665&    4.3&  &E                               \\ 
 {\rm RXCJ0607.0-4928}&{\rm A3380       }&06~07~01.4& -49~28~60&0.0553&  3&  3.592&  14.4&  0.258& 14.0&  0.266&    4.6&B &E                               \\ 
 {\rm RXCJ0616.5-3948}&{\rm S0579       }&06~16~31.4& -39~48~01&0.1520&  7&  5.180&   9.7&  3.004& 11.0&  3.097&    8.2&  &E                               \\ 
 {\rm RXCJ0616.8-4748}&{\rm             }&06~16~53.6& -47~48~18&0.1164&  1&  4.813&   9.8&  1.597& 14.0&  1.613&    4.8&  &E                               \\ 
 {\rm RXCJ0621.7-5242}&{\rm             }&06~21~43.6& -52~42~11&0.0511&  2&  5.058&   8.8&  0.308& 14.0&  0.324&    5.2&  &70,71                           \\ 
 {\rm RXCJ0624.6-3720}&{\rm A3390       }&06~24~36.7& -37~20~09&0.0333& 15&  5.608&  11.3&  0.142& 17.5&  0.148&    7.1&D &146                             \\ 
 {\rm RXCJ0626.3-5341}&{\rm A3391       }&06~26~22.8& -53~41~44&0.0514& 85& 19.583&   3.8&  1.198& 19.0&  1.261&    5.4&  &1                               \\ 
 {\rm RXCJ0627.1-3529}&{\rm A3392       }&06~27~08.2& -35~29~20&0.0554&  9&  9.969&   6.3&  0.715&  9.0&  0.841&    7.1&  &146                             \\ 
 {\rm RXCJ0627.2-5428}&{\rm A3395       }&06~27~14.4& -54~28~12&0.0506&159& 23.679&   3.2&  1.403& 18.0&  1.509&    8.5&PL&1                               \\ 
 {\rm RXCJ0628.8-4143}&{\rm A3396       }&06~28~49.8& -41~43~42&0.1759& 10&  5.743&   9.2&  4.521& 12.0&  4.613&    6.3&  &E                               \\ 
 {\rm RXCJ0631.3-5610}&{\rm             }&06~31~20.7& -56~10~20&0.0540&  3&  7.695&   9.0&  0.525& 20.0&  0.536&    7.9&L &E                               \\ 
 {\rm RXCJ0637.3-4828}&{\rm A3399       }&06~37~18.9& -48~28~42&0.2026& 11&  3.400&  10.2&  3.641&  6.5&  3.958&    6.9&  &E                               \\ 
 {\rm RXCJ0638.7-5358}&{\rm S0592       }&06~38~46.5& -53~58~18&0.2266& 11&  7.531&   8.1& 10.085&  9.5& 10.616&    6.6&  &E                               \\ 
 {\rm RXCJ0645.4-5413}&{\rm A3404       }&06~45~29.3& -54~13~08&0.1644&  2& 10.597&   7.9&  7.139& 13.0&  7.360&    6.6&  &E                               \\ 
 {\rm RXCJ0658.5-5556}&{\rm 1ES0657^h   }&06~58~31.1& -55~56~49&0.2965& 78&  9.079&   7.9& 21.646&  9.0& 23.028&    6.3&  &E,73,79                         \\ 
 {\rm RXCJ0712.0-6030}&{\rm AM 0711^i   }&07~12~05.4& -60~30~06&0.0322&  2&  3.515&  17.7&  0.083& 15.0&  0.087&    9.7&  &E                               \\ 
 {\rm RXCJ0738.1-7506}&{\rm             }&07~38~09.0& -75~06~24&0.1110&  3&  3.083&  15.9&  0.938&  8.5&  0.998&   17.4&  &E                               \\ 
 {\rm RXCJ0821.8+0112}&{\rm A0653       }&08~21~51.7& +01~12~42&0.0822&  6&  4.142&  19.2&  0.673& 12.0&  0.701&    4.2&B &E                               \\ 
 {\rm RXCJ0909.1-0939}&{\rm A0754       }&09~09~08.4& -09~39~58&0.0542& 92& 57.001&   3.3&  3.879& 19.0&  4.310&    4.6&  &146                             \\ 
 {\rm RXCJ0910.6-1034}&{\rm A0761       }&09~10~36.3& -10~34~52&0.0916&  2&  5.161&  11.5&  1.042&  7.5&  1.171&    4.6&  &53,81                           \\ 
 {\rm RXCJ0918.1-1205}&{\rm A0780       }&09~18~06.5& -12~05~36&0.0539&  4& 39.461&   9.1&  2.659& 15.0&  3.022&    4.9&BX&1                               \\ 
 {\rm RXCJ0937.9-2020}&{\rm S0617       }&09~37~59.7& -20~20~45&0.0344& 11&  3.592&  11.5&  0.098& 12.0&  0.105&    4.0&  &25                              \\ 
 {\rm RXCJ0944.1-2116}&{\rm             }&09~44~10.3& -21~16~40&0.0077&  1&  2.480&  43.5&  0.003& 11.5&  0.004&    4.4&  &E,12,84                         \\ 
 {\rm RXCJ0944.6-2633}&{\rm             }&09~44~36.6& -26~33~56&0.1421&  3&  6.032&  10.0&  3.019&  6.0&  3.510&    6.0&  &E                               \\ 
 {\rm RXCJ0945.4-0839}&{\rm A0868       }&09~45~24.4& -08~39~15&0.1535&  2&  4.726&  10.8&  2.814&  8.5&  2.994&    4.2&  &1,56                            \\ 
 {\rm RXCJ0948.6-8327}&{\rm             }&09~48~39.4& -83~27~56&0.1982&  1&  3.025&  25.9&  3.085&  7.5&  3.282&    9.2&D &E                               \\ 
 {\rm RXCJ0953.2-1558}&{\rm             }&09~53~12.1& -15~58~52&0.0302&  2&  3.675&  13.9&  0.077& 15.0&  0.081&    5.3&  &E                               \\ 
 {\rm RXCJ0956.4-1004}&{\rm A0901       }&09~56~26.4& -10~04~12&0.1634&  9&  9.115&   9.6&  6.077& 17.0&  6.077&    5.1&D &E                               \\ 
 {\rm RXCJ0958.3-1103}&{\rm A0907       }&09~58~22.1& -11~03~35&0.1669&  2&  7.833&   8.3&  5.472&  8.5&  5.948&    5.1&  &E                               \\ 
 {\rm RXCJ1013.5-1350}&{\rm             }&10~13~36.0& -13~50~33&0.1517&  5&  3.776&  14.6&  2.191&  6.5&  2.434&    7.6&  &H                               \\ 
 {\rm RXCJ1013.6-0054}&{\rm A0957       }&10~13~40.3& -00~54~52&0.0445& 48&  9.014&  12.8&  0.412& 16.5&  0.434&    3.8&  &3,41,55,96                      \\ 
 {\rm RXCJ1013.7-0006}&{\rm A0954       }&10~13~44.8& -00~06~31&0.0927& 19&  3.393&  14.9&  0.703&  8.0&  0.764&    3.8&  &55,119                          \\ 
 {\rm RXCJ1017.3-1040}&{\rm A0970       }&10~17~23.4& -10~40~39&0.0586& 51& 11.231&   8.8&  0.905& 12.5&  0.995&    5.1&  &85                              \\ 
 {\rm RXCJ1020.4-0631}&{\rm A0978       }&10~20~28.8& -06~31~11&0.0540& 63&  3.704&  13.2&  0.253& 11.0&  0.269&    4.7&  &2                               \\ 
 {\rm RXCJ1023.8-2715}&{\rm A3444       }&10~23~50.8& -27~15~31&0.2542&  7&  7.055&   9.7& 12.109&  6.0& 13.760&    5.6&  &H                               \\ 
 {\rm RXCJ1027.9-0647}&{\rm A1023       }&10~27~59.6& -06~47~46&0.1176&  6&  3.146&  15.6&  1.075&  6.0&  1.208&    4.5&X\ast &H                           \\ 

            \noalign{\smallskip}
            \hline
         \end{array}
      \]
\end{table*}
%
%

}
{\footnotesize
\begin{table*}
         \label{TabR1d}
      \[
         \begin{array}{llrrrrrrrrrrrl}
            \hline
            \noalign{\smallskip}
 {\rm name}& {\rm alt. name}& R.A. & Decl. & z & N_{gal} & F_x & Error 
& L_x & R_{ap}& L_x^* & N_H & Cm&Ref.  \\
(1) & (2) & (3) & (4) & (5) & (6) & (7) & (8) & (9) & (10) & (11) & (12) & (13) & (14) \\
            \noalign{\smallskip}
            \hline
            \noalign{\smallskip}
 {\rm RXCJ1036.6-2731}&{\rm A1060       }&10~36~41.8& -27~31~28&0.0126&154& 84.353&   6.5&  0.297& 35.5&  0.334&    4.9&L &146                             \\ 
 {\rm RXCJ1038.4-2454}&{\rm             }&10~38~24.1& -24~54~10&0.1230& 10&  4.134&  13.5&  1.545&  8.5&  1.661&    5.5&  &H                               \\ 
 {\rm RXCJ1039.7-0841}&{\rm A1069       }&10~39~44.5& -08~41~01&0.0650& 35&  6.444&  15.2&  0.639& 13.0&  0.673&    3.9&  &2                               \\ 
 {\rm RXCJ1041.5-1123}&{\rm             }&10~41~34.2& -11~23~19&0.0839&  7&  3.178&  15.3&  0.544& 10.0&  0.573&    4.1&  &H                               \\ 
 {\rm RXCJ1044.5-0704}&{\rm A1084       }&10~44~33.0& -07~04~22&0.1342&  6&  9.451&  12.2&  4.213&  7.0&  4.899&    3.4&  &H                               \\ 
 {\rm RXCJ1050.4-1250}&{\rm USGC S152^j }&10~50~25.5& -12~50~47&0.0155&  6& 10.929&   6.8&  0.059& 13.0&  0.072&    5.0&  &E,20,31                         \\ 
 {\rm RXCJ1050.5-0236}&{\rm A1111       }&10~50~35.5& -02~36~00&0.1651&  9&  3.636&  15.0&  2.512&  7.0&  2.730&    4.0&BX&H                               \\ 
 {\rm RXCJ1050.6-2405}&{\rm             }&10~50~36.3& -24~05~46&0.2037&  1&  3.601&  18.4&  3.904&  8.5&  4.067&    5.9&X\ast &E                           \\ 
 {\rm RXCJ1058.1+0135}&{\rm A1139       }&10~58~10.4& +01~35~11&0.0398& 27&  2.379&  20.0&  0.087& 15.0&  0.089&    4.0&B &146                             \\ 
 {\rm RXCJ1101.2-2243}&{\rm A1146       }&11~01~16.8& -22~43~41&0.1416& 58&  3.464&  15.4&  1.740&  7.0&  1.891&    6.2&  &1,86                            \\ 
 {\rm RXCJ1107.3-2300}&{\rm S0651       }&11~07~19.0& -23~00~08&0.0639&  2&  4.570&  16.2&  0.441& 15.0&  0.450&    5.2&  &H                               \\ 
 {\rm RXCJ1114.1-3811}&{\rm  MS1111.8^k }&11~14~12.0& -38~11~21&0.1306&  9&  6.979&  10.9&  2.942& 12.0&  3.065&    9.9&  &E                               \\ 
 {\rm RXCJ1115.8+0129}&{\rm             }&11~15~54.0& +01~29~44&0.3499&  3&  3.848&  22.2& 13.172&  9.0& 13.579&    4.4&B &E                               \\ 
 {\rm RXCJ1130.3-1434}&{\rm A1285       }&11~30~19.5& -14~34~59&0.1068&  8&  9.929&  13.1&  2.754& 14.5&  2.839&    4.1&  &H                               \\ 
 {\rm RXCJ1131.9-1955}&{\rm A1300       }&11~31~56.3& -19~55~37&0.3075& 60&  5.117&  14.7& 13.270&  8.5& 13.968&    4.5&  &87,88                           \\ 
 {\rm RXCJ1135.2-1331}&{\rm A1317       }&11~35~17.2& -13~31~08&0.0722&  6&  4.046&  17.6&  0.505& 11.0&  0.532&    3.8&  &19,89                           \\ 
 {\rm RXCJ1139.4-3327}&{\rm             }&11~39~27.3& -33~27~14&0.1076&  5&  4.313&  55.2&  1.220&  7.5&  1.341&    6.9&  &H                               \\ 
 {\rm RXCJ1141.4-1216}&{\rm A1348       }&11~41~24.3& -12~16~20&0.1195&  6&  5.344&  12.3&  1.877&  7.0&  2.109&    3.3&X &H,16                            \\ 
 {\rm RXCJ1145.3-3425}&{\rm A3490       }&11~45~19.1& -34~25~43&0.0697&  7&  6.414&  15.6&  0.737&  9.0&  0.828&    6.4&  &H                               \\ 
 {\rm RXCJ1149.7-1219}&{\rm A1391       }&11~49~47.8& -12~19~04&0.1557&  6&  4.792&  15.2&  2.930& 11.0&  3.021&    3.1&X &19,31                           \\ 
 {\rm RXCJ1151.6-1619}&{\rm             }&11~51~38.1& -16~19~14&0.0722&  4&  5.182&  12.3&  0.641&  7.0&  0.745&    3.5&  &H                               \\ 
 {\rm RXCJ1200.0-3124}&{\rm A3497       }&12~00~05.0& -31~24~21&0.0685&  2&  6.467&  32.4&  0.716& 12.5&  0.762&    5.7&  &H                               \\ 
 {\rm RXCJ1202.9-0650}&{\rm A1448       }&12~02~54.3& -06~50~07&0.1268&  6&  3.890&  14.1&  1.552& 12.0&  1.584&    3.0&  &16,31,80                        \\ 
 {\rm RXCJ1203.2-2131}&{\rm A1451       }&12~03~17.0& -21~31~22&0.1992& 16&  6.271&  21.4&  6.408&  5.0&  7.629&    4.5&  &H                               \\ 
 {\rm RXCJ1204.4+0154}&{\rm MKW4        }&12~04~25.2& +01~54~02&0.0199& 25& 17.188&   6.0&  0.153& 17.0&  0.176&    1.9&  &64,94,95,97                     \\ 
 {\rm RXCJ1206.2-0848}&{\rm             }&12~06~12.5& -08~48~22&0.4414&  1&  3.231&  15.8& 18.325&  5.0& 20.590&    4.2&  &E                               \\ 
 {\rm RXCJ1212.3-1816}&{\rm             }&12~12~18.9& -18~16~43&0.2690&  2&  3.121&  45.6&  6.073& 10.0&  6.197&    4.5&X &E                               \\ 
 {\rm RXCJ1215.4-3900}&{\rm             }&12~15~29.0& -39~00~55&0.1190&  6&  5.534&  35.5&  1.916& 10.5&  2.017&    6.3&  &H                               \\ 
 {\rm RXCJ1219.3-1315}&{\rm A1520       }&12~19~19.8& -13~15~36&0.0688&  5&  4.081&  35.0&  0.460&  2.5&  0.821&    4.0&  &E                               \\ 
 {\rm RXCJ1234.2-3856}&{\rm             }&12~34~17.0& -38~56~34&0.2373&  2&  3.995&  14.0&  5.979&  4.0&  7.291&    6.2&X\ast &E                           \\ 
 {\rm RXCJ1236.7-3354}&{\rm S0700       }&12~36~44.7& -33~54~10&0.0796&  4&  4.932&  19.8&  0.749&  8.5&  0.832&    5.6&  &H                               \\ 
 {\rm RXCJ1236.7-3531}&{\rm S0701       }&12~36~46.5& -35~31~58&0.0736&  2&  3.913&  16.4&  0.508&  8.5&  0.558&    5.3&  &H                               \\ 
 {\rm RXCJ1244.6-1159}&{\rm A1606       }&12~44~38.0& -11~59~07&0.0963&  9&  6.344&  13.1&  1.416& 10.5&  1.506&    3.7&  &31                              \\ 
 {\rm RXCJ1247.7-0247}&{\rm A1612       }&12~47~43.2& -02~47~32&0.1797&  2&  3.111&  34.0&  2.595& 11.5&  2.621&    1.8&  &E                               \\ 
 {\rm RXCJ1248.7-4118}&{\rm A3526       }&12~48~47.9& -41~18~28&0.0114&287&250.983&   2.4&  0.721& 80.0&  0.751&    8.2&L &158                             \\ 
 {\rm RXCJ1252.5-3116}&{\rm             }&12~52~34.1& -31~16~04&0.0535&  3& 12.910&   7.8&  0.861&  9.5&  1.013&    5.5&  &H                               \\ 
 {\rm RXCJ1253.0-0912}&{\rm HCG 62      }&12~53~05.5& -09~12~01&0.0146& 45&  6.538&  13.2&  0.031& 13.0&  0.037&    2.9&  &48                              \\ 
 {\rm RXCJ1253.2-1522}&{\rm A1631       }&12~53~14.4& -15~22~48&0.0462& 71&  6.720&  14.9&  0.332& 18.5&  0.342&    4.0&  &1,148                           \\ 
 {\rm RXCJ1253.6-3931}&{\rm             }&12~53~40.9& -39~31~55&0.1794&  1& 11.009&   8.3&  8.971&  6.0& 10.680&    7.9&X\ast &E                           \\ 
 {\rm RXCJ1254.0-0642}&{\rm A1634       }&12~54~02.7& -06~42~04&0.1962&  2&  4.049&  22.5&  4.016& 12.5&  4.057&    2.5&B &E                               \\ 
 {\rm RXCJ1254.3-2901}&{\rm A3528N      }&12~54~23.5& -29~01~22&0.0542& 69&  8.801&  20.0&  0.603&  8.0&  0.727&    6.1&PB&2                               \\ 
 {\rm RXCJ1254.6-2913}&{\rm A3528S      }&12~54~41.4& -29~13~24&0.0544& 52& 15.402&  16.0&  1.064& 12.0&  1.196&    6.1&PB&H,3,64,75,98                    \\ 
 {\rm RXCJ1254.7-1526}&{\rm             }&12~54~46.9& -15~26~40&0.1506& 14&  3.737&  18.9&  2.134& 10.5&  2.200&    4.0&  &H                               \\ 
 {\rm RXCJ1255.5-3019}&{\rm A3530       }&12~55~34.5& -30~19~50&0.0541& 46& 10.314&  10.4&  0.704& 16.5&  0.741&    6.0&B &75,77,76,98                     \\ 
 {\rm RXCJ1255.7-1239}&{\rm             }&12~55~42.1& -12~39~18&0.0585& 13&  2.989&  22.2&  0.241& 10.5&  0.256&    3.6&  &16,31,99                        \\ 
 {\rm RXCJ1256.9-3119}&{\rm             }&12~56~59.8& -31~19~19&0.0561& 16&  5.378&  16.0&  0.397& 12.5&  0.422&    6.0&  &12,75                           \\ 
 {\rm RXCJ1257.1-1724}&{\rm A1644       }&12~57~09.8& -17~24~01&0.0473& 92& 37.967&  10.0&  1.952& 22.0&  2.077&    5.3&L &1                               \\ 
 {\rm RXCJ1257.1-1339}&{\rm             }&12~57~10.1& -13~39~20&0.0151&  2&  2.302&  18.0&  0.012&  6.5&  0.016&    3.5&  &20                              \\ 
 {\rm RXCJ1257.2-3022}&{\rm A3532       }&12~57~16.9& -30~22~37&0.0554& 44& 18.733&  24.0&  1.340& 15.5&  1.457&    6.0&B &1,104                           \\ 
 {\rm RXCJ1258.6-0145}&{\rm A1650       }&12~58~41.1& -01~45~25&0.0845&  2& 20.909&   6.1&  3.563& 14.0&  3.873&    1.5&  &1                               \\ 
 {\rm RXCJ1258.8-2640}&{\rm A1648       }&12~58~49.8& -26~40~03&0.0767&  2&  5.119&  31.2&  0.719&  8.5&  0.799&    6.9&  &H                               \\ 
 {\rm RXCJ1259.3-0411}&{\rm A1651       }&12~59~21.5& -04~11~41&0.0845& 34& 23.207&   6.0&  3.946& 14.0&  4.289&    1.7&  &78,105                          \\ 
 {\rm RXCJ1301.6-0650}&{\rm             }&13~01~36.3& -06~50~00&0.0898&  4&  4.008&  76.4&  0.776& 14.0&  0.792&    2.6&  &31                              \\ 
 {\rm RXCJ1302.8-0230}&{\rm A1663       }&13~02~50.7& -02~30~22&0.0847&  3&  4.460&  21.9&  0.772& 12.0&  0.804&    1.7&  &E,19,31                         \\ 
 {\rm RXCJ1303.7-2415}&{\rm A3541       }&13~03~44.0& -24~15~03&0.1288&  8&  8.656&  12.5&  3.541&  7.5&  4.024&    9.2&  &H                               \\ 
 {\rm RXCJ1304.2-3030}&{\rm             }&13~04~16.7& -30~30~55&0.0117&  8&  8.280&  13.6&  0.025& 20.0&  0.028&    6.2&L &5,82,83,98                      \\ 
 {\rm RXCJ1305.9-3739}&{\rm S0721       }&13~05~54.5& -37~39~41&0.0497& 25&  8.181&  23.9&  0.470& 16.0&  0.495&    5.6&DB&14                              \\ 
 {\rm RXCJ1309.2-0136}&{\rm MS1306.7^l  }&13~09~17.0& -01~36~45&0.0880&  0&  5.334&  15.1&  0.989& 12.5&  1.030&    1.8&  &39                              \\ 
 {\rm RXCJ1311.4-0120}&{\rm A1689       }&13~11~30.0& -01~20~07&0.1832& 66& 15.332&   8.0& 13.088& 10.5& 14.073&    1.8&  &1                               \\ 
 {\rm RXCJ1314.2-0659}&{\rm A1698       }&13~14~13.5& -06~59~36&0.1892&  7&  3.942&  19.9&  3.624& 11.0&  3.698&    2.8&  &E                               \\ 

            \noalign{\smallskip}
            \hline
         \end{array}
      \]
\end{table*}
%
%

}
{\footnotesize
\begin{table*}
         \label{TabR1e}
      \[
         \begin{array}{llrrrrrrrrrrrl}
            \hline
            \noalign{\smallskip}
 {\rm name}& {\rm alt. name}& R.A. & Decl. & z & N_{gal} & F_x & Error 
& L_x & R_{ap}& L_x^* & N_H & Cm&Ref.  \\
(1) & (2) & (3) & (4) & (5) & (6) & (7) & (8) & (9) & (10) & (11) & (12) & (13) & (14) \\
            \noalign{\smallskip}
            \hline
            \noalign{\smallskip}
 {\rm RXCJ1314.4-2515}&{\rm             }&13~14~28.0& -25~15~41&0.2439&  9&  6.794&  16.5& 10.615& 11.5& 10.943&    6.7&  &H                               \\ 
 {\rm RXCJ1315.3-1623}&{\rm NGC 5044    }&13~15~24.0& -16~23~23&0.0087& 16& 58.129&   4.0&  0.097& 36.0&  0.110&    4.9&L &15,20,64,74,90,91,92,93         \\ 
 {\rm RXCJ1317.1-3821}&{\rm             }&13~17~09.6& -38~21~42&0.2553&  2&  4.062&  14.8&  7.059&  4.5&  8.305&    4.7&  &E                               \\ 
 {\rm RXCJ1320.7-4102}&{\rm S0727       }&13~20~42.7& -41~02~23&0.0495&  7&  4.906&  17.8&  0.280& 13.0&  0.298&    7.3&  &E                               \\ 
 {\rm RXCJ1325.1-2014}&{\rm A1732       }&13~25~06.8& -20~14~11&0.1926& 10&  4.160&  14.8&  3.959&  9.0&  4.124&    7.8&  &H                               \\ 
 {\rm RXCJ1326.2+0013}&{\rm             }&13~26~17.8& +00~13~32&0.0826& 16&  5.548&  11.7&  0.911&  9.0&  1.001&    1.8&X &E,55,96,55,96                   \\ 
 {\rm RXCJ1326.9-2710}&{\rm A1736       }&13~26~54.0& -27~10~60&0.0458&109& 36.893&   6.5&  1.778& 34.0&  1.796&    5.4&L &146                             \\ 
 {\rm RXCJ1327.9-3130}&{\rm A3558       }&13~27~57.5& -31~30~09&0.0480&341& 58.526&   3.9&  3.096& 18.0&  3.518&    3.6&P &1,14,44,102,141,148             \\ 
 {\rm RXCJ1329.7-3136}&{\rm A3558 (B)   }&13~29~42.9& -31~36~09&0.0488& 57& 14.434&  30.0&  0.796& 14.5&  0.865&    3.6&PB&3,14,44,63,101,102              \\ 
 {\rm RXCJ1330.8-0152}&{\rm A1750       }&13~30~49.9& -01~52~22&0.0852& 46& 13.211&  11.6&  2.291& 24.0&  2.291&    2.5&DL&1,41,96                         \\ 
 {\rm RXCJ1331.5-3148}&{\rm A3558 (C)   }&13~31~32.4& -31~48~55&0.0448& 55&  4.901&  32.0&  0.226&  9.0&  0.257&    4.0&PB&H,3,14,74,76,100,101,102        \\ 
 {\rm RXCJ1332.3-3308}&{\rm A3560       }&13~32~22.6& -33~08~22&0.0487& 29& 14.129&   9.3&  0.775& 15.5&  0.833&    3.9&  &H,76,87,103,108                 \\ 
 {\rm RXCJ1332.9-2519}&{\rm             }&13~32~55.8& -25~19~26&0.1199&  3&  3.427&  27.0&  1.212& 12.0&  1.237&    6.1&X &E                               \\ 
 {\rm RXCJ1333.6-3139}&{\rm A3562       }&13~33~36.3& -31~39~40&0.0490&114& 24.485&   6.6&  1.357& 17.0&  1.475&    3.9&P &1,141                           \\ 
 {\rm RXCJ1333.6-2317}&{\rm A1757       }&13~33~42.0& -23~17~02&0.1264&  4&  6.442&  15.0&  2.540& 12.0&  2.646&    7.1&  &H                               \\ 
 {\rm RXCJ1336.6-3357}&{\rm A3565       }&13~36~38.8& -33~57~30&0.0123& 45&  2.348&  18.0&  0.008&  9.5&  0.010&    4.1&  &146                             \\ 
 {\rm RXCJ1337.4-4120}&{\rm             }&13~37~28.1& -41~20~01&0.0519&  2&  4.813&  16.3&  0.303&  8.5&  0.344&    6.3&  &E                               \\ 
 {\rm RXCJ1342.0+0213}&{\rm A1773       }&13~42~05.5& +02~13~39&0.0765& 14&  5.788&  12.9&  0.808& 11.5&  0.860&    1.8&  &1,10                            \\ 
 {\rm RXCJ1346.2-1632}&{\rm             }&13~46~13.6& -16~32~42&0.0557&  1&  3.064&  16.8&  0.223&  6.0&  0.265&    7.5&  &E                               \\ 
 {\rm RXCJ1346.8-3752}&{\rm A3570       }&13~46~52.5& -37~52~28&0.0377&  8&  5.527&  33.8&  0.180& 14.0&  0.191&    4.4&  &4,12,113                        \\ 
 {\rm RXCJ1347.2-3025}&{\rm A3574W      }&13~47~12.3& -30~25~10&0.0145&  5&  2.479&  50.6&  0.012& 10.0&  0.014&    4.4&PX&12,64,74,103,107,112,114,115    \\ 
 {\rm RXCJ1347.4-3250}&{\rm A3571       }&13~47~28.4& -32~50~59&0.0391& 84&115.471&  10.0&  3.996& 36.0&  4.206&    3.9&L &146                             \\ 
 {\rm RXCJ1347.5-1144}&{\rm             }&13~47~30.0& -11~44~56&0.4516&  2&  6.468&  10.4& 38.732&  5.5& 44.520&    4.9&  &E                               \\ 
 {\rm RXCJ1348.9-2526}&{\rm A1791       }&13~48~55.8& -25~26~31&0.1269&  8&  3.842&  28.8&  1.536& 11.0&  1.584&    5.8&  &E                               \\ 
 {\rm RXCJ1349.3-3018}&{\rm A3574E      }&13~49~19.3& -30~18~34&0.0160& 55&  7.549&  25.0&  0.043& 20.0&  0.047&    4.4&PX&146                             \\ 
 {\rm RXCJ1350.7-3343}&{\rm             }&13~50~43.9& -33~43~17&0.1142&  1&  4.021&  17.3&  1.280&  7.0&  1.422&    4.8&  &E                               \\ 
 {\rm RXCJ1352.2-0945}&{\rm A1807       }&13~52~14.0& -09~45~33&0.2021&  2&  3.031&  16.6&  3.225& 10.0&  3.291&    3.5&  &E                               \\ 
 {\rm RXCJ1353.4-2753}&{\rm             }&13~53~29.3& -27~53~19&0.0468&  3&  4.993&  20.8&  0.253& 10.0&  0.281&    4.7&B &E                               \\ 
 {\rm RXCJ1401.6-1107}&{\rm A1837       }&14~01~36.7& -11~07~28&0.0698& 38&  5.517&  13.4&  0.634&  9.5&  0.697&    4.7&  &1                               \\ 
 {\rm RXCJ1403.5-3359}&{\rm S0753       }&14~03~35.9& -33~59~16&0.0132& 12&  8.081&  14.4&  0.032& 17.0&  0.036&    5.6&  &5,91,98,109,114                 \\ 
 {\rm RXCJ1407.4-2700}&{\rm A3581       }&14~07~28.1& -27~00~55&0.0230& 24& 26.535&   5.8&  0.316& 15.0&  0.381&    4.3&  &E,146                           \\ 
 {\rm RXCJ1408.1-0904}&{\rm CAN 40^b    }&14~08~07.0& -09~04~16&0.0354&  4&  4.781&  13.9&  0.137& 11.5&  0.151&    3.7&  &E,120                           \\ 
 {\rm RXCJ1415.2-0030}&{\rm A1882       }&14~15~14.2& -00~30~04&0.1403& 20&  4.217&  20.0&  2.063& 15.5&  2.063&    3.2&X &55                              \\ 
 {\rm RXCJ1416.8-1158}&{\rm             }&14~16~51.5& -11~58~34&0.0982& 10&  3.976&  17.7&  0.931&  5.5&  1.108&    6.0&X &16,31                           \\ 
 {\rm RXCJ1421.9-2009}&{\rm             }&14~21~57.3& -20~09~36&0.1208&  4&  3.078&  25.7&  1.115& 10.0&  1.149&    7.5&  &H                               \\ 
 {\rm RXCJ1435.0-2821}&{\rm A3605       }&14~35~00.6& -28~21~56&0.0689&  2&  4.358&  50.3&  0.491& 17.5&  0.496&    6.4&  &E                               \\ 
 {\rm RXCJ1436.8-0900}&{\rm             }&14~36~53.5& -09~00~13&0.0842&  2&  3.776&  30.3&  0.646&  5.0&  0.798&    6.2&  &E                               \\ 
 {\rm RXCJ1455.2-3325}&{\rm             }&14~55~13.9& -33~25~36&0.1158&  1&  3.815&  16.7&  1.252&  6.0&  1.423&    7.1&  &E                               \\ 
 {\rm RXCJ1456.3-0549}&{\rm A1994       }&14~56~19.6& -05~49~39&0.2200&  5&  3.059&  16.1&  3.900&  5.5&  4.333&    6.3&  &E                               \\ 
 {\rm RXCJ1459.0-0843}&{\rm             }&14~59~03.5& -08~43~30&0.1043&  3&  3.837&  16.0&  1.022&  8.5&  1.099&    7.1&  &E                               \\ 
 {\rm RXCJ1459.4-1811}&{\rm S0780       }&14~59~29.3& -18~11~13&0.2357&  2& 10.187&  18.0& 14.910& 11.5& 15.531&    7.8&  &E                               \\ 
 {\rm RXCJ1501.1+0141}&{\rm NGC5813     }&15~01~11.9& +01~41~53&0.0050&  6& 12.324&   6.3&  0.007&  8.5&  0.014&    4.2&  &12,15                           \\ 
 {\rm RXCJ1504.1-0248}&{\rm             }&15~04~07.7& -02~48~18&0.2153&  6& 22.111&   5.3& 26.389& 12.0& 28.073&    6.0&X &E                               \\ 
 {\rm RXCJ1506.4+0136}&{\rm NGC5846     }&15~06~29.7& +01~36~08&0.0066&  8&  8.028&  13.8&  0.008& 13.0&  0.011&    4.2&  &48,64,123,125,126               \\ 
 {\rm RXCJ1511.5+0145}&{\rm             }&15~11~33.5& +01~45~51&0.0384&  2&  2.459&  58.5&  0.084& 17.0&  0.085&    4.0&B &E,123                           \\ 
 {\rm RXCJ1512.2-2254}&{\rm             }&15~12~12.6& -22~54~59&0.3152&  1&  3.452&  17.3&  9.371&  6.0& 10.186&    8.5&  &E                               \\ 
 {\rm RXCJ1512.8-0128}&{\rm             }&15~12~51.0& -01~28~47&0.1223&  2&  3.354&  18.5&  1.238&  5.5&  1.423&    5.2&  &E                               \\ 
 {\rm RXCJ1514.9-1523}&{\rm             }&15~14~58.0& -15~23~10&0.2226&  3&  5.270&  16.7&  6.802&  9.0&  7.160&    8.5&  &E                               \\ 
 {\rm RXCJ1516.3+0005}&{\rm A2050       }&15~16~19.2& +00~05~52&0.1181& 17&  4.956&  13.8&  1.697&  7.5&  1.886&    4.6&  &E,55                            \\ 
 {\rm RXCJ1516.5-0056}&{\rm A2051       }&15~16~34.0& -00~56~56&0.1198&  7&  3.918&  61.7&  1.383& 13.0&  1.397&    5.5&  &55                              \\ 
 {\rm RXCJ1524.2-3154}&{\rm             }&15~24~12.8& -31~54~14&0.1028&  6& 11.346&  15.4&  2.891& 11.0&  3.142&    8.3&  &E                               \\ 
 {\rm RXCJ1539.5-8335}&{\rm             }&15~39~33.9& -83~35~32&0.0728&  2& 16.844&   8.5&  2.102&  9.0&  2.502&    8.5&  &E                               \\ 
 {\rm RXCJ1540.1-0318}&{\rm A2104       }&15~40~07.5& -03~18~29&0.1533& 52&  7.434&  10.1&  4.375&  9.0&  4.704&    9.2&  &128                             \\ 
 {\rm RXCJ1548.7-0259}&{\rm A2128       }&15~48~45.9& -02~59~47&0.1010&  2&  3.943&  16.6&  0.980& 11.0&  1.021&    8.4&B &111,116                         \\ 
 {\rm RXCJ1558.3-1410}&{\rm             }&15~58~23.2& -14~10~04&0.0970&  1& 14.131&   7.5&  3.181& 10.0&  3.574&   11.1&B &117                             \\ 
 {\rm RXCJ1615.7-0608}&{\rm A2163       }&16~15~46.9& -06~08~45&0.2030&  1& 19.686&   6.4& 20.853&  9.5& 23.170&   12.3&  &118                             \\ 
 {\rm RXCJ1633.8-0738}&{\rm             }&16~33~53.9& -07~38~42&0.0974& 10&  3.445&  15.0&  0.793& 11.0&  0.818&   11.8&  &E                               \\ 
 {\rm RXCJ1655.9-0113}&{\rm             }&16~55~56.2& -01~13~45&0.0408&  1&  3.792&  33.0&  0.145& 15.0&  0.151&    8.0&  &93                              \\ 
 {\rm RXCJ1657.7-0149}&{\rm             }&16~57~45.4& -01~49~54&0.0313&  3&  4.194&  16.0&  0.094& 17.5&  0.097&    8.8&  &E                               \\ 
 {\rm RXCJ1705.1-8210}&{\rm S0792       }&17~05~10.3& -82~10~26&0.0737&  4&  7.092&  17.7&  0.916& 12.0&  0.974&    7.6&  &E                               \\ 

            \noalign{\smallskip}
            \hline
         \end{array}
      \]
\begin{list}{}{}
\item
This catalogue is continued in Table~\ref{tab6} where also all the references for
the reference codes in this table are given.
\end{list}
\end{table*}
%
%

}

Further more detailed X-ray properties on the REFLEX clusters as
determined with the GCA method are given in Table~\ref{tab7}, where
the following columns are listed: (1) name, 
{(2) and (3) repeat the J2000 sky coordinates, but now in units of decimal
degrees,} (4) the count rate as
measured with the GCA method for the aperture size given in column 10
in Table~\ref{tab5}, (5) X-ray flux, $F_n$, determined in the first
step for an assumed temperature of 5 keV used for the source
selection, (6) the X-ray luminosity in the rest frame 0.1 to 2.4 keV
band uncorrected for missing (unabsorbed) flux, 
(7) the number of source photons detected
within the reference aperture (column 14), (8) the probability for the X-ray
source to be a point source in values of $-log_{10}(P)$ (determined as
described below), (9) and (10) the best fitted core radius for the
$\beta$-model fit and a minimal core radius still consistent within
$2\sigma$ error limits, respectively.  Note that the core radii
determined here are only a qualitative measure for the source extent,
since the errors are very large and the fitting grid was coarsely
spaced. Therefore core radii and their lower limiting values
can give a further qualitative feeling for the extent of the X-ray 
sources, but {\bf we recommend not to use these results for
$r_c$ as a quantitative measure of cluster shapes at this
point.} Columns (11) and (12) give the spectral hardness ratio defined by
the equation given below and its Poisson error. Column (13)
indicates the deviation of the measured hardness ratio from the
expectation value calculated for given $N_H$ and for an assumed
temperature of 5 keV as factors of $\sigma$.
The last column (14) repeats from Table 6 the
aperture radius within which the source count rate was measured,
but now in physical units of Mpc. This is the 
radius where the cumulative source count rate profile reaches the 
plateau value.
The print version of this paper gives only a few
example lines of this table. The full table as well as a second
version of Table~\ref{tab5} for an Einstein-de Sitter Universe model 
is provided in the electronic version.

\begin{table*}
      \caption{X-ray properties of the REFLEX clusters}
         \label{tab7}
      \[
         \begin{array}{lrrrrrrrrrrrrr}
            \hline
            \noalign{\smallskip}
 {\rm name}&~~RA~~&~~DEC~&~{\rm count}~& F_n~~~~ &L_x~~~~~&N_{ph}~ &P_{ext}~ & r_c~~~~~ & r_c(min)
&~~~~~ HR & HR_{err} & \Delta HR & R_{ap}\\
  &(2000)&(2000)& {\rm rate}~~&  10^{-12}{\rm erg}~~ & 10^{44}~~~ &  &  &  &  &  & &( \sigma)~~ &  \\
  &&& s^{-1}~~   & {\rm s^{-1} cm^{-2}} & {\rm erg s^{-1}} & &  & {\rm arcmin} & {\rm arcmin} &   &  & &{\rm Mpc}\\ 
(1) & (2) & (3) & (4) & (5) & (6) & (7) & (8) & (9) & (10) & (11) & (12) & (13) & (14) \\
            \noalign{\smallskip}
            \hline
            \noalign{\smallskip}
 {\rm RXCJ0003.1-0605}&  0.7991&  -6.0860&0.218&  4.473&  6.395&   76.3&  3.8&  1.5&  1.0&  0.7&  0.1& 0.06&  1.664\\ 
 {\rm RXCJ0003.2-3555}&  0.8006& -35.9271&0.400&  7.762&  0.421&   56.8&  7.5&  1.5&  0.5&  0.5&  0.2& 0.70&  0.575\\ 
 {\rm RXCJ0003.8+0203}&  0.9607&+  2.0634&0.205&  4.192&  0.855&   74.2&  3.3&  1.0&  0.5&  0.8&  0.1& 1.18&  0.877\\ 
 {\rm RXCJ0006.0-3443}&  1.5126& -34.7241&0.301&  5.853&  1.875&   68.0&  5.4&  1.5&  1.0&  0.2&  0.1&-1.05&  1.248\\ 
 {\rm RXCJ0011.3-2851}&  2.8364& -28.8551&0.612& 12.129&  1.089&  155.4& 13.9&  1.5&  1.5&  0.4&  0.1&-1.53&  0.896\\ 
 {\rm RXCJ0013.6-1930}&  3.4094& -19.5021&0.306&  6.089&  1.285&   95.8&  6.1&  1.5&  1.5&  0.5&  0.1&-0.61&  1.203\\ 
 {\rm RXCJ0014.3-6604}&  3.5768& -66.0775&0.220&  4.474&  2.907&   60.9&  0.4&  0.0&  0.0&  0.6&  0.1&-0.80&  1.240\\ 
 {\rm RXCJ0014.3-3023}&  3.5783& -30.3834&0.250&  4.929& 12.787&   30.5&  1.3&  0.5&  0.0&  0.2&  0.2&-1.67&  3.255\\ 
 {\rm RXCJ0015.4-2350}&  3.8500& -23.8450&0.175&  3.526&  0.336&   48.5&  1.8&  6.0&  2.0&  0.4&  0.2&-1.36&  0.854\\ 
 {\rm RXCJ0017.5-3509}&  4.3904& -35.1650&0.166&  3.241&  0.729&   58.8&  6.7&  3.0&  2.0&  0.3&  0.2&-0.73&  1.236\\ 

            \noalign{\smallskip}
            \hline
         \end{array}
      \]
The complete table is given in electronic form at CDS and our home page.
\end{table*}
%
%


The probability of an X-ray source to be a point source, as given in
column (5) was determined by means of a Kolmogorov-Smirnov test
comparing the normalized radially cumulative photon distribution 
within an aperture radius of 6 arcmin with
the expectations for a point source and background. In this test we
consider a source to be significantly extended if the probability is
less than 1\%, corresponding to an entry value in Table~\ref{tab7}
larger than 2. The source hardness ratio, as given in column (8), was
determined from the source count rate in the soft (ROSAT PSPC channel
11 to 40) and hard (PSPC channel 52 to 201) band by the formula:

$$ HR = {H - S \over H + S} \eqno$$    

where H is the hard band and S the soft band count rate.

Figure 1 shows the sky distribution of the REFLEX clusters with five
different X-ray flux classes marked by different symbols. The empty
region around RA = $75\deg$ (5 hr) and DEC = $-70\deg$ is due to 
the excision of the regions of
the Large and Small Magellanic Clouds. Fig. 2 shows the same sky
distribution but now with the clusters sorted in five different
redshift classes. This illustrates the large scale clustering of the
clusters, and several known superclusters are marked.

\section{Selection properties of the cluster sample}

\begin{figure}
\psfig{figure=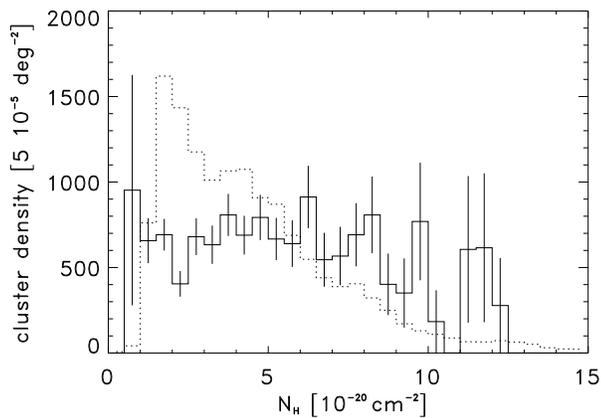,height=6.2cm}
\caption{Sky density of the REFLEX clusters as a function of 
the Galactic hydrogen column density (solid line) in units of deg$^2$
multiplied by a factor of 20000. The dotted line gives the sky area distribution
of $N_H$ values in units of deg$^2$ for $0.5 \cdot 10^{20}$ cm$^{-2}$ bins.}
\label{fig4}
\end{figure}

A number of tests in papers I through III demonstrate the quality
of the catalogue, in particular the completeness and homogeneity. 
For example through the
LogN-logS distribution, a $V/V_{max}$-test, and the photon number
distribution, do not indicate a significant
deficiency at faint fluxes or low source photon numbers.  The sample
shows a large scale density homogeneity at least out to $z = 0.15$
with pronounced traces of large scale structure on scales smaller than
the sample size (paper III). The two-point correlation function
(paper II) shows
a good overall isotropy (except for possibly small dynamical redshift
distortion effects which are expected and currently investigated)
 and the number count distribution on large scales
($\ge 50 h^{-1}$ Mpc) is approximately Gaussian (paper IV).  We do not
see any significant bias in the cluster density as a function of
Galactic latitude or longitude (papers II and III). Here we provide
another test, showing the cluster density as a function of
interstellar absorption. Fig.~\ref{fig4} shows the 
surface density of clusters averaged over
$0.5\,10^{20}$ cm$^{-2}$ column density intervals and compares it to the
distribution of $N_H$ over the sky.
The density is
almost constant within the formal Poissonian errors (which do not
include fluctuations from large-scale structure and cosmic variance)
up to the region where the statistics becomes too poor and the
expectation value drops below one or two.  Since the column density as
well as the stellar density, which hampers the cluster identification,
increases with decreasing $|b|$, we
are also testing here the bias in the optical cluster identification
due to crowding of the sky fields near the Galactic plane and find no
major effect. These plots have to be compared with similar tests
conducted with the optically identified clusters in the Abell
catalogue (e.g. Ebeling et al. 1996).

   \begin{table}
      \caption{Statistics of the sky coverage as a function of the
      sensitivity. A graphical representation of this table can be
      found in B\"ohringer et al. 2001, Fig. 23}
         \label{tab8}
      \[
         \begin{array}{ll}
            \hline
            \noalign{\smallskip}
 {\rm sensitivity~ limit} & {\rm fraction~ of~ the} \\
  10^{-13} {\rm erg~s^{-1}~cm^{-2}~count^{-1}} & {\rm survey~ area} \\
            \noalign{\smallskip}
            \hline
            \noalign{\smallskip}
  0.102 & 8.6~10^{-4} \\
  0.300 & 0.032 \\
  0.500 & 0.192 \\
  0.700 & 0.574 \\
  1.000 & 0.780 \\
  2.000 & 0.940 \\
  3.002 & 0.970 \\
  5.000 & 0.984 \\
            \noalign{\smallskip}
            \hline
         \end{array}
      \]
The complete table is given in electronic form at CDS and our home page.
   \end{table}

   \begin{table}
      \caption{Sensitivity as a function of the sky position used to
      produce the sensitivity map shown in B\"ohringer et al. 2001,
      Fig.22. Also given are the sky distribution of the RASS 2
      exposure (shown in B\"ohringer et al. 2001, Fig.2) and the
      interstellar column density distribution in the REFLEX area.}
         \label{tab9}
      \[
         \begin{array}{lllll}
            \hline
            \noalign{\smallskip}
 {\rm RA } & {\rm DEC } & {\rm exposure} & N_H & {\rm sensitivity~ limit} \\
 {\rm (J2000)} & {\rm (J2000) } & (sec) & 10^{20}*{\rm cm^{-2}} &  
   10^{-13} {\rm erg~s^{-1}~cm^{-2}}\\
 &  &  &  & {\rm ~~~~~~count^{-1}}\\
            \noalign{\smallskip}
            \hline
            \noalign{\smallskip}
  0.0000 & 2.0000 &  401.09 &  3.300 & 0.514 \\
  1.0006 & 2.0000 &  360.85 &  3.000 & 0.567 \\
  2.0012 & 2.0000 &  250.08 &  3.000 & 0.818 \\
  3.0018 & 2.0000 &  332.16 &  3.040 & 0.616 \\
  4.0024 & 2.0000 &  398.57 &  3.070 & 0.514 \\
  5.0030 & 2.0000 &  404.37 &  3.150 & 0.508 \\
  6.0037 & 2.0000 &  442.88 &  3.150 & 0.464 \\
  7.0043 & 2.0000 &  662.26 &  3.040 & 0.309 \\
  8.0049 & 2.0000 &  464.21 &  2.730 & 0.438 \\
  9.0055 & 2.0000 &  291.09 &  2.660 & 0.696 \\
 10.0061 & 2.0000 &  323.05 &  2.320 & 0.622 \\
             \noalign{\smallskip}
            \hline
         \end{array}
      \]
The complete table is given in electronic form at CDS and our home page.
   \end{table}

\begin{figure}                                                                  
\psfig{figure=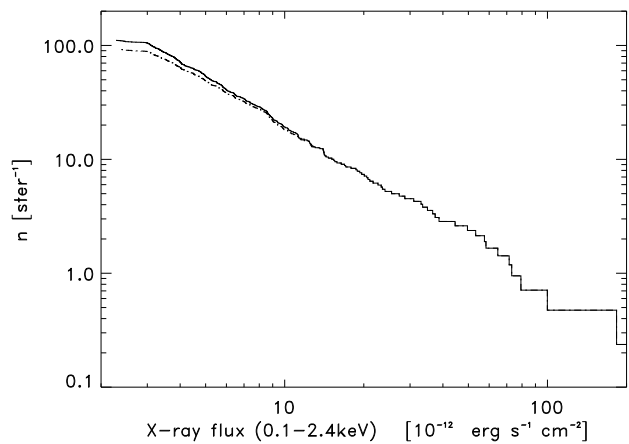,height=6cm}
\caption{LogN-LogS distribution of the REFLEX clusters. The upper curve
shows all the clusters while the lower, dashed curve shows the 79.5\% fraction 
of significantly extended sources as measured with the KS test for
the survey point spread function.}
\label{fig5}
\end{figure}

Another mark of quality of the REFLEX cluster identification is the fact that the 
vast majority of the clusters, 79.5\%, have been identified as extended sources.
Fig.~\ref{fig5} shows the LogN-logS distribution of the subsample 
of clusters detected as significantly extended in 
the GCA.

To reproduce the observational results described here in theoretical
modeling the precise selection function of the REFLEX survey has to be
known. Two major factors modify the depth of our survey as a function
of sky position: the RASS exposure time and the interstellar column
density in the line-of-sight. We model the sensitivity function
which depends on these two parameters in the following way: we
calculate the flux limit reached per number of photons specified to be
required for secure source detection. In this sense Table~\ref{tab8}
gives the sky area in the REFLEX survey for a given flux limit for the
detection of one photon. This sky coverage table is for example used
in the calculation of the luminosity function in paper IV,
where we assume conservatively that at least 30 photons are
detected for the cluster. The flux limit as listed in the first column
of Table~\ref{tab8} has therefore to be multiplied by 30 and we note
for example that 78\% of the survey area reaches the nominal flux
limit of the REFLEX survey of $3 \cdot 10^{-12}$ erg s$^{-1}$ cm
$^{-2}$.  Note also, that this excludes some clusters in the catalogue
with less than 30 source photon counts from the analysis.  For the
large-scale structure analysis in paper III we have included more
clusters for a better three-dimensional coverage and relaxed the
detection requirement to 10 source photons. In this case the REFLEX
survey reaches the nominal flux limit in 97\% of the survey area,
which corresponds to a more simple survey window function in k-space
for the determination of the density fluctuation power spectrum. 

For the large-scale structure modeling, e.g. for the construction of
the comparison Poisson sample used in the evaluation of the two-point
correlation function or the density distribution power spectrum, the
sensitivity limit as a function of the sky position has to be
known. This has been constructed from the exposure map of the RASS2
product and the 21 cm maps from Dickey \& Lockman (1990) and Stark et
al. (1992), and is given in a similar way in Table~\ref{tab9}.  In fact
Table~\ref{tab8} is only a cumulative account of Table~\ref{tab9}.
The table gives the sensitivity distribution on a roughly one square
degree grid.  It lists also the exposure and interstellar column
density distribution used in the calculation. For both tables only a
few example lines are given in the print version of the paper, but the
full tables (which in case of Table~\ref{tab9} consists of 13902
lines) are provided in the electronic version of the paper and on our
home page given above.

{\footnotesize
\begin{table*}
      \caption{The REFLEX cluster catalogue - continued}
         \label{tab6}
      \[
         \begin{array}{llrrrrrrrrrrrl}
            \hline
            \noalign{\smallskip}
 {\rm name}& {\rm alt. name}& R.A. & Decl. & z &  N_{gal} & F_x & Error 
& L_x & R_{ap}& L_x^* & N_H & Cm&Ref.  \\
(1) & (2) & (3) & (4) & (5) & (6) & (7) & (8) & (9) & (10) & (11) & (12) & (13) & (14) \\
            \noalign{\smallskip}
            \hline
            \noalign{\smallskip}
 {\rm RXCJ1706.4-0132}&{\rm Zw1703.8^m  }&17~06~26.6& -01~32~23&0.0912&  4&  6.081&  19.5&  1.209&  9.5&  1.314&   10.2&  &E                               \\ 
 {\rm RXCJ1840.6-7709}&{\rm             }&18~40~37.2& -77~09~20&0.0194&  1& 10.295&  10.4&  0.087& 10.0&  0.112&    8.1&  &34                              \\ 
 {\rm RXCJ1847.3-6320}&{\rm S0805       }&18~47~20.0& -63~20~13&0.0146& 24&  5.968&  26.1&  0.029& 13.0&  0.034&    6.0&  &12                              \\ 
 {\rm RXCJ1855.8-6654}&{\rm             }&18~55~53.2& -66~54~48&0.1797&  2&  3.444&  25.2&  2.856&  8.5&  3.006&    6.1&  &E                               \\ 
 {\rm RXCJ1912.6-7517}&{\rm S0810       }&19~12~40.3& -75~17~30&0.0726& 13&  9.451&  18.1&  1.177& 10.5&  1.293&    6.7&B &E                               \\ 
 {\rm RXCJ1925.4-4256}&{\rm A3638       }&19~25~29.6& -42~56~57&0.0774&  1&  6.550&  21.5&  0.938& 10.0&  1.020&    6.6&  &E,1,121                         \\ 
 {\rm RXCJ1926.9-5342}&{\rm             }&19~26~58.3& -53~42~11&0.0570&  2&  5.504&  26.6&  0.420&  4.0&  0.618&    3.9&  &E                               \\ 
 {\rm RXCJ1928.2-5055}&{\rm A3639       }&19~28~14.0& -50~55~48&0.1496& 10&  3.675&  19.7&  2.069&  6.5&  2.299&    5.3&  &133                             \\ 
 {\rm RXCJ1931.6-3354}&{\rm             }&19~31~38.7& -33~54~47&0.0972&  2&  6.392&  16.8&  1.456&  9.5&  1.583&    7.6&  &E                               \\ 
 {\rm RXCJ1934.7-5053}&{\rm S0821       }&19~34~47.3& -50~53~49&0.2371&  2&  4.040&  21.1&  6.035&  9.0&  6.286&    5.5&  &E                               \\ 
 {\rm RXCJ1938.3-4748}&{\rm             }&19~38~19.5& -47~48~16&0.2665&  1&  3.143&  20.6&  5.987&  8.0&  6.236&    6.1&  &E                               \\ 
 {\rm RXCJ1946.5-4256}&{\rm S0827       }&19~46~34.3& -42~56~32&0.1128&  4&  3.219&  20.8&  1.006&  9.0&  1.059&    5.9&  &E                               \\ 
 {\rm RXCJ1947.3-7623}&{\rm             }&19~47~19.3& -76~23~32&0.2170&  3&  3.527&  20.3&  4.335&  4.0&  5.287&    7.0&  &E                               \\ 
 {\rm RXCJ1952.2-5503}&{\rm A3651       }&19~52~16.5& -55~03~42&0.0600& 79&  7.498&  17.0&  0.635& 12.5&  0.683&    4.9&  &2                               \\ 
 {\rm RXCJ1953.0-5201}&{\rm A3653       }&19~53~00.9& -52~01~51&0.1069& 11&  5.838&  14.3&  1.629& 10.5&  1.733&    4.8&  &E                               \\ 
 {\rm RXCJ1958.2-3011}&{\rm             }&19~58~14.8& -30~11~22&0.1171&  4& 11.271&   9.5&  3.753& 12.0&  3.993&    8.4&X\ast &E                           \\ 
 {\rm RXCJ1959.1-3454}&{\rm A3654       }&19~59~10.7& -34~54~36&0.1728& 12&  3.103&  22.2&  2.373&  5.5&  2.666&    7.5&B &E                               \\ 
 {\rm RXCJ2003.5-2323}&{\rm             }&20~03~30.4& -23~23~05&0.3171&  4&  3.124&  18.2&  8.601&  6.5&  9.248&    8.6&  &E                               \\ 
 {\rm RXCJ2009.0-5421}&{\rm S0849       }&20~09~04.0& -54~21~28&0.0516&  2&  3.140&  20.1&  0.196&  7.0&  0.228&    5.4&  &E                               \\ 
 {\rm RXCJ2009.9-4823}&{\rm S0851       }&20~09~54.1& -48~23~35&0.0097&  4&  3.390&  21.4&  0.007&  8.5&  0.010&    5.0&  &74,135,136                      \\ 
 {\rm RXCJ2011.3-5725}&{\rm             }&20~11~23.1& -57~25~39&0.2786&  5&  3.153&  23.0&  6.649&  6.0&  7.227&    3.3&  &E                               \\ 
 {\rm RXCJ2012.0-4129}&{\rm A3668       }&20~12~03.0& -41~29~30&0.1496&  4&  4.181&  16.0&  2.353&  8.5&  2.503&    5.0&  &E                               \\ 
 {\rm RXCJ2012.5-5649}&{\rm A3667       }&20~12~30.5& -56~49~55&0.0556&162& 70.892&   5.3&  5.081& 26.0&  5.405&    4.6&L &2,3,138,140,141                 \\ 
 {\rm RXCJ2014.2-8038}&{\rm A3664       }&20~14~13.4& -80~38~30&0.1380&  8&  3.525&  19.9&  1.674&  7.0&  1.820&    9.9&P &E                               \\ 
 {\rm RXCJ2014.8-2430}&{\rm             }&20~14~49.7& -24~30~30&0.1612&  2& 14.040&  13.4&  9.157&  6.5& 11.033&    7.4&  &E                               \\ 
 {\rm RXCJ2016.2-8047}&{\rm A3666       }&20~16~14.0& -80~47~57&0.1309&  2&  3.184&  20.3&  1.362&  6.5&  1.497&    9.9&P &E                               \\ 
 {\rm RXCJ2018.4-4102}&{\rm IC4992      }&20~18~25.6& -41~02~48&0.0192&  2&  3.871&  16.6&  0.032&  9.5&  0.039&    4.7&  &5,15                            \\ 
 {\rm RXCJ2018.7-5242}&{\rm S0861       }&20~18~45.7& -52~42~22&0.0505& 13& 11.735&   9.6&  0.694& 14.0&  0.754&    4.7&  &E                               \\ 
 {\rm RXCJ2020.8-3002}&{\rm A3674       }&20~20~49.9& -30~02~01&0.2073&  5&  3.132&  28.4&  3.531&  9.5&  3.640&    6.7&  &E                               \\ 
 {\rm RXCJ2021.9-5256}&{\rm A3675       }&20~21~54.7& -52~56~52&0.1383&  3&  3.359&  16.5&  1.603& 11.5&  1.636&    4.3&  &E                               \\ 
 {\rm RXCJ2023.0-2056}&{\rm S0868       }&20~23~01.6& -20~56~55&0.0564&  2&  5.497&  17.8&  0.411&  8.5&  0.467&    5.6&  &E                               \\ 
 {\rm RXCJ2023.4-5535}&{\rm             }&20~23~24.5& -55~35~32&0.2320&  3&  3.351&  18.6&  4.788&  6.5&  5.204&    5.2&  &E                               \\ 
 {\rm RXCJ2030.7-3532}&{\rm             }&20~30~47.9& -35~32~40&0.1398&  2&  3.220&  20.4&  1.573&  7.0&  1.710&    3.9&  &E                               \\ 
 {\rm RXCJ2031.8-4037}&{\rm             }&20~31~51.5& -40~37~14&0.3416&  2&  3.310&  21.8& 10.715&  5.0& 12.039&    3.9&  &E                               \\ 
 {\rm RXCJ2032.1-5627}&{\rm A3685       }&20~32~08.8& -56~27~09&0.1380&  5&  3.668&  17.7&  1.742&  7.5&  1.893&    5.1&  &E                               \\ 
 {\rm RXCJ2034.3-3429}&{\rm A3693       }&20~34~19.5& -34~29~15&0.1240&  9&  4.385&  22.0&  1.668& 10.0&  1.756&    3.9&  &E,3                             \\ 
 {\rm RXCJ2034.7-3404}&{\rm A3694       }&20~34~42.1& -34~04~26&0.0936&  2&  8.730&  14.2&  1.835& 10.5&  1.995&    3.9&  &40                              \\ 
 {\rm RXCJ2034.7-3548}&{\rm A3695       }&20~34~47.9& -35~48~48&0.0894& 81& 16.940&  12.4&  3.206& 28.0&  3.238&    3.6&L &2,140,141                       \\ 
 {\rm RXCJ2034.9-2143}&{\rm             }&20~34~55.8& -21~43~02&0.1947&  2&  3.310&  42.3&  3.244& 11.0&  3.277&    4.1&  &E                               \\ 
 {\rm RXCJ2035.7-2513}&{\rm A3698       }&20~35~44.3& -25~13~04&0.0200&  7&  2.388&  28.1&  0.022& 12.5&  0.024&    4.5&X &146                             \\ 
 {\rm RXCJ2043.2-2144}&{\rm             }&20~43~12.6& -21~44~06&0.2041&  3&  3.628&  16.4&  3.930&  5.5&  4.416&    4.2&  &E                               \\ 
 {\rm RXCJ2043.2-2629}&{\rm S0894       }&20~43~14.4& -26~29~24&0.0408&  1&  4.647&  31.6&  0.178& 13.0&  0.189&    5.3&  &9,28                            \\ 
 {\rm RXCJ2048.1-1750}&{\rm A2328       }&20~48~10.6& -17~50~38&0.1475&  3&  5.930&  15.2&  3.215& 10.5&  3.349&    4.8&  &E                               \\ 
 {\rm RXCJ2055.8-5455}&{\rm A3718       }&20~55~51.5& -54~55~12&0.1390&  8&  4.087&  44.9&  1.960&  4.0&  2.481&    4.5&  &E                               \\ 
 {\rm RXCJ2058.2-0745}&{\rm A2331       }&20~58~15.2& -07~45~41&0.0793& 16&  3.486&  19.3&  0.530& 13.5&  0.541&    5.3&  &10                              \\ 
 {\rm RXCJ2101.4-4100}&{\rm S0915       }&21~01~28.1& -41~00~06&0.1694&  4&  3.112&  16.2&  2.278&  6.0&  2.531&    3.5&  &E                               \\ 
 {\rm RXCJ2101.5-1317}&{\rm             }&21~01~34.7& -13~17~31&0.0282&  2&  4.250&  20.7&  0.077& 10.0&  0.089&    4.1&  &E                               \\ 
 {\rm RXCJ2101.8-2802}&{\rm A3733       }&21~01~48.7& -28~02~06&0.0382& 91&  8.842&  11.5&  0.294& 20.0&  0.303&    6.9&L &13,14,146                       \\ 
 {\rm RXCJ2102.1-2431}&{\rm RBS 1712^c  }&21~02~09.9& -24~31~53&0.1880&  1&  5.238&  10.9&  4.718&  8.0&  5.073&    5.3&  &19,27,169                       \\ 
 {\rm RXCJ2103.4-4319}&{\rm A3736       }&21~03~26.0& -43~19~46&0.0487&  4&  4.599&  50.1&  0.253& 15.0&  0.264&    3.2&  &132                             \\ 
 {\rm RXCJ2104.3-4120}&{\rm A3739       }&21~04~20.1& -41~20~58&0.1651&  3&  5.988&  15.3&  4.093& 13.5&  4.134&    3.6&  &E                               \\ 
 {\rm RXCJ2104.9-5149}&{\rm             }&21~04~54.7& -51~49~35&0.0491&  1&  8.926&  12.3&  0.500& 11.0&  0.562&    3.1&  &5,145                           \\ 
 {\rm RXCJ2104.9-8243}&{\rm A3728       }&21~04~58.7& -82~43~22&0.0969&  4&  2.997&  20.1&  0.682&  6.0&  0.775&    8.2&  &E                               \\ 
 {\rm RXCJ2106.0-3846}&{\rm A3740       }&21~06~04.3& -38~46~21&0.1521&  2&  4.360&  14.1&  2.545&  9.0&  2.679&    3.8&DB&182                             \\ 
 {\rm RXCJ2107.2-2526}&{\rm A3744       }&21~07~12.3& -25~26~17&0.0381& 71&  4.854&  11.8&  0.162& 10.5&  0.180&    5.5&  &2                               \\ 
 {\rm RXCJ2111.6-2308}&{\rm AM2108^n    }&21~11~39.9& -23~08~59&0.0333& 15&  2.472&  20.7&  0.063& 13.5&  0.066&    4.4&  &17                              \\ 
 {\rm RXCJ2116.8-5929}&{\rm S0927       }&21~16~48.5& -59~29~53&0.0602&  0&  4.534&  17.3&  0.388& 10.5&  0.417&    4.0&  &9                               \\ 
 {\rm RXCJ2124.3-7446}&{\rm             }&21~24~22.8& -74~46~25&0.0586&  2&  5.009&  19.8&  0.405&  8.0&  0.466&    5.8&  &E                               \\ 
 {\rm RXCJ2125.2-0657}&{\rm             }&21~25~12.4& -06~57~56&0.1153&  2&  3.285&  36.7&  1.076& 11.0&  1.109&    5.9&  &E                               \\ 
 {\rm RXCJ2125.9-3443}&{\rm A3764       }&21~25~55.0& -34~43~28&0.0757& 38&  2.931&  53.7&  0.404& 14.0&  0.408&    4.9&  &140,141                         \\ 

            \noalign{\smallskip}
            \hline
         \end{array}
      \]
\end{table*}
%
%

{\footnotesize
\begin{table*}
         \label{tab6g}
      \[
         \begin{array}{llrrrrrrrrrrrl}
            \hline
            \noalign{\smallskip}
 {\rm name}& {\rm alt. name}& R.A. & Decl. & z & N_{gal} & F_x & Error 
& L_x & R_{ap}& L_x^* & N_H & Cm&Ref.  \\
(1) & (2) & (3) & (4) & (5) & (6) & (7) & (8) & (9) & (10) & (11) & (12) & (13) & (14) \\
            \noalign{\smallskip}
            \hline
            \noalign{\smallskip}
 {\rm RXCJ2127.1-1209}&{\rm A2345       }&21~27~11.0& -12~09~33&0.1760&  1&  5.271&  13.6&  4.154& 11.0&  4.282&    4.8&  &81                              \\ 
 {\rm RXCJ2129.8-5048}&{\rm A3771       }&21~29~51.0& -50~48~04&0.0796&  2&  5.051&  66.2&  0.767& 11.5&  0.807&    2.2&  &181,121                         \\ 
 {\rm RXCJ2133.4-7156}&{\rm             }&21~33~24.3& -71~56~09&0.0559&  2&  2.998&  23.1&  0.220&  8.5&  0.242&    2.0&  &E                               \\ 
 {\rm RXCJ2134.2-1328}&{\rm A2351       }&21~34~16.5& -13~28~48&0.0897&  2&  8.918&  13.8&  1.712&  7.5&  1.991&    5.2&B &E                               \\ 
 {\rm RXCJ2135.2+0125}&{\rm A2355       }&21~35~17.2& +01~25~54&0.1244&  2&  4.493&  26.0&  1.721& 13.5&  1.738&    4.8&  &1                               \\ 
 {\rm RXCJ2139.8-2228}&{\rm S0963       }&21~39~51.8& -22~28~24&0.0328& 16&  3.154&  20.8&  0.078& 10.0&  0.087&    3.4&B &E,25                            \\ 
 {\rm RXCJ2143.9-5637}&{\rm APMCC 699^f }&21~43~58.3& -56~37~35&0.0824&  2& 10.658&  10.2&  1.733& 13.0&  1.844&    3.4&  &E                               \\ 
 {\rm RXCJ2145.9-1006}&{\rm A2377       }&21~45~54.8& -10~06~16&0.0808&  1&  7.190&  41.6&  1.124& 12.5&  1.183&    4.0&P &1,1                             \\ 
 {\rm RXCJ2146.3-5717}&{\rm A3806       }&21~46~20.9& -57~17~19&0.0760& 99&  7.072&  34.6&  0.974& 17.5&  0.994&    2.6&B &2                               \\ 
 {\rm RXCJ2146.9-4354}&{\rm A3809       }&21~46~57.8& -43~54~36&0.0620& 94&  9.190&  11.0&  0.833& 14.0&  0.886&    1.8&B &2                               \\ 
 {\rm RXCJ2147.0-1019}&{\rm             }&21~47~00.5& -10~19~04&0.0780&  2&  2.959&  18.1&  0.434&  5.5&  0.517&    4.0&PX&E                               \\ 
 {\rm RXCJ2147.9-4600}&{\rm S0974       }&21~47~55.5& -46~00~19&0.0593&  4&  6.906&  11.9&  0.573& 13.5&  0.603&    2.7&  &24,30,32                        \\ 
 {\rm RXCJ2149.1-3041}&{\rm A3814       }&21~49~07.4& -30~41~55&0.1184& 19&  6.182&  12.9&  2.117&  9.5&  2.276&    2.3&  &33,96                           \\ 
 {\rm RXCJ2151.3-5521}&{\rm             }&21~51~22.7& -55~21~12&0.0385& 31&  3.137&  19.4&  0.107& 13.0&  0.113&    2.9&B &E,129                           \\ 
 {\rm RXCJ2151.8-1543}&{\rm A2382       }&21~51~52.9& -15~43~02&0.0614&  4&  4.028&  17.8&  0.359&  9.0&  0.395&    4.0&  &27,81                           \\ 
 {\rm RXCJ2152.2-1942}&{\rm A2384 (B)   }&21~52~14.2& -19~42~20&0.0963&  4&  4.059&  30.0&  0.912&  6.0&  1.060&    3.0&PB&E                               \\ 
 {\rm RXCJ2152.4-1933}&{\rm A2384 (A)   }&21~52~24.0& -19~33~54&0.0943&  1&  7.717&  20.0&  1.648&  8.5&  1.852&    3.0&PB&1                               \\ 
 {\rm RXCJ2154.1-5751}&{\rm A3822       }&21~54~09.2& -57~51~19&0.0760& 84& 15.994&   6.8&  2.185& 16.0&  2.300&    2.1&  &2                               \\ 
 {\rm RXCJ2154.2-0400}&{\rm A2389       }&21~54~12.1& -04~00~19&0.1509&  3&  3.540&  34.5&  2.031& 13.0&  2.052&    5.0&  &E                               \\ 
 {\rm RXCJ2157.4-0747}&{\rm A2399       }&21~57~25.8& -07~47~41&0.0579&  8&  5.851&  19.0&  0.462& 14.0&  0.481&    3.5&DB&E,10,122                        \\ 
 {\rm RXCJ2158.3-2006}&{\rm A2401       }&21~58~20.1& -20~06~16&0.0570& 23&  3.662&  19.0&  0.280& 11.0&  0.298&    2.6&  &2,3                             \\ 
 {\rm RXCJ2158.4-6023}&{\rm A3825       }&21~58~27.2& -60~23~58&0.0750& 61&  8.097&  10.3&  1.085& 13.0&  1.142&    2.8&  &3                               \\ 
 {\rm RXCJ2158.5-0948}&{\rm A2402       }&21~58~30.5& -09~48~28&0.0809&  1&  5.663&  12.2&  0.890& 10.0&  0.957&    4.0&  &111,157,181                     \\ 
 {\rm RXCJ2201.8-2226}&{\rm S0987       }&22~01~50.9& -22~26~40&0.0691& 10&  5.737&  15.0&  0.647& 14.0&  0.674&    2.6&  &S,24,33                         \\ 
 {\rm RXCJ2201.9-5956}&{\rm A3827       }&22~01~56.0& -59~56~58&0.0980& 20& 18.577&   6.3&  4.264& 11.5&  4.686&    2.8&  &3                               \\ 
 {\rm RXCJ2202.0-0949}&{\rm A2410       }&22~02~05.9& -09~49~28&0.0809& 10&  5.863&  13.6&  0.921& 13.5&  0.959&    4.2&D &E,10                            \\ 
 {\rm RXCJ2205.6-0535}&{\rm A2415       }&22~05~40.5& -05~35~36&0.0582&  3& 14.289&  16.2&  1.135& 15.5&  1.207&    4.7&B &E                               \\ 
 {\rm RXCJ2209.3-5148}&{\rm A3836       }&22~09~23.3& -51~48~54&0.1065&  2&  5.942&  15.6&  1.645& 13.5&  1.696&    2.1&  &E,29                            \\ 
 {\rm RXCJ2210.3-1210}&{\rm A2420       }&22~10~19.7& -12~10~34&0.0846&  9& 15.613&   7.5&  2.674& 11.0&  2.971&    3.9&  &S,10                            \\ 
 {\rm RXCJ2211.7-0350}&{\rm             }&22~11~43.4& -03~50~07&0.2700^{est}&  0&  3.326&  16.7&  6.528&  5.0&  7.418&    5.9&  &-                               \\ 
 {\rm RXCJ2213.0-2753}&{\rm             }&22~13~05.2& -27~53~59&0.0610&  6&  3.346&  26.8&  0.294& 10.0&  0.316&    1.4&  &S,96                            \\ 
 {\rm RXCJ2214.5-1022}&{\rm A2426       }&22~14~32.6& -10~22~18&0.0980& 15& 12.458&   8.7&  2.867& 12.5&  3.050&    3.9&  &S,2                             \\ 
 {\rm RXCJ2216.2-0920}&{\rm A2428       }&22~16~15.5& -09~20~24&0.0825&  2&  8.593&  10.1&  1.400&  8.5&  1.591&    4.5&  &S                               \\ 
 {\rm RXCJ2216.9-1725}&{\rm RBS 1842^c  }&22~16~56.4& -17~25~34&0.1301&  2&  5.515&  14.5&  2.315& 11.0&  2.411&    2.3&X\ast &S                           \\ 
 {\rm RXCJ2217.7-3543}&{\rm A3854       }&22~17~43.3& -35~43~34&0.1486& 44&  6.406&  10.7&  3.535&  8.5&  3.842&    1.1&  &33,37,38                        \\ 
 {\rm RXCJ2218.0-6511}&{\rm RBS 1847^c  }&22~18~05.6& -65~11~06&0.0951&  5&  9.021&   8.9&  1.953& 14.0&  2.034&    2.8&  &E                               \\ 
 {\rm RXCJ2218.2-0350}&{\rm MS2215^o    }&22~18~17.1& -03~50~03&0.0901&  3&  9.355&  11.1&  1.813& 14.5&  1.889&    5.7&D &S                               \\ 
 {\rm RXCJ2218.6-3853}&{\rm A3856       }&22~18~40.2& -38~53~51&0.1411& 10&  7.132&  10.2&  3.516&  9.0&  3.781&    1.3&  &16,31,33                        \\ 
 {\rm RXCJ2218.8-0258}&{\rm             }&22~18~49.1& -02~58~07&0.0902&  8&  4.044&  32.4&  0.790& 10.0&  0.840&    5.8&B &E                               \\ 
 {\rm RXCJ2220.5-3509}&{\rm A3866       }&  ~  ~00.0&    ~  ~00&0.1544&  1&  9.489&   8.8&  5.656&  8.5&  6.215&    1.1&X &146                             \\ 
 {\rm RXCJ2223.8-0138}&{\rm A2440       }&22~23~53.0& -01~38~16&0.0906& 48&  9.875&  10.9&  1.929& 10.5&  2.097&    5.3&  &S,42                            \\ 
 {\rm RXCJ2224.4-5515}&{\rm APMCC 772^f }&22~24~27.5& -55~15~22&0.0791&  2&  5.965&  15.6&  0.894& 12.0&  0.941&    3.5&  &E                               \\ 
 {\rm RXCJ2224.7-5632}&{\rm S1020       }&22~24~43.6& -56~32~05&0.0355&  0&  3.057&  16.2&  0.089&  7.5&  0.105&    4.1&  &9                               \\ 
 {\rm RXCJ2225.8-0636}&{\rm A2442       }&22~25~51.0& -06~36~12&0.0897& 12&  5.129&  14.3&  0.991&  8.5&  1.089&    5.1&  &E                               \\ 
 {\rm RXCJ2227.8-3034}&{\rm A3880       }&22~27~52.4& -30~34~12&0.0579& 28& 10.035&   8.2&  0.789&  8.5&  0.939&    1.1&  &14,33                           \\ 
 {\rm RXCJ2228.8-6053}&{\rm             }&22~28~51.3& -60~53~56&0.0423&  2&  3.829&  15.5&  0.158& 13.0&  0.166&    2.2&  &E                               \\ 
 {\rm RXCJ2234.5-3744}&{\rm A3888       }&22~34~31.0& -37~44~06&0.1510& 70& 11.225&   8.9&  6.363&  7.5&  7.314&    1.2&X &1,44                            \\ 
 {\rm RXCJ2235.6+0128}&{\rm A2457       }&22~35~40.6& +01~28~18&0.0594& 18& 11.139&  15.7&  0.924& 18.0&  0.953&    5.8&B &10                              \\ 
 {\rm RXCJ2243.0-2009}&{\rm A2474       }&22~43~04.6& -20~09~59&0.1359&  3&  4.977&  40.0&  2.297& 14.5&  2.320&    2.6&  &E                               \\ 
 {\rm RXCJ2246.3-5243}&{\rm A3911       }&22~46~18.6& -52~43~46&0.0965&  2& 12.436&   9.7&  2.778& 26.0&  2.806&    1.5&L &132                             \\ 
 {\rm RXCJ2248.5-1606}&{\rm A2485       }&22~48~32.9& -16~06~23&0.2472&  1&  3.063&  31.4&  4.998&  9.5&  5.100&    3.3&  &E                               \\ 
 {\rm RXCJ2248.7-4431}&{\rm S1063       }&22~48~43.5& -44~31~44&0.3475&  3&  8.590&  10.9& 28.939&  9.0& 30.786&    1.8&  &E                               \\ 
 {\rm RXCJ2249.9-6425}&{\rm A3921       }&22~49~57.0& -64~25~46&0.0940& 32& 14.033&   8.5&  2.948& 14.0&  3.103&    2.8&  &E,3                             \\ 
 {\rm RXCJ2251.0-1624}&{\rm A2496       }&22~51~00.6& -16~24~24&0.1221&  3&  5.238&  20.4&  1.917&  5.5&  2.282&    3.2&  &S                               \\ 
 {\rm RXCJ2251.7-3206}&{\rm             }&22~51~47.6& -32~06~12&0.2460&  1&  3.562&  18.8&  5.718&  6.5&  6.215&    1.4&X\ast &E                           \\ 
 {\rm RXCJ2253.5-3343}&{\rm A3934       }&22~53~34.2& -33~43~27&0.2240&  1&  3.036&  18.7&  4.008&  3.5&  5.010&    1.2&  &7,124                           \\ 
 {\rm RXCJ2254.0-6315}&{\rm AM2250^p    }&22~54~03.2& -63~15~15&0.2112&  7&  4.714&  11.7&  5.490& 11.5&  5.602&    2.2&  &E                               \\ 
 {\rm RXCJ2305.5-4513}&{\rm A3970       }&23~05~34.6& -45~13~14&0.1253&  3&  4.327&  42.8&  1.683& 11.0&  1.735&    1.7&  &S                               \\ 
 {\rm RXCJ2306.6-1319}&{\rm             }&23~06~36.0& -13~19~12&0.0659&  2&  5.890&  34.4&  0.601& 15.0&  0.620&    3.1&  &E                               \\ 
 {\rm RXCJ2307.2-1513}&{\rm A2533       }&23~07~15.3& -15~13~42&0.1110&  1&  4.904&  26.1&  1.484&  6.0&  1.726&    2.8&  &19                              \\ 

            \noalign{\smallskip}
            \hline
         \end{array}
      \]
\begin{list}{}{}
\item[$^{est}$ this redshift is highly uncertain due to the currently very poor spectra.
The observation will be repeated]
This table is also available in electronic form at CDS and at our home page (see footnote to abstract).
\end{list}
\end{table*}
%
%

}
{\footnotesize
\begin{table*}
         \label{TabR1h}
      \[
         \begin{array}{llrrrrrrrrrrrl}
            \hline
            \noalign{\smallskip}
 {\rm name}& {\rm alt. name}& R.A. & Decl. & z & N_{gal} & F_x & Error 
& L_x & R_{ap}& L_x^* & N_H & Cm &Ref.  \\
(1) & (2) & (3) & (4) & (5) & (6) & (7) & (8) & (9) & (10) & (11) & (12) & (13) & (14) \\
            \noalign{\smallskip}
            \hline
            \noalign{\smallskip}
 {\rm RXCJ2308.3-0211}&{\rm A2537       }&23~08~23.2& -02~11~31&0.2966&  2&  4.264&  14.3& 10.174& 13.5& 10.174&    4.3&  &S                               \\ 
 {\rm RXCJ2312.3-2130}&{\rm A2554       }&23~12~20.7& -21~30~02&0.1108& 35&  4.209&  24.1&  1.268&  4.5&  1.585&    2.0&  &38,127                          \\ 
 {\rm RXCJ2313.0-2137}&{\rm A2556       }&23~13~00.9& -21~37~55&0.0871&  9&  7.764&  18.0&  1.419&  7.0&  1.669&    2.0&  &S,127                           \\ 
 {\rm RXCJ2313.9-4244}&{\rm S1101       }&23~13~58.6& -42~44~02&0.0564&  4& 23.412&   7.2&  1.738& 12.0&  1.998&    1.9&  &S                               \\ 
 {\rm RXCJ2315.7-3746}&{\rm A3984       }&23~15~44.2& -37~46~56&0.1786&  3&  5.819&  20.5&  4.738&  8.0&  5.150&    1.5&  &S                               \\ 
 {\rm RXCJ2315.7-0222}&{\rm             }&23~15~45.2& -02~22~37&0.0267&  3&  8.332&  10.6&  0.134& 20.0&  0.141&    4.2&L &S,27,51                         \\ 
 {\rm RXCJ2316.1-2027}&{\rm A2566       }&23~16~07.5& -20~27~19&0.0822& 11&  7.415&  18.2&  1.199&  9.0&  1.332&    2.1&  &S,127                           \\ 
 {\rm RXCJ2319.2-6750}&{\rm A3990       }&23~19~12.0& -67~50~24&0.0286&  0&  2.434&  14.5&  0.045& 10.5&  0.049&    2.8&D &47                              \\ 
 {\rm RXCJ2319.6-7313}&{\rm A3992       }&23~19~41.8& -73~13~51&0.0984&  3&  3.993&  17.9&  0.937&  7.5&  1.030&    1.9&  &E                               \\ 
 {\rm RXCJ2321.4-2312}&{\rm A2580       }&23~21~24.3& -23~12~20&0.0890& 17&  3.270&  25.6&  0.621&  7.0&  0.690&    2.0&  &S,127                           \\ 
 {\rm RXCJ2321.5-4153}&{\rm A3998       }&23~21~33.4& -41~53~56&0.0894& 16&  8.720&  13.3&  1.662& 11.0&  1.787&    2.0&  &S,16,31,33,48                   \\ 
 {\rm RXCJ2321.8-6941}&{\rm A3995       }&23~21~48.6& -69~41~55&0.1846& 14&  4.334&  10.5&  3.815&  5.5&  4.385&    3.4&  &E                               \\ 
 {\rm RXCJ2325.3-1207}&{\rm A2597       }&23~25~20.0& -12~07~38&0.0852&  3& 20.558&   5.8&  3.556& 11.5&  3.996&    2.5&  &S,1                             \\ 
 {\rm RXCJ2326.7-5242}&{\rm             }&23~26~46.9& -52~42~36&0.1074&  3&  3.370&  20.3&  0.955& 11.0&  0.985&    1.6&  &E                               \\ 
 {\rm RXCJ2331.2-3630}&{\rm A4010       }&23~31~12.7& -36~30~24&0.0957& 30& 11.013&  22.1&  2.417&  9.5&  2.686&    1.4&  &E,140,141                       \\ 
 {\rm RXCJ2336.2-3136}&{\rm S1136       }&23~36~17.0& -31~36~37&0.0643&  2&  5.273&  20.5&  0.516& 12.0&  0.549&    1.2&  &15,52,26                        \\ 
 {\rm RXCJ2337.6+0016}&{\rm A2631       }&23~37~40.6& +00~16~36&0.2779&  5&  3.555&  19.8&  7.420& 10.0&  7.571&    3.8&  &55                              \\ 
 {\rm RXCJ2340.1-8510}&{\rm A4023       }&23~40~10.3& -85~10~42&0.1934&  3&  3.176&  18.9&  3.066&  8.5&  3.194&    7.5&  &E                               \\ 
 {\rm RXCJ2341.2-0901}&{\rm A2645       }&23~41~16.8& -09~01~39&0.2510&  5&  3.414&  26.1&  5.731& 11.5&  5.789&    2.5&  &1                               \\ 
 {\rm RXCJ2344.2-0422}&{\rm             }&23~44~16.0& -04~22~03&0.0786&  2& 12.638&   7.4&  1.855& 11.0&  2.061&    3.5&  &S                               \\ 
 {\rm RXCJ2347.4-0218}&{\rm HCG 97      }&23~47~24.4& -02~18~52&0.0223& 10&  2.629&  38.8&  0.030& 12.5&  0.033&    3.6&  &58,59                           \\ 
 {\rm RXCJ2347.7-2808}&{\rm A4038       }&23~47~43.2& -28~08~29&0.0300&157& 49.578&   3.8&  1.014& 22.5&  1.127&    1.5&L &146                             \\ 
 {\rm RXCJ2351.6-2605}&{\rm A2667       }&23~51~40.7& -26~05~01&0.2264&  1&  9.295&  10.1& 12.422&  8.0& 13.651&    1.7&X &S,65,129                        \\ 
 {\rm RXCJ2354.2-1024}&{\rm A2670       }&23~54~13.4& -10~24~46&0.0765&219&  9.622&  12.1&  1.337& 10.5&  1.469&    2.9&B &66,86,105                       \\ 
 {\rm RXCJ2357.0-3445}&{\rm A4059       }&23~57~02.3& -34~45~38&0.0475& 45& 32.663&   5.5&  1.698& 24.0&  1.787&    1.1&L &146                             \\ 
 {\rm RXCJ2359.3-6042}&{\rm A4067       }&23~59~19.2& -60~42~00&0.0989& 30&  4.544&  20.5&  1.080&  9.5&  1.149&    2.4&  &141                             \\ 
 {\rm RXCJ2359.9-3928}&{\rm A4068       }&23~59~55.7& -39~28~47&0.1024& 12&  5.024&  18.4&  1.279& 11.0&  1.346&    1.3&  &16,31,67                        \\ 

            \noalign{\smallskip}
            \hline
         \end{array}
      \]
\begin{list}{}{}
\item[$^{\rm a}$ Cl0053-37, $^{\rm b}$ cluster from the catalogue of Wegner et al. (1996, 1999),
$^{\rm c}$ from the ROSAT Bright Source Catalogue]
\item[(Fischer et al. 1998, Schwope et al. 2000),$^{\rm d}$ RASSCALS145, $^{\rm e}$ ZwCl0258.9+0142,
$^{\rm f}$ cluster from the catalogue by]
\item[ Dalton et al. (1994, 1997), $^{\rm g}$ Eridanus group, 
$^{\rm h}$ ES0657-558, $^{\rm i}$ AM 0711-602, 
$^{\rm j}$ USZ-SSRS2-group from the catalogue] 
\item[of Ramella et al. (2002), $^{\rm k}$ MS1111.8-3754, $^{\rm l}$ MS1306.7-0121, 
$^{\rm m}$ ZwCl1703.8-0129,
$^{\rm n}$ AM 2108-232, $^{\rm o}$ MS 2215.7-0404, $^{\rm p}$ AM 2250-633]

{\tiny
{key to the redshift references:} (E,H) ESO-Key Programme, (S) South Galactic Pole Project (part of
Ph.D. thesis K.A. Romer), 
(1) Struble \& Rood (1999), (2) Katgert et al. (1996), (3)  Katgert et al. (1998),
(4)  Postman \& Lauer (1995), (5) Lauberts \& Valentijn (1989), 
(6)  Ebeling et al. (1996), (7)  Bauer et al. (2000), (8) Kapahi et
     al. (1998), (9) Abell et al. (1989),
(10) Slinglend et al. (1998), (11) M\"{u}ller et al. (1999), (12) Smith et al. (2000), 
(13) Solanes \& Stein (1998), (14) Stein (1996), (15) NED information prior to 1992 without reference, 
(16) Tucker et al. (2000), (17) Dalton et al. (1994), (18) Dale et al. (1998), 
(19) Quintana \& Ramirez (1994), (20) Da Costa et al. (1998), (21) Patten et al. (2000),
(22) Barton et al. (1996), (23) Mahdavi et al. (2000), (24) Loveday et al. (1996),
(25) Garilli et al. (1993), (26) Ratcliffe et al. (1998), (27) De Grandi et al. (1999),
(28) Green et al. (1998), (29) Dalton et al. (1997), (30) Muriel et al. (1995),
(31) Shectman et al. (1996), (32) Fairall (1984), (33) Collins et al. (1995), 
(34) Mould et al. (1991),
(35) Ledlow \& Owen (1995), (36) Slinglend et al. (1998), (37) West \& Frandsen (1981),
(38) Colless \& Hewett (1987), (39) Stocke et al. (1991), (40) Muriel et al. (1991), 
(41) Beers et al. (1991), (42) Mohr et al. (1996), (43) Olowin et al. (1988), 
(44) Teague et al. (1990),
(45) Reimers et al. (1996), (46) Maurogordato et al. (1997), (47) Huchra priv. com, 
(48) Zabludoff \& Mulchaey (1998), (49)  Andernach (2002), (50) Batuski et al. (1999),
(51) Huchra et al. (1993), (52) Di Nella et al. (1996), (53) Crawford et al. (1995),
(54) Crawford et al. (1999), 55. NED information based on SDSS, York et al. (2001), (56) Kristian et al. (1978),
(57) Owen et al. (1999), (58) Hickson et al. (1992), (59) De Carvalho et al. (1997), 
(60) Jorgensen et al. (1995), (61) Ettori et al. (1995), (62) Green et al. (1990),
(63) Dantas et al. (1997), AJ, 118, 1468, (64) de Vaucouleurs et al. (1991), 
(65) Caccianiga et al. (2000), (66) Sharples et al. (1998), (67) Vettolani et al. (1998),
(68) Andernach \& Tago (1998), (69) Vettolani et al. (1989), (70) Tritton (1972),
(71) Tittley \& Henriksen (2001), (72) Dressler \& Shectman (1988), (73) Barrena et al.
(2002), (74) Mathewson \& Ford (1996), (75) Bardelli et al. (2001), (76) Quintana et al. (1995),
(77) Vettolani et al. (1990), (78) Zabludoff et al. (1990), (79) Tucker et al. (1998),
(80) Mamon et al. (2001), (81) Owen et al. (1995), (82) Huchtmeier (1994), (83) da Costa et al. (1987),
(84) Matthews \& van Driel (2000), (85) Sodre et al. (2001), (86) den Hartog \& Katgert (1996),
(87) Pierre et al. (1997), (88) Lemonon et al. (1997), (89) Bade et al. (1995), 
(90) Jones \& Forman (1999), (91) Fairall et al. (1992), (92) Mulchaey et al. (1996),
(93) Strauss et al. (1992), (94) Beers et al. (1995), (95)  Koranyi \& Geller (2002),
(96) NED information based on 2dF Survey, Colless et al. (2001), (97) Ramella et al. (2002), (98) Kaldare et al. (2003),
(99) Quintana et al. (1996), (100)  Richter (1987), (101) Bardelli et al. (1998), 
(102) Bardelli et al. (1994), (103) Drinkwater et al. (1999), (104) Christiani et al. (1987),
(105) Oegerle \& Hill (2001), (106)  Henry \& Mullis (1997) (priv. com.),
(107) Willmer et al. (1999), (108) Heckman et al. (1994), (109) Bernardi et al. (2002),
(110) Goto et al. (2002), (111) Huchra et al. (1991), (112) Quintana \& de Souza (1993),
(113) Schindler (2000), (114) Willmer et al. (1991), (115) Pena et al. (1991),
(116) Thompson et al. (1992), (117) Peterson et al. (1997), (118) Arnaud et al. (1992),
(119) Batuski et al. (1991), (120) Wegner et al. (1999), (121) Peacock \& West (1992),
(122) Blakeslee \& Tonry (1992), (123) Falco et al. (1999), (124) Beers et al. (1992),
(125) Trager et al. (2000), (126) Davoust et al. (1995), (127) Caretta et al. (2002),
(128) Liang et al. (2000), (129) Rizza et al. (1998), (130) Malamuth et al. (1992),
(131) Caldwell \& Rose (1997), (132) Lauer \& Postman (1994), (133) Garilli et al. (1991),
(134) Whiteoak (1972), (135) Ramella et al. (1996), (136) Mathewson et al. (1992),
(137) Melnick \& Quintana (1981), (138) Sodre et al. (1992), (139) Cruddace et al. (2002),
(140) den Hartog (1995), (141) Mazure et al. (1996), (142) Couch et al. (1998),
(143) Couch \& Sharples (1987), (144) da Costa et al. (1991), (14). Schwope et al. (2000),
(146) Chen et al. (1998), (147) Crawford et al. (1993), (148) Fadda et al. (1996),
(149) Durret et al. (1998), (150) Lumbsden et al. (1992), (151) Alonso et al. (1999),
(152) Merrifield \& Kent (1991), (153) Way et al. (1998), (154) Fetisova et al. (1993),
(155) Quintana et al. (1996), (156) Trasarti-Battistoni (1998), (157) Postman et al. (1992),
(158) Ramirez \& de Souza (1998), (159) B\"ohringer et al. (2000), (160) Dressler et al. (1986),
(161) Ebeling et al. (2002), (162) Zabludoff et al. (1993), (163) Huchra et al. (1983),
(164) Lucey et al. (1983), (165) Rose et al. (2002), (166) Drinkwater et al. (2001),
(167) Mieske et al. (2002), (168) Minniti et al. (1998), (169)  Edge \& Stewart (1991),
(170) Hilker et al. (1999), (171) Carter \& Malin (1983), (172) Quintana et al. (1994),
(173) Metcalfe et al. (1989), (174) Marzke et al. (1996), (175) Edge et al. (1992),
(176) Hewitt \& Burbidge  (1991), (177) Ellis et al. (1984), (178) Tonry (1985),
(179) Quintana \& Ramirez (1990), (180) Maurogordato et al. (2000), 
(181) Lebedev \& Lebedeva (1992), (182) Ebeling \& Maddox (1995), 
}   
\end{list}
\end{table*}
%
%

}}

\section{Information on individual Objects}

The REFLEX data base used for the cosmology work in the REFLEX papers
up to number VIII is based on the catalogue version of 1999.  As we
have collected more information since, we went through a further
critical inspection of all the catalogue entries prior to the release
of this catalogue, in addition to the revision of redshifts as
described above.  A major route of approach was to carefully assess
the nature of those objects for which the X-ray criteria left us with
a not very satisfactory means of identification. Thus we inspected
especially all those X-ray sources again which appeared point-like in
the RASS.  The fact that an X-ray source appears point-like at the
resolution of the All-Sky Survey is not yet a criterion to reject a
source.  An illustrative example for this is the massive, and
optically very impressive cluster RXCJ1206.2-0848 at redshift
$z=0.4414$ which appears in the RASS as a point source.  But we use
the point-like appearance here as a flag for sources to be doubly checked.
On the other hand, if an X-ray source is clearly well extended,
we can be sure that most of the X-ray emission is diffuse and the 
contribution to the X-ray flux from AGNs or
radio galaxies found at other wavelength can at most be a minor fraction.  
In addition to the
X-ray based reinspection we have also screened again the literature
for any information that could cast doubts on the cluster
identification, with a special eye on contaminating AGN. One source 
of further observational
information that was used for some clusters are pointed observations
with the ROSAT PSPC and HRI detector as well as new observations with
XMM-Newton and CHANDRA.

As a result of this screening procedure we could improve our
identification and discarded 6 objects from the REFLEX list as
described in the subsection below. We should also point out that 
most of the critical cases commented below were known as problematic targets
to our observing team at the time of the observation. 
One of the observational requirements was therefore
to take a spectrum of suspicious objects in the center of the X-ray
emission (often bright galaxies) to critically 
check for the presence of an AGN. More details about the spectra of these
objects will be given in Guzzo et al. (in preparation).
Thus for some of the objects where the identification is still 
not completely certain the standard optical means have already been used 
and a final decision can only come from a 
high angular resolution X-ray observation or for example 
from optical polarimetric measurements. In total we are left with  
14 cases where the identification is not definitive. Following our previous 
philosophy to keep the objects in our working list until we can positively
rule them out as cluster identifications, we have conservatively kept them in our catalogue
but we have marked them in the table. We realistically expect that
less than 10 are non-cluster objects. The effect this small fraction
(less than 2.3 \%) has on any of the cosmological applications of
REFLEX is completely negligible.

In the following we provide notes on the results of our inspection
including the critica cases mentioned as well as positive confirmations.
These remarks should be of help for any further
identification or follow-up work that may be conducted on these REFLEX
objects.  The comments given for each object illustrate some aspects
of the manual screening process.
 
{\bf RXCJ0014.3-6604}, A2746 (also included in XBACS) is a typical
example of a point-like X-ray source which was inspected in more
detail. The X-ray spectral hardness ratio is not outside the
acceptable limit for cluster emission.  The source appears very
compact with a core radius $2\sigma$ upper limit of about $ 120
h_{50}^{-1}$ kpc at a redshift of $z = 0.156$.  This does not
rule out that the X-ray emission comes from a cluster.  We have
five concordant redshifts including the cD galaxy, with no spectrum
indicating an AGN and the optical image is clearly showing a galaxy
cluster.  Therefore, even though we cannot establish the extent of the
X-ray source, we conclude that the source is most likely a cluster.

{\bf RXCJ0015.4-2350}, A14, has an X-ray emission which is very faint
and diffuse: about 48 source photons spread in a clumpy distribution
over a region with a radius of about 12 arcmin. The cluster position
was not taken as the center of the large-scale distribution but at a
local maximum.  The cD galaxy, ESO 473-G 005 at the position 00 15
10.6, -23 52 57.0 is located at this highest local maximum. This is
surely a good example of a dynamically young cluster with a size
slightly smaller than the Virgo cluster. Even though there is a slight
probability that the X-ray emission originates from a collection of
point sources, by far the most likely interpretation is that of a poor
galaxy cluster. This is supported by the finding of four coincident
galaxy redshift in the optical follow-up.

{\bf RXCJ0117.8-5455} is another source with X-ray emission consistent
with a point source origin. The source is also detected as a bright radio
source in the SUMSS survey (Mauch et al. 2003).
Since the cluster is found to be at a
redshift of $z = 0.2510$ as determined from 6 galaxy velocities (where
none of the spectra indicate an AGN), an upper limit on the X-ray core
radius of about $270~ h_{50}^{-1}$ kpc is consistent with a
cluster source and also the spectral hardness ratio is within the
limits expected for cluster emission. Therefore this point-like X-ray
source is kept in the catalogue until we have better data for a more 
secure identification.

{\bf RXCJ0132.6-0804} is classified as Seyfert 1.5 in the cross
correlation work of the ROSAT All-Sky Survey and the NVSS sources by
Bauer et al. (2000). We find narrow emission lines in the central
galaxy, PKS 0130-083, of this cluster at $z = 0.1485$. The OIII lines 
are more prominent than the H$\beta$ lines which makes the AGN identification
more likely than that of the typcail emission lines often found in central 
cluster galaxies in cooling core clusters 
(e.g. Crawfrord et al. 1999). The X-ray
luminosity of the object is about $3.6~ 10^{44}$ erg s$^{-1}$ which is
quite high for a Seyfert galaxy. The Digital Sky Survey image shows
clearly a central dominant galaxy surrounded by a collection of 
galaxies and we recorded three coincident galaxy redshifts. The spectral 
hardness ratio is within the limits expected for thermal emission from a cluster. 
Therefore the identification of this X-ray source is uncertain at
this stage. The X-ray emission could come from the cluster or the AGN.
A high resolution X-ray observation is required for a definitive decision. 

{\bf RXCJ0250.2-2129} has been classified as a BL Lac 
($z = 0.4980$, off-set 0.4 arcmin) in the ROSAT
Bright Source Catalogue by Schwope et al. (2000) and is listed as a
radio source in Bauer et al. (2000). The BL Lac is
not at the same redshift as the cluster. The X-ray
source shows a marginal extent and is too soft by about 1.3$\sigma$
compared to the expectation. Our deeper CCD image clearly shows an
optical cluster and we found two coincident galaxy redshifts.  We
therefore take the cluster identification as more likely, but cannot
rule out a contribution to the X-ray emission by the AGN.
The cluster is also coverd by a ROSAT PSPC archive observation.
But the offset from the pointing center is so large that the possibility
to better resolve a point source is not improved over the survey data.

\begin{figure}                                                                 
\psfig{figure=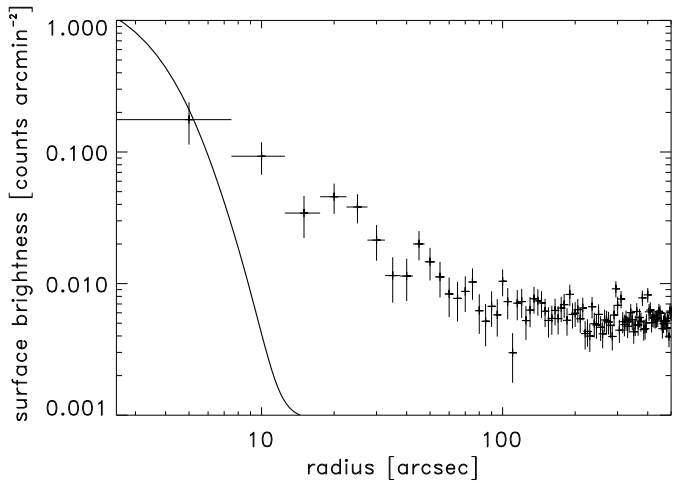,height=6.3cm}
\caption{Surface brightness profile of the REFLEX cluster
RXCJ0301.6+0155 as observed in a pointed observation with the ROSAT
HRI (data points and Poissonian errors).  The profile is compared to
the point spread function of the ROSAT HRI (solid line).}
\label{fig6}
\end{figure}

{\bf RXCJ0301.6+0155} This source, which has previously been
identified as a cluster in the NORAS Survey (B\"ohringer et al. 2000)
and which is coinciding with a Zwicky cluster (ZwCl0258.9+01), appears
point-like.  A ROSAT HRI observation in the archive showed
that the source is very compact but definitely not a point source.
Therefore the cluster
identification is well justified.  The comparison between the surface
brightness profile and the point spread function of the ROSAT HRI is
shown in Fig.~\ref{fig6}.

{\bf RXCJ0331.1-2100} is coincident within 0.5 arcmin with a Seyfert
1.9 galaxy identified by Bauer et al. (2000). The source is listed in
Schwope et al. (2000) as RBS 0436.  The X-ray emission is only
marginally extended, but the spectral hardness ratio is consistent
with thermal intracluster medium emission. We have nine coincident
galaxy redshifts for this cluster.  Therefore it is possible that the
AGN is contaminating the X-ray radiation, but there is also definitely
a cluster at this position.

{\bf RXCJ0336.3-0347} This is a compact X-ray source with a peaked
center and a weak extent. The central galaxy is listed as an AGN, as
2MASS and radio source, probably a BL Lac at $z = 0.1595$ (Bauer et
al. 2000, Veron-Cetty \& Veron 2001). It appears to have a bright
central spot in sky images, but our spectra also easily show the underlaying
galaxy continuum with the Balmer break. The optical image also shows
an optical cluster for which we have
three concordant redshifts at the same redshift as the AGN.  
The most likely interpretation of this
source is that it is indeed a galaxy cluster with some (less than
half) contamination of the X-ray emission by the AGN.  The hardness
ratio of the X-ray emission is consistent with a cluster, also.

{\bf RXCJ0425.8-0833} is coincident within 1.3 arcmin with a Seyfert 2
galaxy identified by Hewitt \& Burbidge (1991). The source is listed
in Schwope et al. (2000) as RBS 0540. The central dominant galaxy has 
the spectrum of a passive ellitpical galaxy without emission lines
as found in our follow-up observations. The X-ray emission is very
extended and the dominant part of the X-ray emission must come from
the cluster, for which we have two concordant redshifts.

{\bf RXCJ0437.1-2027}, A499, is detected with only 26 photons and the
result for the extent is not completely convincing. A ROSAT HRI
observation from the archive 
shows that the source is quite compact but clearly extended. Two
coincident galaxy redshifts are available for this cluster.

{\bf RXCJ0528.9-3927} also identified as RBS0653 (Schwope et
al. 2000) is listed in NED as a QSO.  The X-ray emission is slightly
extended but the spectral hardness ratio deviates by 5.2$\sigma$ from the value
expected for thermal cluster emission. A deeper X-ray observatio with XMM-Newton
shows this object to be an X-ray luminous cluster with a less than 20\% 
contamination (in the ROSAT hard band) by a bright, soft AGN point source.

{\bf RXCJ0918.1-1205}, A780 or Hydra A, is a well known X-ray
cluster. Thus even though one finds a coincidence within 0.2 arcmin with a
Seyfert galaxy listed by de Vaucouleurs et al. (1991), there is no
doubt about the clear dominance of the thermal cluster emission.

{\bf RXCJ1027.9-0647}, A1023, is coincident within 0.2 arcmin with an
AGN at $z=0.1165$ identified by Grazian et al. (2002). The X-ray emission is
marginally extended and the spectral hardness ratio is consistent with
thermal cluster emission.  There is a residual possibility that the
X-ray emission of this source is contaminated by an AGN. Six
coincident galaxy redshifts support the existence of a galaxy cluster
at this position.

{\bf RXCJ1050.5-0236}, A1111, is coincident within 0.6 arcmin with a
Seyfert 2 galaxy identified by Machalski \& Condon (1999). The X-ray
emission is found to be extended and the dominant part of the X-ray
emission comes from the cluster.  There is an X-ray emitting star in
the south of the cluster whose X-ray emission was deblended.  Nine
coincident galaxy redshifts leave no doubt about the existence of a
cluster at this position (see also the ASCA study of Matsumoto et al.
2001).

{\bf RXCJ1050.6-2405} features a point source and an extended fainter
halo in the RASS.  About one third of the flux seems to come from the
central compact emission. The hardness ratio gives no indication for
two source components, however, and is perfectly consistent with
thermal emission from a cluster. At the cluster redshift of $z=0.2037$
the compact center could still be a bright compact cluster cooling
core as well as a contaminating central AGN.  Near the center (0.4
arcmin offset) is a radio source, PKS B1048-238 (MRC1048-238), at a
redshift of $z=0.2060$ (McCarthy et al. 1996) and identified with a broad
line radio galaxy Kapahi et al. (1998). We expect that most of the X-ray
emission is due to the cluster. But so far we have made no effort to 
subtracted the possible AGN contribution.

{\bf RXCJ1141.4-1216}, A1348, is coincident within 0.2 arcmin with a
Seyfert 1.8 galaxy listed by Machalski \& Condon (1999). The X-ray
emission shows a small extent and the X-ray emission is slightly
softer ($\sim 2.9\sigma$) than expected. From the extended X-ray
emission we still conclude that most of the emission comes from the
cluster. The X-ray luminosity of the object is $3.3 \cdot 10^{44}$ erg
s$^{-1}$ which would be towards the upper end of the luminosity
distribution of Seyfert galaxies. Six coincident galaxy redshifts
confirm the existence of a cluster at this position.

{\bf RXCJ1149.7-1219}, A1391, is coincident within 0.9 arcmin with an
AGN listed by Machalski \& Condon (1999). The X-ray emission is found
to be clearly extended and the dominant part of the X-ray emission
must come from the cluster.  Six coincident galaxy redshifts confirm
the existence of a cluster at this position.

{\bf RXCJ1212.3-1816} was detected with only 12.5 source photons with
a significance of just about 3$\sigma$. This signal is detected in an
aperture significantly larger than the point spread function and
the significance
decreases for a smaller detection cell. Therefore, if the detection is
accepted, the source has to be extended. We have two coincident cluster galaxy
redshifts.

{\bf RXCJ1234.2-3856} appears as a point source in RASS with a spectral
hardness ratio consistent with thermal cluster emission.  It is also
listed as a radio source without classification by Bauer et
al. (2000). Two coincident galaxy redshifts give some further support
for the cluster identification.

{\bf RXCJ1253.6-3931} appears as point-like source in the RASS with a
cluster like hardness ratio and coincides within 0.5 arcmin with the
unclassified radio source PMN J1253-3932 (see also Bauer et al. 2000).
We have a spectrum as well as a redshift from the central galaxy which
does not show an AGN or BL Lac signature. A CCD image shows a galaxy cluster.
Therefore we keep this source in
our list as a likely cluster candidate.

{\bf RXCJ1326.2+0013} is coincident within 0.2 arcmin with a BL Lac
found in the 2dF survey. The X-ray emission is clearly extended and
the emission from the intracluster medium certainly dominates.  The
spectral hardness ratio is as expected and there are 
16 coincident galaxy redshift.  Therefore there is no doubt
about the cluster identification of this source.

{\bf RXCJ1332.9-2519} is detected in the RASS with very low surface
brightness, but with 50 counts and an about 4$\sigma$ detection. An
archival PSPC observation, with an exposure of only 982 sec shows a 
similar very low surface brightness structure providing a good
confirmation of this very diffuse structure.  We have three coincident
galaxy redshifts from the follow-up observations confirming the
existence of a cluster at this position.

\begin{figure}                                                                  
\psfig{figure=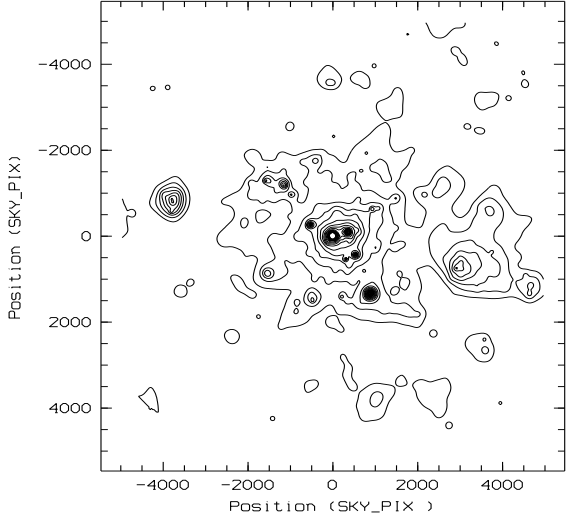,height=7cm}
\caption{ROSAT PSPC image of Abell 3574 (RXCJ1347.2-3025 \&
 RXCJ1349.3-3018) from a
pointed PSPC observation with an exposure of 8.3 ksec. The image is
exposure and vignetting corrected.  It has been treated with a
variable Gaussian filter. The contours are logarithmically spaced. We
can see two bright point sources in the center and another one to the
south which carry most of the flux received from this object. The
diffuse emission of the cluster shows some clumpiness and amounts to
about 20\% of the total emission. The central part and the Western
subcluster are tabulated as two different catalogue entries.
One sky pixel is 0.5 arcsec, thus the image size is 42 $\times$ 42
arcmin$^2$. The image is centered on the 
sky coordinate 13 49 19.2, -30 18 36 (J2000).}
\label{fig7}
\end{figure}

{\bf RXCJ1347.2-3025 \& RXCJ1349.3-3018}, A 3574 or Klemola 27 at a
redshifts of 0.0145 and 0.0160, respectively, is a cluster with two components. 
It presents a particular problem to its
detection and identification. The central galaxy is the X-ray luminous
Seyfert 1 galaxy, IC 4329, and two further AGNs contribute to the X-ray
emission from the cluster region. While the central detection could
easily have been dismissed because of its almost point-like appearance
in the RASS and its identification with a
known X-ray AGN, the additional detection of fragments of the diffuse
outer emission in the RASS analysis triggered a further
inspection of this cluster. A pointed observation shows clearly the
diffuse X-ray halo of the cluster in addition to the three cluster AGN
point sources.  We have used this pointed observation to
subtract the contribution of the three AGN from the total
emission. The central AGN IC4329A carries about 75\% of the flux, the
two other AGNs contribute about 4 to 5\% each and the cluster emission
amounts to 15 - 20\%.  Therefore, with a net flux of $\sim 1.35~
10^{-11}$ erg s$^{-1}$ cm$^{-2}$ in the ROSAT band, the cluster is well above the flux
limit of the REFLEX and the XBACS survey.  Fig.~\ref{fig7} shows the surface
brightness distribution of this cluster. In the catalogue we have
listed the central cluster region and the western concentration (about
25 arcmin, $\sim 0.6$ Mpc distance) as separate sources identified
with A3574E and A3574W, respectively, recognizing that they are two
separate virializing clumps. Also the diffuse Western component is, with
a nominal flux, $F_n \sim 3.5~ 10^{-12}$ erg s$^{-1}$ cm $^{-2}$, above
the REFLEX flux limit.

The cluster properties were determined also with the help
of the pointed observation, which allows a better contamination
subtraction.  The cluster was dismissed in XBACS as an AGN. 
At a larger distance it would probably have escaped its
identification as a cluster in REFLEX. This shows that there
is probably not only a small fraction of clusters which are
erroneously included in the sample because of an AGN contribution to the
X-ray flux, but also a small fraction of clusters which are erroneously
discarded because the cluster X-ray emission is not easily visible if
it is blended by a bright AGN.

{\bf RXCJ1350.7-3343} has a central galaxy with a Seyfert 1 spectrum but
the X-ray emission in RASS is extended and therefore indicates a galaxy cluster.

{\bf RXCJ1415.2-0030}, tentatively identified with A1882 (offset 13.3 arcmin), 
is coincident within 1.6 arcmin with a
QSO identified in the Sloan Digital Sky Survey. The X-ray emission is
very extended and diffuse, however, and we do not detect a
contribution by a point source. 20 concordant redshifts confirm the
existence of a cluster.

{\bf RXCJ1416.8-1158} appears as a point source in RASS and has a
spectrum that is 1.7$\sigma$ too soft. 10 coincident galaxy redshift
give good support for the existence of a cluster at this position.

{\bf RXCJ1504.1-0248} is a very peculiar object. We have obtained a
spectrum of the central galaxy (with an offset of 0.3 arcmin from our
reference position of the cluster) which shows Liner like emission
lines. It was identified as an AGN by Shectman et al. (1996).  
The X-ray source is very compact but
significantly extented (the probability that the emission comes from a
point source is estimated to be 0.0016). The X-ray luminosity is
extremely high with $L_x \sim 4.3 \cdot 10^{45}$ erg s$^{-1}$ in the
0.1 - 2.4 keV band and not typical for a narrow line AGN. 

A short CHANDRA observation that was finally performed early 2004 shows
a perfect galaxy cluster image without any significant AGN contamination.
This is in our sample the most spectacular example of an object 
which was for long time marked as very suspicious for its unusual 
compactness for given brightness and distance but was finally revealed
to be a prominet X-ray cluster. The Liner-like spectrum observed for the
central galaxy is also typical for cooling flow clusters (e.g. Crawford
et al. 1999 and references theirin).

{\bf RXCJ1958.2-3011} is an X-ray source that appears point-like in
RASS with a hardness ratio consistent with cluster emission. It is
listed as an unclassified radio source in Bauer et al. (2000). The
optical image shows a galaxy with a bright nucleus and our ESO spectrum
indicates an enhanced blue continuum.  There
is clearly a cluster visible with a likely Bright Cluster Galaxy (BCG)
that is about 1 arcmin offset to the NEE from the radio source and the
center of the cluster emission.  The cluster detection is further
supported by four concordant redshifts.  Therefore, we classify this
object as a cluster-AGN combination whose fluxes must be determined
by a higher angular resolution X-ray image.  Since low X-ray luminosity
AGN appear very frequently in clusters, with an X-ray luminosity well
below that of the cluster emission, we keep this object in the REFLEX
catalogue at present.

{\bf RXCJ2035.7-2513} shows diffuse X-ray emission around a large
early type galaxy with a luminosity of about $4~ 10^{42}$ erg
s$^{-1}$. An archival PSPC image also shows the diffuse emission but
also some contamination of 10-20\% by point sources, which is within
the error of the flux determination in the catalogue and no correction
has been made. The object is probably associated with A3698
whose center has an offset of 4.6 arcmin assumed to coincide with NGC
6936. Seven coincident galaxy redshifts confirm the existence of a
cluster at the X-ray source position.

{\bf RXCJ2147.0-1019} contains a BL Lac identified  by Bauer et
al. (2000) within 0.2 arcmin of our reference position. The X-ray source
is marginally extended (probability 97\%) and has a hardness ratio
well consistent with thermal cluster emission. We cannot rule out that
the BL Lac could substantially contribute to the cluster emission,
however.

{\bf RXCJ2216.9-1725} contains a Seyfert galaxy identified by 
Schwope et al. (2000) within 0.2 arcmin of our reference position. The
X-ray source is marginally extended (probability 90\%) and the X-ray
hardness ration shows that the source is about 2$\sigma$ too
soft. Therefore, it is not impossible that the Seyfert galaxy
contributes to the X-ray emission. The cluster is with an X-ray
luminosity of $L_x \sim 4 \cdot 10^{44}$ erg s$^{-1}$ so luminous that
the Seyfert galaxy has to be relatively bright to affect the total flux.

{\bf RXCJ2220.5-3509}, A3866, has an AGN found in our spectroscopic
follow-up 40 arcsec from the X-ray center at
$z=0.0754$. The X-ray emission shows a small but significant extent,
but the core of the emission looks like a point source. The cluster
redshift is measured as $z = 0.1544$ and the optical image shows a nice
cluster with a dominant giant galaxy close to the X-ray maximum. We
expect that the X-ray emission from the cluster is partly contaminated
by the AGN.

{\bf RXCJ2234.5-3744}, A3888, is a cluster in which we also detected a
Seyfert 1 galaxy within 2 arcmin from the center (A. K. Romer,
Ph.D. Thesis). Therefore, there was some concern about the X-ray
contamination from the AGN in this source. An XMM-Newton observation
(P.I. A.C. Edge) allows us to distinguish the cluster and AGN X-ray
emission. We find that the AGN contribution is about 10\% of the total
emission in the 0.5 - 2.0 keV band. 70 coincident galaxy redshifts
confirm the existence of a cluster at this position.

{\bf RXCJ2251.7-3206} is an X-ray source that appears point-like in
RASS with a hardness ratio which is too soft by about $2.9\sigma$
compared to the expectation for cluster emission. It has been found
to be a radio source (Bauer et al. 2000) classified 
as a BL Lac (Schwope et al. 2000).  We have one galaxy redshift 
for this cluster.
The question if it is an X-ray BL Lac or an X-ray
cluster is undecided. This is to be clarified by further
observations.
  
{\bf RXCJ2351.6-2605} contains an AGN within 0.2 arcmin of our
reference position identified by Caccianiga et al. (2000). The X-ray
source is, however, extended with high significance, and the X-ray
hardness ratio is perfectly consistent with thermal emission from a
cluster.  We do not detect a signature of point source
contribution. Therefore, the identification of this X-ray source as a
cluster is safe inspite of the coincidence with the AGN.

For some of the objects in this list, where the central cluster 
galaxy is identified with an AGN because of the observation of emission 
lines, the emission lines could also be associated to the cluster cooling core.
Emission lines with preferentially low excitation energies are frequently
observed in cooling core clusters (e.g. Crawford et al. 1999 and references 
therein). In these cases the dominant X-ray emission comes from the cluster 
and its cooling core. Therefore the observation of emission lines in the 
central cluster galaxies does not cast doubts on the cluster identification
in general.

\subsection{Objects removed from the present catalogue at final inspection}

In the final two observing runs a number of cluster candidates were
observed which had been flagged to be weak
cluster candidates. A large fraction of them turned out to be clusters
at the telescope and were therefore included in the catalogue.  Now,
during this final inspection a few of them turn out to be most
probably optical clusters with a dominant X-ray AGN. In the following
we list these 7 objects which were excluded from our catalogue.

{\bf RXCJ0730.8-6602} looks like a point source. The central galaxy
with a redshift of $z = 0.1063$ could be an AGN; it has a bright core
in the optical image.  While the DSS image shows a trace of a
promising galaxy grouping at the center of the X-ray position, a
deeper R-band CCD did not confirm the presence of a rich enough
cluster.  The Parks 4.85 GHz survey lists a 84 mJy radio source, PMN0730-6602,
(Griffith \& Wright 1993) and the SUMSS survey (Mauch et al. 2003)
as a 81.7 mJy source at 485 MHz.  Therefore we have removed this object from
the final catalogue and classify it tentatively as an X-ray AGN,
possibly a BL Lac, within a galaxy group, where the X-ray emission
comes preferentially from the AGN.

{\bf RXCJ0934.4-1721} is a source with a marginal X-ray extent, but a
hardness ratio consistent with thermal cluster emission
within 1$\sigma$.  There is no striking galaxy overdensity in deeper optical
images and no clear central BCG. A galaxy close to the center with a
redshift of $z = 0.2499$ was identified as a BL Lac candidate by Bauer
et al. (2000) in the ROSAT-NVSS correlation sample. Therefore, this
object is most probably not a cluster and was removed from
our cluster catalogue. 

{\bf RXCJ1046.8-2535} looks like a point source in the RASS and is
confirmed to be a point source in a ROSAT HRI observation. Therefore,
a cluster identification is ruled out for the dominant fraction of the
X-ray emission.  Nevertheless we find an optical cluster at redshift
$z = 0.2426$ with 8 coincident galaxy velocities.  The source is also
listed as a radio source in Condon et al. (1998) and Bauer et
al. (2000). There is an indication in the HRI observation that there
is a faint halo underneath the point source with a flux of at most
5\% of that of the central source, which could be the emission from
the cluster. The flux is, however, more than an order of magnitude
below the REFLEX limit and therefore the cluster was removed from the
REFLEX sample. This object falls most probably into the category of
X-ray AGN in a galaxy group or cluster.

{\bf RXCJ1213.3-2617} is coincident within 0.5 arcmin with a BL Lac
identified by Fischer et al. (1998). Our spetrum of the central galaxy does
not show an AGN or BLLac signature, however, and a CCD image provides 
some indication of a cluster. The X-ray emission is not significantly extended and the
spectral hardness ratio is consistent with thermal cluster emission. 
A short ROSAT HRI exposure (2.7 ksec) shows only a point source, whose
flux corresponds only to about 0.13 PSPC counts s$^{-1}$ compared to
0.244 observed in the survey. There is no signature of further extended emission
in the HRI image. Therefore the most likely interpretation of this X-ray source
is an AGN which has shown a dimming by a factor of 1.8 between the 
two ROSAT observations.

{\bf RXCJ1545.7-2339} appears as point-like in RASS, but has a
reasonable hardness ratio for cluster emission. The archival HRI data
show that more than 90\% of the flux comes from a point
source. Since the total flux of the source is only about twice above
the REFLEX flux limit and since the cluster emission for this object
at a redshift of 0.1205 should well be extended at the HRI resolution,
we remove this source from the REFLEX sample. Bauer et al. (2000) list
this as a radio source without further classification, making an AGN
counterpart likely, but our ESO spectroscopic observations provided no
evidence for an AGN. Two
coincident galaxy redshifts found make it likely that the AGN resides
in a group or cluster.

{\bf RXCJ2040.0-7114}, tentatively identified with the cluster A3701 at
redshift $z = 0.1607$, is an X-ray source that appears point-like in
the RASS with a hardness ratio consistent with cluster emission. An
archival ROSAT HRI observation shows a surface brightness distribution
with a bright point source and a faint, small halo with an upper limit
on the flux contribution of only a few percent. This brings this X-ray
source well below the REFLEX flux limit and we removed this source
from the REFLEX catalogue. An R-band CCD image shows a nice galaxy
cluster with two bright central galaxies, which is no surprise as it
was already classified as a cluster by Abell et al. (1989).  Our
spectroscopic follow-up provides 12 coincident galaxy redshifts
confirming the cluster detection.  Therefore, this object falls most
probably into the category of X-ray luminous AGN in a galaxy cluster.

{\bf RXCJ2041.8-3733}, tentatively identified with the cluster S892
at redshift $z = 0.0997$ (offset $\sim 5$ arcmin), 
is an X-ray source that appears point-like
in the RASS with a hardness ratio which is too soft by about $1.8\sigma$
compared to the expectation for cluster emission.  
An archival ROSAT HRI observation shows only a point source.
Our spectroscopic flollow-up observations show a Seyfert 1 spectrum
for the central galaxy and provide 8 further coincident galaxy refdshifts.
A deep R-band image shows a cluster with an
appearance consistent with the determined redshift around $z = 0.1$
and the central galaxy appears to have a bright core.
At the measured redshift this cluster should appear clearly 
as an extended X-ray source. in the RASS and definitely in the HRI image.
Therefore we identify the main X-ray emission with the AGN and removed this
object from the REFLEX catalogue.

The source is also listed
in the bright RASS - NVSS correlation by Bauer et al. (2000) as an
unclassified radio source and as a cluster in the ROSAT Bright
Survey (Schwope et al.  2000).

\section{Notes on double clusters}

One problem in defining an X-ray cluster catalogue, as well as in comparing
different cluster catalogues, is the identification of double clusters and
of single clusters with substructure. This becomes obvious when we
compare our results to other compilations in the next section.  For
the present catalogue we have taken a very pragmatic approach and made
the distinction on the basis of a visual inspection as to how well the
different parts can be separated. If the X-ray halos are hardly connected 
in the RASS images, given the short exposure, we treated them a separate units,
while different maxima still engulfed by a common X-ray halo where treated as
multiple-maxima clusters. This is a subjective criterion because the distinction
is exposure and distance dependend. Also due to the short exposure in
the RASS only a smaller fraction of the objects in this category will be found 
having enough source counts. Therefore a catalogue with any kind of completeness
is extremely difficult to produce. A more viable approach is possibly a statistical 
characterization of the type used by Schuecker et al. (2001b) to describe
the substructure frequencey in the brighter REFLEX clusters. Therefore
the present compilation is merely pointing out the most obvious examples
of the two kinds, which could also be interesting targets for follow-up
studies.

We distinguish between two classes of objects in the catalogue: 
close cluster pairs and clusters with two pronounced X-ray maxima
in the surface brightness distribution which were treated as single
units. In total we find 10 close cluster pairs (including one triplet
and one quadruplet) at closely concordant redshifts, listed in 
Table~\ref{tab10}. We have not included pairs seen only in projection 
where the redshift of the two components is clearly different. 
Table~\ref{tab11} lists those 14 clusters which feature two or
several distinct X-ray maxima.

Another complication arises in the cluster redshift determination as
quite frequently several redshift groupings are found in the
line-of-sight towards the cluster, indicating that several clusters
and groups of galaxies are seen in projection, or narrow features of
the large-scale structure (like walls and filaments) are threaded by
the line-of-sight. In this case we have assigned the redshift to the
cluster, which is derived from the largest number galaxies (including
the bright central galaxy).
In most cases the assignment is quite obvious, as illustrated by 
Table~\ref{tab12} for the combined data of our follow-up observations 
and the literature search. The table is ordered such that the first component
refers to the chosen cluster identification while the remaining components
are ordered by their redshift.
The table also gives a helpful reference for those cases were
discrepancies may be found in future observations. Most of the
detailed information listed in the table comes from the ESO key
program of the ENACS survey by Katgert et al. (1996) and Mazure et
al. (1996). For each line-of-sight component we list the redshift and
the number of known coincident galaxies.

   \begin{table}
      \caption{List of cluster pairs and close groupings in the REFLEX catalogue}
         \label{tab10}
      \[
         \begin{array}{lrl}
            \hline
            \noalign{\smallskip}
 {\rm name} & {\rm redshift} & {\rm alternative name} \\
            \noalign{\smallskip}
            \hline
            \noalign{\smallskip}
{\rm RXCJ0229.3-3332} & 0.0779 &   \\
{\rm RXCJ0230.7-3305} & 0.0760 & {\rm A3027} \\
           \hline
{\rm RXCJ0542.1-2607} & 0.0380 &   \\
{\rm RXCJ0545.4-2556} & 0.0424 & {\rm A548W} \\
{\rm RXCJ0548.6-2527} & 0.0410 & {\rm A548E} \\
            \hline
{\rm RXCJ0626.3-5341} & 0.0531 & {\rm 3391} \\
{\rm RXCJ0627.2-5428} & 0.0512 & {\rm 3395} \\
           \hline
{\rm RXCJ1254.3-2901} & 0.0553 & {\rm A3528a} \\
{\rm RXCJ1254.6-2913} & 0.0532 & {\rm A3528b} \\
           \hline
{\rm RXCJ1255.5-3019} & 0.0544 & {\rm A3530} \\
{\rm RXCJ1257.2-3022} & 0.0555 & {\rm A3532} \\
           \hline
{\rm RXCJ1327.9-3130} & 0.0482 & {\rm A3558,~Shapley~ center}\\
{\rm RXCJ1329.7-3136} & 0.0495 & {\rm Shapley center}\\
{\rm RXCJ1331.5-3148} & 0.0429 & {\rm Shapley~ center}\\
{\rm RXCJ1333.6-3139} & 0.0502 & {\rm A3562,~Shapley center}\\
           \hline
{\rm RXCJ1347.2-3025} & 0.0141 & {\rm A3574W,~ substructure} \\
{\rm RXCJ1349.3-3018} & 0.0141 & {\rm A3574E,~ main~ cluster} \\
           \hline
{\rm RXCJ2014.2-8038} & 0.1373 & {\rm A3664} \\  
{\rm RXCJ2016.2-8047} & 0.1309 & {\rm A3666} \\ 
           \hline
{\rm RXCJ2145.9-1006} & 0.0808 & {\rm A2377} \\
{\rm RXCJ2147.0-1019} & 0.0780 &  \\
           \hline
{\rm RXCJ2152.2-1942} & 0.0963 & {\rm A2384b} \\
{\rm RXCJ2152.4-1933} & 0.0943 & {\rm A2384a} \\
           \hline
            \noalign{\smallskip}
            \hline
         \end{array}
      \]
   \end{table}
%
%


   \begin{table}
      \caption{List of REFLEX clusters with two or more clearly visible X-ray maxima}
         \label{tab11}
      \[
         \begin{array}{lll}
            \hline
            \noalign{\smallskip}
 {\rm name} & {\rm morphology} & {\rm orientation} \\
            \noalign{\smallskip}
            \hline
            \noalign{\smallskip}
{\rm RXCJ0034.6-0208} & {\rm two~maxima} & {\rm  East-West}   \\
{\rm RXCJ0152.7-0100} & {\rm two~maxima} & {\rm  East-West}   \\
{\rm RXCJ0157.4-0550} & {\rm two~maxima} & {\rm  NE-SW}   \\
{\rm RXCJ0330.0-5235} & {\rm two~maxima} & {\rm  NE-SW}   \\
{\rm RXCJ0624.6-3720} & {\rm two~maxima} & {\rm  East-West}   \\
{\rm RXCJ0948.6-8327} & {\rm two~maxima} & {\rm  East-West}   \\
{\rm RXCJ0956.4-1004} & {\rm three~maxima} & {\rm  NE-W-S~(diffuse)}   \\
{\rm RXCJ1305.9-3739} & {\rm two~maxima} & {\rm  NE-SW}   \\
{\rm RXCJ1330.8-0512} & {\rm large~elongation} & {\rm  towards~NE}   \\
{\rm RXCJ2106.0-3846} & {\rm two~maxima} & {\rm  NE-SW}   \\
{\rm RXCJ2157.4-0747} & {\rm two~maxima} & {\rm  East-West}   \\
{\rm RXCJ2202.0-0949} & {\rm two~maxima} & {\rm  NE-SW}   \\
{\rm RXCJ2218.2-0350} & {\rm two~maxima} & {\rm  NEE-SWW}   \\
{\rm RXCJ2319.2-6750} & {\rm two~maxima} & {\rm  North-South}   \\
           \hline
            \noalign{\smallskip}
            \hline
         \end{array}
      \]
   \end{table}
%
%


\section{Comparison to other catalogues}

We have inspected the previously published catalogues of clusters detected in
the RASS, to further check the completeness of our sample. Of these previous
surveys the RASSB1 (De Grandi et al. 1999), the XBACS (Ebeling et al. 1998), 
the SGP Survey (Cruddace et al. 2002, 2003), and HIFLUGCS
(Reiprich \& B\"ohringer 2002) surveys have made use of the material
that was compiled during the ongoing REFLEX survey. Thus these samples are
not independent. But since we have used a very strict automated selection 
criterion for the primary selection of the cluster candidates and have
not included arbitrarily all we know about clusters in the REFLEX region,
this is a very important completeness test and a test of the galaxy overdensity
detection method applied to the COSMOS data base as described in paper I.

   \begin{table*}
      \caption{Multiple redshift clustering in the line-of-sight of REFLEX clusters}
         \label{tab12}
      \[
         \begin{array}{lllllllll}
            \hline
            \noalign{\smallskip}
{\rm Name}  & {\rm alt. name} &\multicolumn{6}{c}{\rm redshifts}   & {\rm   references} \\
            &                 &{\rm Comp 1}&{\rm Comp 2}&{\rm Comp 3}&{\rm Comp 4}&{\rm Comp 5}&{\rm Comp 6}& \\
            \noalign{\smallskip}
            \hline
            \noalign{\smallskip}
{\rm RXCJ0003.2-3555}& {\rm A2717}      & 0.0490(40)& 0.0720(5)&       &       &       &       & 2\\
{\rm RXCJ0011.3-2851}& {\rm A2734}      & 0.0620(83)& 0.0260(5)& 0.1190(4)& 0.1410(6)&       &       & 2\\
{\rm RXCJ0013.6-1930}& {\rm A0013}      & 0.0940(37)& 0.0270(4)&       &       &       &       & S,2,140,141\\
{\rm RXCJ0017.5-3509}& {\rm A2755}      & 0.0968(23)& 0.1210(10)&       &       &       &       & E,2,3,33,140,141\\
{\rm RXCJ0041.8-0918}& {\rm A0085}      & 0.0555(308)& 0.0762&       &       &       &       & 130,148,149\\
{\rm RXCJ0042.1-2832}& {\rm A2811}      & 0.1082(29)& 0.0540(6)&       &       &       &       & E,33,96\\
{\rm RXCJ0056.3-0112}& {\rm A0119}      & 0.0442(104)& 0.1400(4)&       &       &       &       & 2\\
{\rm RXCJ0108.8-1524}& {\rm A0151A}     & 0.0533(63)& 0.0410(25)& 0.1000(35)&       &       &       & 1,2\\
{\rm RXCJ0110.0-4555}& {\rm A2877}      & 0.0238(58)& 0.2470(97)&       &       &       &       & 12,130,131,146\\
{\rm RXCJ0115.2+0019}& {\rm A0168}      & 0.0450(76)& 0.0176(4)& 0.0720(4)& 0.0890(7)&       &       & S,2\\
{\rm RXCJ0137.2-0912}&                  & 0.0409(5)& 0.0700&       &       &       &       & S,120\\
{\rm RXCJ0152.7+0100}& {\rm A0267}      & 0.2300(1)& 0.0592(8)&       &       &       &       & 55,147,159\\
{\rm RXCJ0152.9-1345}& {\rm NGC0720}    & 0.0050(3)& 0.8348(6)&       &       &       &       & 12,160,161\\
{\rm RXCJ0157.4-0550}& {\rm A0281}      & 0.1289(4)& 0.088    &       &       &       &       & E,1\\
{\rm RXCJ0202.3-0107}& {\rm A0295}      & 0.0427(47)& 0.1020(5)&       &       &       &       & 2,148,162\\
{\rm RXCJ0231.9+0114}& {\rm RCS145^d}   & 0.0221(10)& 0.2881&       &       &       &       & 55,64,110\\
{\rm RXCJ0337.0-3949}& {\rm A3142}      & 0.1030(21)& 0.0660(12)&       &       &       &       & 2\\
{\rm RXCJ0338.4-3526}& {\rm FORNAX}     & 0.0051(32)& 0.1124(48)&       &       &       &       & 166,167,168,170,171\\
{\rm RXCJ0342.8-5338}& {\rm A3158}      & 0.0590(105)& 0.0740(4)& 0.1020(4)&       &       &       & 2\\
{\rm RXCJ0408.2-3053}& {\rm A3223}      & 0.0600(81)& 0.1100(8)& 0.1370(8)&       &       &       & E,2\\
{\rm RXCJ0448.2-2028}& {\rm A0514}      & 0.0720(90)& 0.0850(4)& 0.1100(8)&       &       &       & 2\\
{\rm RXCJ0525.5-3135}& {\rm A3341}      & 0.0380(64)& 0.0780(15)& 0.1150(18)& 0.1310(7)& 0.1540(5)&       & 2\\
{\rm RXCJ0525.8-4715}& {\rm A3343}      & 0.1913(5)& 0.1626(6)&       &       &       &       & E\\
{\rm RXCJ0530.6-2226}& {\rm A0543}      & 0.1706(11)& 0.0850(10)&       &       &       &       & E,1\\
{\rm RXCJ0542.1-2607}& {\rm CID 36}     & 0.0390(4)& 0.0292(2)& 0.0429(69)&       &       &       & E,120\\
{\rm RXCJ0548.6-2527}& {\rm A0548E}     & 0.0420(237)& 0.0310(4)& 0.0630(9)& 0.0870(14)& 0.1010(21)& 0.1380(4)& 2,3\\
{\rm RXCJ0637.3-4828}& {\rm A3399}      & 0.2026(11)& 0.1180(3)& 0.3796(2)&       &       &       & E\\
{\rm RXCJ0645.4-5413}& {\rm A3404}      & 0.1644(2)& 0.3377(2)&       &       &       &       & E\\
{\rm RXCJ0658.5-5556}& {\rm 1ES0657^h}& 0.2965(78)& 0.0790(3)&       &       &       &       & E,73,79\\
{\rm RXCJ0944.1-2116}&                  & 0.0077(1)& 0.0152(1)&       &       &       &       & E,12,84\\
{\rm RXCJ1039.7-0841}& {\rm A1069}      & 0.0650(35)& 0.1140(4)&       &       &       &       & 2\\
{\rm RXCJ1050.4-1250}& {\rm USGC S152}  & 0.0155(6)& 0.0760(5)&       &       &       &       & E,20,31\\
{\rm RXCJ1141.4-1216}& {\rm A1348}      & 0.1195(6)& 0.1392&       &       &       &       & H,16\\
{\rm RXCJ1254.3-2901}& {\rm A3528(A)}   & 0.0542(69)& 0.0730(9)&       &       &       &       & 2\\
{\rm RXCJ1327.9-3130}& {\rm A3558}      & 0.0480(341)& 0.0320(4)& 0.0771(6)& 0.1285(7)&       &       & 1,14,44,102,141,148\\
{\rm RXCJ1329.7-3136}& {\rm A3558(B)}   & 0.0488(57)& 0.1823(4)&       &       &       &       & 3,14,44,63,101,102\\
{\rm RXCJ1330.8-0152}& {\rm A1750}      & 0.0852(46)& 0.1492(3)&       &       &       &       & 1,41,96\\
{\rm RXCJ1333.6-3139}& {\rm A3562}      & 0.0490(114)& 0.0367(4)&       &       &       &       & 1,141\\
{\rm RXCJ1347.5-1144}&                  & 0.4516(2)& 0.2090(3)&       &       &       &       & E\\
{\rm RXCJ1401.6-1107}& {\rm A1837}      & 0.0698(38)& 0.0372(14)&       &       &       &       & 1\\
{\rm RXCJ1952.2-5503}& {\rm A3651}      & 0.0600(79)& 0.1010(5)&       &       &       &       & 2\\
{\rm RXCJ2012.5-5649}& {\rm A3667}      & 0.0556(162)& 0.0350(2)& 0.0990(5)&       &       &       & 2,3,138,140,141\\
{\rm RXCJ2032.1-5627}& {\rm A3685}      & 0.1380(5)& 0.2852(6)&       &       &       &       & E\\
{\rm RXCJ2034.3-3429}& {\rm A3693}      & 0.1240(6)& 0.0910(16)&       &       &       &       & E,3\\
{\rm RXCJ2034.7-3548}& {\rm A3695}      & 0.0894(81)& 0.0934(18)& 0.1310(7)&       &       &       & 2,140,141\\
{\rm RXCJ2101.8-2802}& {\rm A3733}      & 0.0382(91)& 0.0739(4)&       &       &       &       & 13,14,146\\
{\rm RXCJ2104.3-4120}& {\rm A3739}      & 0.1651(2)& 0.0820(16)&       &       &       &       & E\\
{\rm RXCJ2107.2-2526}& {\rm A3744}      & 0.0381(71)& 0.0650(5)&       &       &       &       & 2\\
{\rm RXCJ2146.3-5717}& {\rm A3806}      & 0.0760(99)& 0.0540(9)& 0.1380(4)&       &       &       & 2\\
{\rm RXCJ2146.9-4354}& {\rm A3809}      & 0.0620(94)& 0.0910(4)& 0.1100(10)& 0.1410(11)& 0.1520(4)&       & 2\\
{\rm RXCJ2154.1-5751}& {\rm A3822}      & 0.0760(84)& 0.0390(4)& 0.0520(4)& 0.1020(4)&       &       & 2\\
{\rm RXCJ2158.3-2006}& {\rm A2401}      & 0.0570(23)& 0.0884(2)& 0.0930(5)&       &       &       & 2,3\\
{\rm RXCJ2158.4-6023}& {\rm A3825}      & 0.0750(61)& 0.1040(17)& 0.1190(4)&       &       &       & 2,3\\
{\rm RXCJ2234.5-3744}& {\rm A3888}      & 0.1510(70)& 0.2077(7)&       &       &       &       & 1,44\\
{\rm RXCJ2249.9-6425}& {\rm A3921}      & 0.0940(32)& 0.1340(4)&       &       &       &       & E,3\\
{\rm RXCJ2312.3-2130}& {\rm A2554}      & 0.1108(35)& 0.0707(5)&       &       &       &       & 38,127\\
{\rm RXCJ2321.5-4153}& {\rm A3998}      & 0.0894(16)& 0.0665(3)&       &       &       &       & S,16,31,33,48\\
{\rm RXCJ2336.2-3136}& {\rm S1136}      & 0.0643(2)& 0.0260(3)&       &       &       &       & 15,52,26\\
{\rm RXCJ2354.2-1024}& {\rm A2670}      & 0.0765(219)& 0.1506(13)&       &       &       &       & 66,86,105\\
           \hline
            \noalign{\smallskip}
            \hline
         \end{array}
      \]
\begin{list}{}{}
\item[The footnotes in the alternative name column (2) are explained in Table~\ref{tab6}]
\end{list}
   \end{table*}
%
%


The XBACS catalogue lists 5 clusters that are not included in REFLEX
after four clusters are removed from the correlation (A3186, A3216,
A3230, A3389), because they are located in the Magellanic cloud regions
that have been excised in the present sample.  For these 5 objects we
have a definite reason for the exclusion.  A3041: the REFLEX flux is 
$2.58~ 10^{-12}$ erg s$^{-1}$ cm$^{-2}$ and below the current flux 
limit; A467: is characterized by a point source which
is by about $5\sigma$ too soft to be consistent with cluster emission;
A3662: the REFLEX flux is $1.67~ 10^{-12}$ erg s$^{-1}$ cm$^{-2}$,
below the flux limit of REFLEX, but there is a $\sim
40 \%$ uncertainty in the flux due a low exposure of only 122 sec in
RASS II; A3701: (RXCJ2040.0-7114) was removed from our cluster list as explained in
section 7; A3716: has no exposure in RASS2 (This is implicitly accounted for in
the construction of the sky coverage map). Thus no cluster has been missed
in REFLEX which is included in XBACS. There are further some
nomenclature problems that are encountered in this comparison: A189 in
XBCAS is a misclassification since the Abell cluster is about half a
degree away and the correct identification should be NGC 533 as in
REFLEX, A1664 is identical with A3541, A3017 in XBACS is identified
with A3016 in REFLEX. Three clusters are listed with two separate
components in XBACS which are treated as one source in the present 
compilation: A901, A1631, and A1750.

A comparison with the SGP sample by Cruddace et al. (2002, 2003) which
is based on the same X-ray source catalogue and source
characterization as REFLEX but used a different optical search method,
shows only two additional clusters with a recently determined flux above the
REFLEX flux limit: RXCJ2356.0-0129 which is here
treated as a double source where each component falls below the flux
limit, and RXCJ0016.3-3121 which was for the REFLEX catalogue
deblended from a soft contaminating component and therefore fell below
the flux limit after this treatment. 
Six clusters listed in the SGP sample catalogue with a flux
above the REFLEX flux limit have slightly lower fluxes in REFLEX
and are scattered below the REFLEX flux limit:
RXCJ2306.8-1324, RXCJ0048.6-2114, RXCJ0108.5-4021, RXCJ0212.8-4707, 
RXCJ0244.1-2611, RXCJ0248.2-0216. They will be included in REFLEX II.
The following 9 clusters in the SGP region are only listed 
in the REFLEX catalogue: RXCJ0034.6-0208,
RXCJ0043.4-2037, RXCJ0132.6-0804, RXCJ0250.2-2129, RXCJ0301.6+0155,
RXCJ2211.7-0350, RXCJ2248.5-1606, RXCJ2251.7-3206,
RXCJ2306.6-1319. Note that already four of these nine objects have a
problematic identification, as discussed in section 7. Therefore the
missing of these clusters is to a large part not a real incompleteness
but a classification problem.

The southern RASS Bright sample, compiled from an earlier version of
the ESO key program X-ray cluster identification list based on the
RASS1 data set (De Grandi et al. 1999), contains two additional
clusters that should be included in REFLEX. One was again found in
our supplementary search for clusters among the sources found to be
extended by the GCA analysis, RXCJ2129.6+0005 at $z = 0.2347$. 
This source and RXJ0600.5-4846 were not flagged by the optical correlation.
Since these objects were not automatically included through the correlation with the
COSMOS data we have not included them in the present list. RXCJ0528.9-3927
listed by De Grandi et al. is removed from the present list as
explained in section 7. Two further objects from this catalogue have been 
removed from our sample in earlier steps of the identification
RXJ2136.4-6224 is a point-like X-ray source coincident with a 
Seyfert 1 galaxy at $z = 0.0588$ and RXJ2253.9-5812 is a point source with a
soft spectrum coincident with the radio source PMNJ2253-5812.   
All the objects listed in the HIFLUGCS sample by Reiprich \&
B\"ohringer (2002) in the REFLEX survey area are included in the 
present catalogue.

\section{Discussion and conclusions}

More than 100 000 X-ray sources were identified in the ROSAT All-Sky Survey
(Voges et al. 1999) and more than 10\% of the X-ray sources in the sky away
from the Galactic plane are expected to be galaxy clusters for the relevant range
of flux limits. Thus taking into account the surely lower detection 
efficiency for extended cluster X-ray sources and the Galactic source population
one can easily expect about 6000 - 8000 clusters among the X-ray sources in the
bright and faint ROSAT All-Sky Survey catalogue. The REFLEX cluster sample with a flux 
limit of $3 \cdot 10^{-12}$ erg s$^{-1}$ cm$^{-2}$ is therefore only the tip of
the iceberg of the RASS cluster population. However, the REFLEX sample
is constructed from a selection with a relatively high flux limit, in order to yield
a sample of high quality. The median number of detected source photons for the REFLEX clusters 
is 79 photons. Relatively safe source detections, e.g. as listed in the RASS faint
source catalogue, can still be obtained from a detection of only 6 photons in a
detection aperture with 2 arcmin radius, which for the typical exposure of about 400 sec
and a typical background in the hard band (channel 52 - 201) of about $3\cdot 10^{-4}$
cts s$^{-1}$ arcmin$^{-2}$ would contain about 1.5 background photons. Thus a
detection of 6 photons corresponds roughly to a 3.7$\sigma$ background enhancement
and roughly to a flux limit of $3 \cdot 10^{-13}$ erg s$^{-1}$ cm$^{-2}$ - a flux
limit one order of magnitude below the REFLEX cut. Thus the RASS provides the prospect of finding
many more clusters than presented here, but the price is a much lower quality of
the X-ray characteristics as well as a much more difficult job for the definitive source
identification. \footnote{Already the number of spurious sources is expected to be of the order
of 1000 in the REFLEX area for the mentioned low flux limit. This is in comparison to a few Thousand 
clusters to be found in the same area not unreasonable. But good extra information as
for example optical or other wavelength data is mandatory for an identification in this case.} 

With the relatively high flux cut the high quality of the REFLEX sample is characterized
by the following most important properties: (i) a flux determination with a typical accuracy 
of 10 - 20\%, (ii) a large fraction of the cluster X-ray sources can be characterized as extended
($\sim 80\%$) which adds very much to a safe identification, (iii) we can in general obtain
a meaningful spectral hardness ratio which allows further discrimination in our identification,
(iv) we arrive at a selection function which is almost homogeneous across the sky. 
At an even higher flux limit,
e.g. $6 \cdot 10^{-12}$ erg s$^{-1}$ cm$^{-2}$, only 9 clusters are not flagged as extended 
X-ray sources by our analysis. The spatially very homogeneous selection function ensures for example 
that, for a detection limit of 10 photons used in
most of our spatial distribution analysis, the flux limit is reached in 97\% of the sky and
only small corrections apply for the remaining region. The selection function
is also characterized by the analysis
of the source photon count distribution shown in Fig. 25 of paper I, from which we conclude
that only about 14 clusters might be missed if the correction for the lower sensitivity in
some regions of the REFLEX survey is neglected. Therefore the cosmological results derived
by us in the series of REFLEX papers could also be reproduced in good approximation if the angular 
modulation of the selection function given in Tables~\ref{tab8} and~\ref{tab9} were neglected.

Since the completion of the optical follow-up observations for REFLEX we have already embarked
on the extension of this survey, REFLEX II which is now close to complete down to a flux limit
of $1.8~ 10^{-12}$ erg s$^{-1}$ cm$^{-2}$ including more than 800
clusters. We stress that in addition to the fact
that the X-ray parameters become less accurate at these lower fluxes, the identification
work becomes significantly harder: the optical counter parts are less striking on for example the
optical DSS images, it is more difficult to detect and catch the signature of contamination
AGN, and the spatial modulation of the selection function can no longer be neglected.
Therefore the gain in the statistics by the increase of the sample
size has a price and there is some optimum quality and size of the cluster sample, 
depending of course on the application, for which the flux limit should not be much lower than that
of REFLEX I.

Further work on the identification of RASS clusters is useful if the RASS data can be combined
with survey information from other wavelengths. A good example is the Sloan Digital Sky Survey.
Based on the parent SDSS cluster sample from Sheldon et al. (2001) we have demonstrated
that we can almost recover the REFLEX X-ray luminosity function for clusters down to a flux limit
in the RASS of $8 \cdot 10^{-13}$ erg s$^{-1}$ cm$^{-2}$. Work in progress on the combined
detection of galaxy clusters in X-rays (RASS) and the optical (SDSS) extends these detections
to even lower fluxes (Schuecker et al., 2003c) recovering an arial density
of about 0.5 clusters deg$^{-2}$.     
  
The high quality of the REFLEX sample makes the present catalogue a
useful basis for the careful selection of targets for detailed
follow-up studies at all wavelengths. A deep XMM-Newton follow-up
study has been targeted at a complete sample of 13 X-ray luminous REFLEX
clusters ($L_x \ge 1\cdot 10^{45}$ erg s$^{-1}$) in the redshift
interval 0.27 to 0.31 (Zhang et al. 2003) to study for example
the evolution of the cluster
population at the high mass end, from $z \sim 0.3$ to the present. One
first result of this study is that all 13 clusters of the
sample show X-ray emission which is dominated by thermal, diffuse
intracluster medium radiation and only two clusters have a point
source contribution of at most 30\%. This provides a very nice
confirmation for the REFLEX cluster sample, since at this upper end of
the redshift distribution of the REFLEX clusters it is already quite
difficult to recognize the point source contamination of the cluster
emission. The same clusters form also the subject of a very detailed
optical study with the VLT in the frame of a Large Program to 
study among other objectives the galaxy 
dynamics, the characteristics of the galaxy population, and the supernova rate.
Recently a Large XMM-Newton follow-up program has been approved, to study in
detail 33 clusters from REFLEX selected in such a way that they form a volume 
complete sample (with a complex but known volume structure), that they cover the
X-ray luminosity range almost homogeneously, and that they are optimally observed with
the field-of-view of the XMM-Newton instruments. The main goal of this survey is
to understand the statistics of cluster structure and the scaling relations of
observable and physical parameters, like the very important X-ray luminosity-mass
and temperature-mass relations. These relations form the backbone of the 
cosmological application of cluster samples like REFLEX.

Further applications of the REFLEX catalogue are in progress involving the study of radio
haloes, Sunyaev-Zeldovich observations, correlations with microwave background data
(e.g. WMAP), and gravitational lensing studies. The cosmological applications 
of the REFLEX sample demonstrates the power of using clusters of galaxies as
cosmological probes in a way complementary to studies based on supernovae and
on the microwave background. Therefore we hope that in a next step a new X-ray all-sky
survey which goes much deeper, covers a larger X-ray band width, and provides a better
angular resolution will enable us with much higher precision to test cosmological 
models and determine cosmological parameters. There is a constant effort in the X-ray community
to make such a project and such an important progress possible.
        
\begin{acknowledgements}                                                        
We like to thank the ROSAT team at MPE for the support with the
data reduction of the ROSAT All-Sky Survey and the staff of ESO La
Silla for the technical support during the numerous observing runs
for the ESO key programme. We also thank G. Vettolani, W.C. Seitter,
K.A. Romer, U.G. Briel, H. Ebeling, R. D\"ummler, T.H. Reiprich, R.A. Schwarz,
S. Molendi, H. Gursky, and D. Yentis for help during the programme.
We also thank the referee for very useful comments.

The production of the COSMOS digital optical data base, based on scans
of the UK Schmidt southern sky survey IIIa-J plates, and the development
of techniques for detecting cluster candidates in the COSMOS data, was a major
project, requiring the coordinated efforts of Harvey MacGillivray (ROE),
Daryl Yentis and Brad Stuart (NRL), and John Wallin (George Mason University).       
 
This research also made use of the
NASA/IPAC Extragalactic Database (NED), which is operated by
the Jet Propulsion Laboratory, California Institute of Technology,
under contract with NASA. P.S. acknowledges the support by
the Verbundforschung under grant No. 50 OR 93065.

\end{acknowledgements}

\end{document}